\catcode`\@=11

\newif\if@fewtab\@fewtabtrue

{\count255=\time\divide\count255 by 60
\xdef\hourmin{\number\count255}
\multiply\count255 by-60\advance\count255 by\time
\xdef\hourmin{\hourmin:\ifnum\count255<10 0\fi\the\count255}}
\def\ps@draft{\let\@mkboth\@gobbletwo
    \def\@oddhead{}
    \def\@oddfoot
       {\hbox to 7 cm{$\scriptstyle Draft\ version:\ \draftdate$
       \hfil}\hskip -7cm\hfil\rm\thepage \hfil}
    \def\@evenhead{}\let\@evenfoot\@oddfoot}

\def\draftcite#1{\ifnum\draftcontrol=1#1\else{}\fi}

\def\@lbibitem[#1]#2{\item{}\hskip -3cm \hbox to 2cm
{\hfil$\scriptstyle\draftcite{#2}$}\hskip
1cm[\@biblabel{#1}]\if@filesw
     {\def\protect##1{\string ##1\space}\immediate
      \write\@auxout{\string\bibcite{#2}{#1}}}\fi\ignorespaces}

\def\@bibitem#1{\item\hskip -3cm \hbox to 2cm
{\hfil $\scriptstyle\draftcite{#1}$}\hskip 1cm
\if@filesw \immediate\write\@auxout
       {\string\bibcite{#1}{\the\value{\@listctr}}}\fi\ignorespaces}

\def\nsection#1{\section{#1}\setcounter{equation}{0}}

\font\tendl=msbm10  scaled \magstep1
\font\sevendl=msbm7 scaled \magstep1
\font\fivedl=msbm5 scaled \magstep1
\font\tengl=eufm10  scaled \magstep1
\font\sevengl=eufm7 scaled \magstep1
\font\fivegl=eufm5 scaled \magstep1

\newfam\dlfam  
\textfont\dlfam=\tendl \scriptfont\dlfam=\sevendl
\scriptscriptfont\dlfam=\fivedl
\newfam\glfam  
\textfont\glfam=\tengl \scriptfont\glfam=\sevengl
\scriptscriptfont\glfam=\fivegl

\def\draftdate{\number\month/\number\day/\number\year\ \ \ \hourmin }

\global\def\draftcontrol{0}
\catcode`\@=12
\documentclass[aps,superscriptaddress,floatfix]{revtex4}
\usepackage{epsfig}
\usepackage{rotating}
\usepackage[all]{xy}
\usepackage[centerlast]{subfigure}
\usepackage{bm}
\usepackage{amssymb}
\usepackage{amsfonts}
\usepackage{amsthm}
\usepackage{amsmath}
\renewcommand{\theequation}{\arabic{section}.\arabic{equation}}

\newcommand{\be}{\begin{eqnarray}}
\newcommand{\en}{\end{eqnarray}\vs 0.5 cm}

\newcommand{\Hol}{{H\hspace{-0.04cm}ol}}
\newcommand{\Grb}{{G\hspace{-0.03cm}rb}}
\newcommand{\Id}{{I\hspace{-0.04cm}d}}

\newcommand{\no}{\noindent}
\newcommand{\vs}{\vskip}

\newcommand{\si}{{\dot{\iota}}}
\newcommand{\Ng}{{\bf g}}
\newcommand{\Nh}{{\bf h}}

\newcommand{\Nt}{{\bf t}}

\newcommand{\NC}{{{\bf C}}}
\newcommand{\NZ}{{{\mathbb Z}}}

\newcommand{\qq}{\begin{eqnarray}}

\newcommand{\ee}{{\rm e}}

\newcommand{\qqq}{\end{eqnarray}}

\newcommand{\tr}{{\rm{tr}}}
\newcommand{\ad}{\hbox{ad}}

\newcommand{\CA}{{\mathcal A}}
\newcommand{\Ch}{{h}}
\newcommand{\Cl}{{\lambda}}

\newcommand{\CF}{{\cal F}}
\newcommand{\CG}{{\cal G}}
\newcommand{\CH}{{\cal H}}
\newcommand{\CI}{{\cal I}}
\newcommand{\CJ}{{\cal J}}
\newcommand{\CK}{{\cal K}}
\newcommand{\CL}{{\cal L}}

\newcommand{\CO}{{\cal O}}
\newcommand{\CP}{{\cal P}}

\newcommand{\CT}{{\cal T}}

\newcommand{\CX}{{\cal X}}

\newcommand{\CZ}{{\cal Z}}

\newcommand{\m}{\hspace{0.025cm}}

\newcommand{\ddt}{{\frac{_d}{^{dt}}}}

\newcommand{\ov}{\overline}

\newcommand{\WZ}{{W\hspace{-0.03cm}Z}}
\def\des#1{D\!es(#1)}

\def\eee#1{\ee^{#1}}
\def\eee#1{\exp\hspace{-0.04cm}\Big( #1\Big)}
\def\limits{}
\def\to{\!\xymatrix@C=0.4cm{\ar[r] &}}
\def\Rightarrow{\!\xymatrix@C=0.4cm{\ar@{=>}[r] &}}
\def\mapsto{\!\xymatrix@C=0.4cm{\ar@{|->}[r] &}\!}
\def\rightarrow{\to}
\def\longrightarrow{\!\xymatrix@C=0.8cm{\ar[r] &}}
\def\longmapsto{\!\xymatrix@C=0.8cm{\ar@{|->}[r] &}}
\def\mapstolabel#1{\!\xymatrix@C=0.8cm{\ar@{|->}[r]^-{#1} &}}
\def\tolabel#1#2{\!\xymatrix@C=#1{\ar[r]^-{#2} &}}
\def\bigddt{\left.{\frac{d}{dt}}\right|_{t=0}}
\def\smallddt{\left.{\textstyle\frac{d}{dt}}\right|_{t=0}}
\def\defn{:=}
\def\quand{\quad\text{ and }\quad}

\def\grb#1{\Grb^\nabla\hspace{-0.05cm}(#1)}
\def\Ggrb#1#2{\grb#1^#2}
\def\Ggrbz#1#2{\Ggrb#1#2_0}
\newcounter{mythm}[section]
\def\themythm{\arabic{section}.\arabic{mythm}}
\newtheorem{theorem}[mythm]{Theorem}
\newtheorem{definition}[mythm]{Definition}
\newtheorem{rem}[mythm]{Remark}
\newtheorem{corollary}[mythm]{Corollary}
\newtheorem{exa}[mythm]{Example}
\newtheorem{proposition}[mythm]{Proposition}
\newtheorem{lemma}[mythm]{Lemma}
\newenvironment{remark}{\begin{rem}\normalfont}{\end{rem}}

\renewenvironment{proof}{\noindent{\bf Proof}.\ }{\\ \vskip -0.5cm\hspace*{13cm}{$\blacksquare$}}

\usepackage{amstext}
\usepackage{color}

\begin{document}


\ 
\vskip -0.9cm

\title{{\Large\bf{\ \\ Global gauge anomalies in two-dimensional bosonic
sigma models\\ \ }}}
\author{Krzysztof Gaw\c{e}dzki}
\affiliation{Laboratoire de Physique, C.N.R.S., ENS-Lyon,
Universit\'e de Lyon, 46 All\'ee d'Italie, 69364 Lyon, France}
\author{Rafa\l\ R. Suszek}
\affiliation{Department Mathematik, Bereich Algebra und Zahlentheorie,
Universit\"at Hamburg, Bundesstra\ss e 55, 20146 Hamburg, Germany}
\author{Konrad Waldorf}
\affiliation{Department of Mathematics, University of California,
Berkeley, 970 Evans Hall \#3840, Berkeley, CA 94720, USA\\ \ \\ \ }
\medskip
\vskip 0.5cm

\begin{abstract}
\vskip -0.3cm
\centerline{\Large$^{\rm\bf Abstract}$}
\vskip 0.3cm

\no We revisit the gauging of rigid symmetries in two-dimensional
bosonic sigma models with a Wess-Zumino term in the action. Such a
term is related to a background closed 3-form $\,H\,$ on the target
space. More exactly, the sigma-model Feynman amplitudes of classical
fields are associated to a {\it bundle gerbe with connection} of
curvature $\,H\,$ over the target space. Under conditions that were
unraveled more than twenty years ago, the classical amplitudes may
be coupled to the topologically trivial gauge fields of the symmetry
group in a way which assures infinitesimal gauge invariance. We show
that the resulting gauged Wess-Zumino amplitudes may, nevertheless,
exhibit global gauge anomalies that we fully classify. The general
results are illustrated on the example of the WZW and the coset
models of conformal field theory. The latter are shown to be
inconsistent in the presence of global anomalies. We introduce a
notion of {\it equivariant gerbes} that allow an anomaly-free
coupling of the Wess-Zumino amplitudes to all gauge fields,
including the ones in non-trivial principal bundles. The
obstructions to the existence of equivariant gerbes and their
classification are discussed. The choice of different equivariant
structures on the same bundle gerbe gives rise to a new type of
discrete-torsion ambiguities in the gauged amplitudes. An explicit
construction of gerbes equivariant with respect to the adjoint
symmetries over compact simply connected simple Lie groups is given.
\end{abstract}

\maketitle
\vskip -1.1cm
\

\tableofcontents

\nsection{Introduction}
\vskip -0.25cm
\no {\it Gauge invariance} constitutes one of the basic principles
underlying the theoretical description of physical reality. The
occurrence of its violations, called {\it gauge anomalies}
\cite{Bertlmann}, in certain models of quantum field theory with
chiral fermions yields a powerful selection principle for the model
building in high energy physics \cite{Weinb}.
Gauge anomalies may describe violations of infinitesimal gauge
invariance, or, if the latter holds, the breakdown of invariance under
{\it large} gauge transformations not homotopic to identity
\cite{WittenGA}. The second type goes under the name of {\it global gauge
anomalies}. Anomalies similar to the ones in theories with chiral fermions
occur also in effective bosonic models describing the low energy sector
\cite{tHooft}. Such effective theories contain Wess-Zumino (WZ) terms in
the action \cite{WessZumino}, see, e.g., the review \cite{Petersen}. The
emergence of global gauge anomalies in bosonic theories with WZ terms
on the Euclidian space-time compactified to the four-dimensional sphere
was extensively analyzed following the work \cite{WittenGA}, see
\cite{Fabbrichesi}.

Starting with Witten's paper \cite{Witt} on non-Abelian bosonization,
the two-dimensional Wess-Zumino actions for bosonic sigma models with
Lie-group targets were studied quite thoroughly in the context of the
Wess-Zumino-Witten (WZW) models of conformal field theory (CFT). In the
latter setting, the problem how to gauge rigid symmetries was solved,
at least in the simplest cases, almost from the very start \cite{DVDP}.
Nevertheless, the general question about the coupling of two-dimensional
Wess-Zumino actions to gauge fields in a way invariant under infinitesimal
gauge transformations was posed and answered only a few years later in
\cite{JJMO} and in \cite{HS}. Besides, this was done only for
topologically trivial gauge fields described by global 1-forms on
the worldsheet. The conditions that permit such gauging and the
obstructions to their fulfillment were subsequently interpreted in
\cite{FS,FS1} in terms of equivariant cohomology, as first indicated
in \cite{WittenHF}, see also \cite{Wu}. The issue of general gauge
invariance of gauged two-dimensional WZ actions was addressed
only very briefly at the end of \cite{FS} and, in the context of the
$T$-duality, in \cite{Hull,Hull1}. We make it the main topic of the
present study.

A convenient tool to treat topological intricacies of Wess-Zumino
actions \cite{Alva,Gaw} on closed two-dimensional worldsheets is provided
by the theory of bundle gerbes with connection \cite{Murr,MurrS}. For
topologically trivial gauge fields, we identify the global gauge anomalies
of gauged WZ actions as the isomorphism classes of certain flat gerbes
over the product of the symmetry group $\,\Gamma\,$ and the target space
$\,M$. \,Such isomorphism classes correspond to the classes in the
cohomology group $\,H^2(\Gamma\times M,U(1))\,$ that may often be
calculated explicitly. \,In particular, we show how to do it in the
case of WZW models. This permits us to prove that, after the
gauging of an adjoint symmetry, some of bulk WZW models with
non-simply connected target groups exhibit global gauge anomalies.
The latter lead to the inconsistency of the corresponding coset models
of CFT \cite{GKO,Godd} realized as gauged WZW models with the gauge
fields integrated out \cite{BRS,GK0,GK,KPSY}. This is the main
surprise resulting from our study.

We also address the problem of the coupling of WZ actions to topologically
non-trivial gauge fields given by connections in non-trivial principal
bundles of the symmetry group. It was indicated in \cite{Hori} that
such a coupling plays an important role in the construction of consistent
coset theories. It seems also important in the $T$-duality \cite{Hull}.
We show that the existence of certain equivariant structures on gerbes,
considered already before for discrete symmetry groups in \cite{GSW1},
enables a non-anomalous coupling to all gauge fields and we analyze
in a cohomological language the obstructions to the existence
of such structures and their classification. Different choices of the
equivariant structure lead to gauge amplitudes that differ by phases
given by characters of the fundamental group of the (connected) symmetry
group. The appearance of such discrete-torsion like phases
in the sector with topologically non-trivial gauge fields
was envisaged in \cite{Hori}. We give an explicit construction of all
non-equivalent equivariant structures relative to the adjoint symmetries
on gerbes relevant for the WZW models with compact simply connected target
groups.

The paper is organized as follows. \,In Sec.\,\ref{sec:WZWampl}, \,we
recall the role of bundle gerbes in the definition of the Feynman
amplitudes of two-dimensional sigma models with a WZ action (in Sec.
\ref{sec:2DWZW}) and we characterize rigid symmetries of such amplitudes
(in Sec.\,\ref{sec:rigid}). \,Sec.\,\ref{sec:trivialgaugefields} is
devoted to the coupling of WZ actions to topologically trivial
gauge fields. In Sec.\,\ref{sec:JJMO-HS}, we recall the old result of
Jack-Jones-Mohammedi-Osborn \cite{JJMO} and Hull-Spence \cite{HS}
describing the coupling of a WZ action to the gauge fields of its
symmetry group. In Sec.\,\ref{sec:eqcoh}, we review the interpretation,
due to Witten \cite{WittenHF} and Figueroa-O'Farrill-Stanciu \cite{FS,FS1},
of the conditions that permit such gauging in terms of the Cartan model
of equivariant cohomology, and, in Sec.\,\ref{sec:lift}, we study further
implications of those conditions. Sec.\,\ref{sec:GGAnomal}.
is devoted to global gauge anomalies in theories with a WZ action coupled
to topologically trivial gauge fields. Sec.\,\ref{sec:gengt}
derives the transformation law of the Feynman amplitudes under general
gauge transformations and identifies, in cohomological terms, the
obstruction to the invariance of the amplitudes under {\it large} gauge
transformations not homotopic to
identity. The general discussion is illustrated in
Sec.\,\ref{sec:expl1} by the example of WZW models with non-simply
connected target groups and gauged adjoint symmetry. In
Sec.\,\ref{sec:WZWextern},
we show that our results are consistent with the known solution for
the partition functions of WZW models and in Sec.\,\ref{sec:coset}, we
examine the toroidal partition functions of the coset models in the
presence of global
anomalies. \,Sec.\,\ref{sec:generalgaugefields} is devoted to the
coupling of WZ actions to topologically non-trivial gauge fields. In
Sec.\,\ref{sec:equivgerbes}, we define gerbes with equivariant
structure. In Sec.\,\ref{sec:topntr}, we describe how to use such
structures to define WZ amplitudes coupled to gauge fields with
arbitrary topology. The general gauge invariance of such amplitudes
is proven in Sec.\,\ref{sec:GGInv}. \,In Sec.\,\ref{sec:obstrclass},
we study subsequently the obstructions to the existence of the three
layers of an equivariant structure on gerbes (in
Secs.\,\ref{sec:alphaobstr}, \ref{sec:betaobstr} and \ref{sec:diagrobstr}).
We use the local-data description of gerbes that is recalled in
Sec.\,\ref{sec:localdata}. The classification of equivariant gerbes
is discussed in Sec.\,\ref{sec:classif}. Sec.\,\ref{sec:ambig} examines
the change of the WZ amplitudes induced by a change of the
equivariant structure of the gerbe. \,Next Sec.\,\ref{sec:expl2}
contains an explicit construction of equivariant structures relative
to the adjoint symmetry on gerbes relevant for the WZW models with
compact simple and simply connected target groups. In
Sec.\,\ref{sec:simpleconn}, we recall the construction of the
corresponding gerbes over the target groups and in
Secs.\,\ref{sec:alpha}, \ref{sec:beta} and \ref{sec:commut}, we
build the different layers of the equivariant structure. \,Finally,
Sec.\,\ref{sec:concl} summarizes our results and discusses
directions for future work. \,More technical proofs are collected in
nine Appendices.

\no{\bf Acknowledgements}. \ The authors acknowledge the support of the
contract ANR-05-BLAN-0029-03 in the early stage of this collaboration.
The work of K.W. was supported by a Feodor-Lynen scholarship granted by
the Alexander von Humboldt Foundation. That of R.R.S. was partially funded
by the Collaborative Research Centre 676 ``Particles, Strings and the
Early Universe - the Structure of Matter and Space-Time''.

\nsection{Wess-Zumino Feynman amplitudes}
\label{sec:WZWampl}
\subsection{2D Wess-Zumino action and gerbes}
\label{sec:2DWZW}

\no Let $\,M\,$ be a smooth manifold and $\,H\,$ a closed 3-form on
$\,M$. \,2-forms $\,B\,$ such that $\,dB=H\,$ provide the background
Kalb-Ramond fields for the two-dimensional sigma model with {\it
target space} $\,M$. \,We shall be mostly interested in situations
when $\,H\,$ is not an exact form so that the 2-forms $\,B\,$ exist
only locally. The classical fields of the sigma model are smooth
maps $\,\varphi:\Sigma\rightarrow M$, \,where $\,\Sigma$,
\,called the {\it worldsheet}, \,is a compact surface, not necessarily
connected, that will be assumed closed and oriented here.
\,The Kalb-Ramond field contributes to the
sigma-model action functional and to the Feynman amplitude of the
field configuration $\,\varphi\,$ the Wess-Zumino terms which, for
the global 2-form $\,B$, \,are equal to
\qq
S_{\WZ}(\varphi)\ \defn\ \int\limits_{\Sigma}\varphi^*B\qquad\quad{\rm and}
\qquad\quad
{\bm A}_{\WZ}(\varphi)\ \defn\ \eee{\,\si\,S_{\WZ}(\varphi)}\
=\ \eee{\,\si\int\limits_{\Sigma}\varphi^*B}\,,
\qqq
respectively, in the units where the Planck constant $\,\hbar=1$. \,The
contribution to the Feynman amplitudes may be defined more
generally if, instead of a global 2-form $\,B$, \,one is given a
{\it bundle gerbe with unitary connection} $\,\CG\,$ over $\,M$,
\,called simply {\it gerbe} below, with curvature equal to the
closed 3-form $\,H\,$ \cite{Murr}. \,Such gerbes are precisely the
geometric objects that allow to define a $\,U(1)$-valued {\it
holonomy} $\,\Hol_{\CG}(\varphi)\,$ of maps
$\,\varphi:\Sigma\rightarrow M$, \,and one sets
\qq
{\bm A}_{\WZ}(\varphi)\ \defn\ \Hol_{\CG}(\varphi)\,.
\label{hol}
\qqq
In particular, \,if $\,H=dB\,$ for a global 2-form on $\,M$, \,there
exists a gerbe $\,\CI_B$ with curvature $\,H$, \,canonically associated
to $\,B$, \,such that
\qq
Hol_{\CI_{B}}(\varphi)\ =\ \eee{\,\si\int\limits_{\Sigma}\varphi^*B}\,.
\qqq
Gerbes with curvature $\,H\,$ exist if and only if the periods of the
closed 3-form $\,H\,$ are in $\,2\pi\NZ$.
\,In particular, $\,H\,$ is not required to be an exact form.

The basic property of the
holonomy of a gerbe $\,\CG\,$ with curvature $\,H\,$ is that it is a
(Cheeger-Simons) differential character. \,This means that if
$\,\tilde\Sigma\,$ is a compact oriented 3-manifold with boundary
$\,\partial\tilde\Sigma=\Sigma$, and if $\,\tilde\varphi:\tilde\Sigma
\rightarrow M$, \,then, \,for $\,\varphi=\tilde\varphi|_\Sigma$,
\qq
\Hol_\CG(\varphi)\ =\
\eee{\,\si\int\limits_{\tilde\Sigma}\tilde\varphi^*H}\,.
\label{ChS}
\qqq
Consequently, \,the gerbe holonomy is fully determined for the boundary
values of fields $\,\tilde\varphi\,$ by the gerbe
curvature $\,H$. \,On the other hand, taking a 3-dimensional ball for
$\,\tilde\Sigma\,$ ones infers easily that the gerbe holonomy
determines the gerbe curvature $\,H$. \,The converse is true only if
the homology  group $\,H_2(M)\,$ is trivial.

The (bundle) gerbes (with unitary connection) $\,\CG\,$ over $\,M\,$ form
a 2-category $\grb M$ with 1-morphisms between gerbes and
2-morphisms between 1-morphisms \cite{Stev}. Below, we shall denote
by $\,\Id\,$ as well the identity maps between spaces as the
identity 1-isomorphisms between gerbes and the identity
2-isomorphisms between 1-isomorphisms, with the meaning of the
symbol that should be clear from the context. \,Gerbes $\,\CG\,$
possess duals $\,\CG^*\,$ with opposite curvature and inverse
holonomy, tensor products $\,\CG_1\otimes\CG_2\,$ with added
curvatures and multiplied holonomies, and pullbacks $\,f^*\CG\,$
under smooth maps $\,f\,$ of the underlying base manifolds
with curvatures related
by the pullback of 3-forms and the same holonomies of maps
$\,\varphi\,$ related by the composition with $\,f$. \,Up to
1-isomorphisms, gerbes are classified by their holonomy. \,Indeed,
\,two gerbes with the same curvature differ, up to a 1-isomorphism,
by a tensor factor that is a flat gerbe (i.e. has vanishing
curvature). Their holonomies differ by the the flat gerbe holonomy
factor that determines a cohomology class in $\ H^2(M,U(1))=Hom(H_2(M),U(1))$.
\,All the elements of $\,H^2(M,U(1))\,$ may be obtained this way.

\subsection{Rigid symmetries of Wess-Zumino amplitudes}
\label{sec:rigid}

\no Rigid symmetries of sigma models are induced by transformations
of the target space. \,Let $\,\Gamma\,$ be a Lie group that,
in general, will not be assumed to
be connected or simply connected. Suppose now that $\,M\,$ is a
$\,\Gamma$-space, \,i.e. that we are given a smooth action
$\,\ell:\Gamma\times M\rightarrow M\,$
of $\,\Gamma\,$ on $\,M$. \,We shall variably write $\ell(\gamma,m)\defn
\ell_\gamma(m)\defn r_m(\gamma)\defn \gamma m$.
\,The infinitesimal action of the Lie algebra $\,\Ng$ of $\,\Gamma\,$
on $\,M\,$ is induced by the vector fields
$\,\bar X\,$ for $\,X\in{\bf g}$, \,where $\,\bar X(m)=\frac{d}{dt}|_{_{t=0}}
\ee^{-tX}m$. \,The assignment preserves the commutators:
$\,[\bar X,\bar Y]=\ov{[X,Y]}$. \,We would like to
determine when the WZ Feynman amplitudes are invariant under this
action. \,Below, $\,\iota_{\CX}\,$ will denote the contraction
with the vector field $\,\CX$, \,and $\,\CL_{\CX}=d\,\iota_{\CX}
+\iota_{\CX}\,d\,$ the Lie derivative with respect to it.

\begin{lemma}
\label{prop:holonomyvariation}
\label{1233}
The variation of the gerbe holonomy
of maps $\,\varphi:\Sigma\rightarrow M\,$ under the infinitesimal
action of $\,X\in{\bf g}\,$ is given by the formula
\qq
\bigddt\,\Hol_\CG(\ee^{-tX}\varphi)
\ =\ \Big(\si\int\limits_\Sigma\varphi^*\iota_{\bar X}H\Big)\,\,
\Hol_\CG(\varphi)\,.
\label{1prop}
\qqq
\end{lemma}

\begin{proof}
The relation (\ref{ChS}) implies that
\qq
\Hol_\CG(\ee^{-tX}\varphi)\ =\ \eee{\,\si\hspace{-0.05cm}\int\limits_{[0,1]
\times\Sigma}\hspace{-0.1cm}
\tilde\varphi_t^*H}\,\,\Hol_\CG(\varphi)\
=\ \eee{\,\si\hspace{-0.05cm}\int\limits_{[0,1]
\times\Sigma}\hspace{-0.1cm}
\tilde\psi_t^*pr_2^*H}\,\,\Hol_\CG(\varphi)
\qqq
for $\,\tilde\varphi_t(s,x)=\ee^{-stX}\varphi(x)$,
$\,\tilde\psi_t(s,x)=(s,\tilde\varphi_t(s,x))\,$  and
$\,pr_2(s,m)=m$. \,Differentiation of the right hand side with
respect to $\,t\,$ gives
\qq
\smallddt\,\Hol_\CG(\ee^{-tX}\varphi)
\ =\ \Big(\,\si\hspace{-0.05cm}\int\limits_{[0,1]\times\Sigma}\hspace{-0.1cm}
\tilde\psi^*_0\,\CL_{\tilde X}\,pr^*_2H\Big)\,\,\Hol_\CG(\varphi)\
=\ \Big(\,\si\hspace{-0.05cm}\int\limits_{[0,1]\times\Sigma}\hspace{-0.1cm}
d\,\tilde\psi^*_0\,\iota_{\tilde X}\,pr_2^*H\Big)\,\,\Hol_\CG(\varphi)\,,
\qqq
where $\,\tilde X\,$ is the vector field on $\,[0,1]\times M\,$ such that
$\,\tilde X(s,m)=\smallddt(s,\ee^{-stX}m)=s\bar{X}(m)$.
\,The Stokes formula applied to the last
integral results in the claim.
\end{proof}

\no Lemma \ref{1233} implies that the left hand side of
Eq.\,(\ref{1prop}) vanishes if and only if
\qq
\int\limits_\Sigma\varphi^*\iota_{\bar X}H\ =\ 0\,.
\qqq
This holds for all $\,\varphi\,$ if and only if $\,\iota_{\bar X}H\,$
is an exact form. \,We obtain this way

\begin{corollary}
The Feynman amplitudes $\,{\bm A}_{\WZ}(\varphi)\,$ are invariant
under the infinitesimal action of the Lie algebra $\,\Ng\,$ (or,
equivalently, of the connected component of unity $\,\Gamma_0\subset
\Gamma$) \,if and only if the 2-forms $\,\iota_{\bar X}H\,$ are
exact for all $\,X\in{\bf g}$.
\end{corollary}

\no Note that the exactness of $\,\iota_{\bar X}H\,$ implies, in
particular, that $\,\CL_{\bar X}H=0$, \,i.e. that the curvature 3-form
$\,H\,$ is invariant under the infinitesimal action
of $\,\Ng$. \,Observe also
that if $\,H=dB\,$ for a global $\,\Ng$-invariant 2-form $\,B$, \,then
$\,\iota_{\bar X}H=-d(\iota_{\bar X}B)\,$ so that the 2-forms
$\,\iota_{\bar X}H\,$ are exact.

If the group $\,\Gamma\,$ is not connected, i.e. $\,\Gamma\not=\Gamma_0$,
\,then the condition for the $\,\Gamma$-invariance of the WZ Feynman
amplitudes is more stringent. Since
\qq
\Hol_{\CG}(\gamma\varphi)\ =\ \Hol_{\ell_\gamma^*\CG}(\varphi)
\qqq
for $\,\gamma\in\Gamma$, \,it follows that
$\ {\bm A}_{\WZ}(\gamma\varphi)={\bm A}_{\WZ}(\varphi)\ $ for all
$\,\varphi\,$ if and only if the the gerbes $\,\ell_\gamma^*\CG\,$
and $\,\CG\,$ have the same holonomy. In particular, they have to
have the same curvature: $\,\ell_\gamma^*H=H$. \,Since the holonomy
determines the 1-isomorphism class of a gerbe, we obtain

\begin{corollary}
The Feynman amplitudes $\,{\bm A}_{\WZ}(\varphi)\,$ are invariant
under the action of $\,\Gamma\,$ if and only if the gerbes
$\,\ell_\gamma^*\CG\,$ and $\,\CG\,$ are 1-isomorphic for all $\,\gamma
\in\Gamma$.
\end{corollary}

\begin{remark}
If the Feynman amplitudes contain also a factor with the standard
sigma-model action $\ S(\varphi)=\Vert d\varphi\Vert^2_{L^2}\ $
defined with the help of Riemannian metrics over the worldsheet and
the target space, then the group $\,\Gamma\,$ of rigid symmetries has to
preserve also the target metric so that, in particular, $\,\bar X\,$
are Killing vector fields for $\,X\in\Ng$.
\end{remark}

\nsection{Coupling to topologically trivial gauge fields}
\label{sec:trivialgaugefields}

\no A natural question arises whether $\,\Ng$-invariant Feynman
amplitudes $\,{\bm A}_\WZ(\varphi)\,$ may be gauged,
\,i.e.\,\,coupled to gauge fields in a gauge-invariant way. \,First,
we shall discuss the case of topologically trivial gauge fields
given by global $\,\Ng$-valued 1-forms $\,A\,$ on the worldsheet
$\,\Sigma$. \,

\subsection{Gauging prescription}
\label{sec:JJMO-HS}

\no In the particular instance when the WZ Feynman amplitudes
are determined by a global $\,\Ng$-invariant
2-form $\,B\,$ with $\,dB=H$, \,one may realize the gauging
by replacing $\,B\,$ with its minimally coupled version
$\,B_A\,$ which is a 2-form on
$\,\Sigma\times M\,$:
\qq
B_A\ \defn\ \exp(-\iota_{\bar A})\,B\ =\ B\,-\,\iota_{\bar A}B\,+\,
\frac{_1}{^2}\iota_{\bar A}^2B\,.
{}
\qqq
Above, for $\,X\in\Ng\,$ and $\alpha\,$ a differential form, \,we
define $\,\iota_{\bar X\otimes\alpha}=\alpha\,\iota_{\bar X}\,$
(omitting the {\it wedge} sign for the exterior product of
differential forms). \,The gauged Wess-Zumino action has then the form
\qq
S_\WZ(\varphi,A)\ \defn\ \int\limits_\Sigma\phi^*B_A\
=\ S_\WZ(\varphi)\,+\int\limits_\Sigma\phi^*\big(-\iota_{\bar A}B\,
+\,\frac{_1}{^2}\iota_{\bar A}^2B\big)\,,
\label{minc}
\qqq
where $\,\phi=(\Id,\varphi):\Sigma\rightarrow\Sigma\times M$.
\,It is well known that the minimal coupling gives an action
that is invariant under infinitesimal gauge transformations
induced by  the maps $\,\Lambda:\Sigma\rightarrow\Ng$. \,This means that
\qq
\ddt\,\,S_\WZ(\ee^{-t\Lambda}
\varphi,\ee^{-t\Lambda}A)\ =\ 0\,,
\label{lginv}
\qqq
where, for $\,x\in\Sigma$,
\qq
\big(\ee^{-t\Lambda}\varphi\big)(x)\ =\ \ee^{-t\Lambda(x)}\varphi(x)\,,\quad
\qquad
\big(\ee^{-t\Lambda}A\big)(x)\ =\ Ad_{\ee^{-t\Lambda(x)}}A(x)\,+\,
\ee^{-t\Lambda(x)}d\,\ee^{\,t\Lambda(x)}\,.
\label{ggtr}
\qqq
The invariance (\ref{lginv}) will also follow from the considerations below.

In the more general case when the Feynman amplitudes $\,{\bm A}_\WZ(\varphi)\,$
are given by the gerbe holonomy, see Eq.\,(\ref{hol}), \,one may still postulate
that the coupling to the gauge fields is realized by terms linear
and quadratic in $\,A$, \,resulting in the replacement of
$\,{\bm A}_\WZ(\varphi)\,$ by
\qq
{\bm A}_\WZ(\varphi,A)\ \defn\
\eee{\,\si\int\limits_\Sigma
\phi^*\big(-v(A)\,+\,\frac{_1}{^2}u(A^2)\big)}\,{\bm A}_\WZ(\varphi)\,,
\label{FAg}
\qqq
where $\,v(X)\,$ are 1-forms on $\,M\,$ linearly dependent on
$\,X\in\Ng$, $\,\,u(X\wedge Y)\,$ are functions on $\,M\,$ linearly
dependent on $\,X\wedge Y\in\Ng\wedge\Ng\,$ and, for a form
$\,\alpha\,$ on $\,\Sigma$, $\,v(X\otimes\alpha):=v(X)\alpha\,$ and
$\,u((X\wedge Y)\otimes\alpha):=u(X\wedge Y)\,\alpha\,$ denote the
induced forms on $\,\Sigma\times M$. \,Necessary conditions for the
consistency of such a coupling were found in \cite{JJMO} and
\cite{HS}. They are summarized in
\vskip 0.4cm

\begin{proposition}
\label{prop:JJMO-HS}
The amplitudes ${\bm A}_\WZ(\varphi,A)$ defined in (\ref{FAg})
are invariant under infinitesimal gauge transformations if and only if
the 1-forms $\,v(X)\,$ satisfy the relations
\qq
\hspace{0.6cm}\iota_{\bar X}H\ =\ dv(X)\,,\qquad \CL_{\bar X}v(Y)\
=\ v([X,Y])\,,\qquad \iota_{\bar X}v(Y)\ =\ -\iota_{\bar Y}v(X)\label{JJMO-HS}
\qqq
for all $\,X,Y\in\Ng$, \,with the functions $\,u\,$ given by
\qq
\hspace{0.6cm}u(X\wedge Y)\ =\ \iota_{\bar X}v(Y)\,.\label{u}
\qqq
\end{proposition}

\no For completeness, we give in Appendix
\ref{app:1} a proof of this result by arguments close to
the original ones of \cite{JJMO} and \cite{HS}.
\vskip 0.4cm

\begin{remark}
\label{rem:3.2}
\item[\ 1.\ ]
The 1-forms $\,v(X)\,$ satisfying Eqs.\,(\ref{JJMO-HS}) may be modified
by 1-forms $\,w(X)\,$ (also linear in $\,X$) \,satisfying the homogeneous
version of these equations.
\item[\ 2.\ ]
To make contact with refs. \cite{JJMO} and \cite{HS} more explicitly, \,let
us introduce a basis $\,(t^a)\,$ of the Lie algebra
$\,\Ng\,$ with $\,[t^a,t^b]=f^{abc}t^c\,$ (the summation
convention!), $\,v(t^a)=:v^a$, \,and $\,u(t^a\wedge t^b)=:u^{ab}$.
\,Denoting by $\,\iota^a\,$ and $\,\CL^a\,$ the contraction with and
the Lie derivative w.r.t. the vector field
$\bar{\hspace{0.05cm}t^a}$, \,the relations (\ref{JJMO-HS}) and
(\ref{u}) may be rewritten as
\qq
\hspace{0.6cm}\iota^a H\ =\ dv^a\,,\qquad\CL^a v^b\ =\ f^{abc}v^c\,,
\qquad \iota^av^b\ =\ -\iota^bv^a\ =\ u^{ab}\,,\label{JJMO-HS1}
{}
\qqq
\end{remark}
\vskip 0.2cm

In view of Proposition \ref{prop:JJMO-HS}, it will be convenient to introduce
a \,2-form $\,\rho_A\,$ on the product manifold
$\,\Sigma\times M\,$ and a gerbe $\,\CG_A\,$
over the same space by the formulae
\qq
\rho_A\ =\ -v(A)\,+\,\frac{_1}{^2}\iota_{\bar A}v(A)\quad\qquad{\rm and}\quad
\qquad \CG_A\ =\ \CI_{\rho_A}\otimes\CG_2\,.
\label{HA1}
\qqq
Eq.\,(\ref{FAg}), together with the conditions
(\ref{JJMO-HS}) and (\ref{u}) on its entries, may then be summarized in
the following

\begin{definition}
\label{def:trivialamplitudes} Let $\,\CG\,$ be a gerbe with
curvature $\,H\,$ over a $\,\Gamma$-space $\,M$, \,and let
$\,v(X)\,$ be 1-forms on $\,M$, \,linearly dependent on $\,X\in\Ng$,
\,satisfying conditions (\ref{JJMO-HS}). \,The Wess-Zumino
contribution of a field $\,\varphi: \Sigma \to M\,$ to the Feynman
amplitude coupled to gauge field 1-form $\,A\,$ on $\,\Sigma\,$
is defined as
\qq
{\bm A}_\WZ(\varphi,A)\ =\ \eee{\,\si\int\limits_\Sigma
\phi^*\rho_A}\,{\bm A}_\WZ(\varphi)\ =\ \Hol_{\CG_A}(\phi)\,,
\label{FAg1}
\qqq
where, as before, $\,\phi=(\Id,\varphi)$.
\end{definition}

\begin{remark}
If the gerbe $\,\CG=\CI_B\,$ for a $\,\Ng$-invariant 2-form $\,B\,$
such that $dB=H$, \,then one may take $\,v(X)=-\iota_{\bar X}B$.
\,In this case, Eq.\,(\ref{FAg1}) agrees with the minimal coupling
(\ref{minc}) of the Wess-Zumino action.
\end{remark}

\no Proposition \ref{prop:JJMO-HS} implies immediately
\vskip 0.4cm

\begin{corollary} Eq.\,(\ref{FAg1}) defines
amplitudes that are invariant under infinitesimal gauge transformations.
\end{corollary}

Below, we shall need two easy implications of relations
(\ref{JJMO-HS}) whose straightforward proof is left to the reader.
They will be employed repeatedly below.

\begin{lemma}
\label{lem:implicationsofhullspence} Relations \,(\ref{JJMO-HS})
\,imply that
\qq
\iota_{\bar X}\iota_{\bar Y}H\ &=&\ v([X,Y])-d\iota_{\bar X}v(Y)\,,
\label{Hv1}\\
\iota_{\bar X}\iota_{\bar Y}\iota_{\bar Z}H\ &=&\
\iota_{\bar X}v([Y,Z])+\iota_{\bar Z}v([X,Y])
+\iota_{\bar Y}v([Z,X]).\label{Hv2}
\qqq
\end{lemma}


\subsection{Equivariant-cohomology interpretation}
\label{sec:eqcoh}

\no In refs. \cite{FS,FS1}, see also \cite{WittenHF} and \cite{Wu},
relations (\ref{JJMO-HS}) were interpreted in terms of equivariant
cohomology. Let $\,\Omega(M)\,$ denote the space of differential
forms on $\,M$. \,Recall that the Cartan complex for equivariant
cohomology is formed of polynomial maps
\qq
{\bf g}\ni X\,\longmapsto\,\hat\omega(X)\in\Omega(M)
\qqq
which satisfy
\qq
\CL_{\bar X}\hat\omega(Y)\,=\smallddt\,
\hat\omega(Ad_{\ee^{tX}}Y)\qquad{\rm for}\quad
X,Y\in{\bf g}\,.
\label{equiv0}
\qqq
We shall call such maps ${\bf g}$-equivariant forms. Note that
relation (\ref{equiv0}) holds if and only if
\qq
\ell_\gamma^*\,\hat\omega(Y)\ =\ \hat\omega(Ad_{\gamma^{-1}}Y)
\label{relg}
\qqq
for $\,\gamma\,$ in the connected component $\,\Gamma_0$ of $\,1\,$
in $\,\Gamma$. \,We shall say that a form $\,\hat\omega\,$ is
$\,\Gamma$-equivariant if the relation (\ref{relg}) is satisfied for
all $\,\gamma\in \Gamma$. \,Of course, the two notions of
equivariance coincide if the group $\,\Gamma\,$ is connected. \,The
$\,\Ng$-equivariant ($\Gamma$-equivariant) forms make up the complex
$\,\Omega_{\bf g}^\bullet(M)\,$ \,($\Omega_{\Gamma}^\bullet(M)$)
\,with the ${\mathbb Z}$-grading that adds twice the degree of the
polynomial to the degree of the form and with the differential of
degree 1 given by the formula
\qq
(\hat d\,\hat\omega)(X)\ =\ d\,\hat\omega(X)\,-\,\iota_{\bar X}\hat\omega(X)\,.
\label{dh}
\qqq
The following result was obtained in \cite{FS,FS1}:

\begin{proposition}
A $\,\Ng$-equivariantly closed 3-form $\hat H=H+v(X)\,$ extends the
closed $\,\Ng$-invariant 3-form $\,H\,$ if and only if the 1-forms
$\,v(X)\,$ satisfy conditions (\ref{JJMO-HS}).
\end{proposition}

\begin{proof}
The $\,\Ng$-equivariance of $\,\hat H\,$ is the relation
\qq
\CL_{\bar X}\hat H(Y)\ =\ \CL_{\bar X}(H+v(Y))\ =\ v([X,Y])
{}
\qqq
that, in view of the $\,\Ng$-invariance of $\,H$, \,reproduces the middle
equality in (\ref{JJMO-HS}).
\,On the other hand, the form $\,\hat H\,$ is $\,\Ng$-equivariantly closed
when
\qq
(\hat d\hspace{0.01cm}\hat H)(X)\ =\ dH\,+\,dv(X)\,-\,\iota_{\bar X}H\,
-\,\iota_{\bar X}v(X)\ =\ 0
{}
\qqq
which, using that $\,dH=0\,$, is equivalent to the left and the right
equalities of (\ref{JJMO-HS}).
\end{proof}

\begin{remark}
\label{rem:frch} The freedom of choice of $\,v(X)\,$ mentioned in
Remark \ref{rem:3.2}(1) consists of the addition of a 1-form
$\,w(X)$ that is $\,\Ng$-equivariantly closed.
\end{remark}

The $\,\Ng$-equivariantly closed 3-form $\,\hat H=H+v(X)\,$ may be
related directly to the curvature of the gerbe $\,\CG_A\,$ of Eq.\,(\ref{HA1})
which is equal to the 3-form
\qq
H_{A}\ =\ H\,+\,d\rho_{A}\,.
\label{HA}
\qqq
on $\,\Sigma\times M$.

\begin{lemma}
\label{lem:HAandF}
\qq
H_{A}\ =\ \exp(-\iota_{\bar A})\big(H+v(F)\big)\,,
{}
\qqq
where $\,F=dA+\frac{1}{2}[A,A]\,$ is the gauge-field strength 2-form.
\end{lemma}

\begin{proof}
Writing $\,A=t^a A^a\,$ and $\,F=t^a F^a\,$ with
$\,F^a=dA^a+\frac{1}{2} f^{bca}A^bA^c$, \,we obtain, \,using the
left one of relations (\ref{JJMO-HS}):
\qq
H_A\,=\,H+d\rho_A\,=\,H
+d\big(-v^aA^a+\frac{_1}{^2}(\iota^av^b)A^aA^b\big)\,=\,
H-\iota^aH\,A^a\,+\,v^a dA^a+\frac{_1}{^2}d(\iota^a v^b)A^aA^b\,.\qquad
{}
\qqq
Eq.\,(\ref{Hv1}) permits to transform the last term on the
right-hand side and to show that
\qq
&H_A\,=\,H\,-\,\iota^aH\,A^a\,+\,v^a dA^a\,+\,\frac{_1}{^2}f^{abc}v^cA^aA^b\,
-\,\frac{_1}{^2}(\iota^a\iota^bH)A^aA^b&\cr
&=\,H\,-\,\iota_{\bar A}H\,+\,v(F)\,+\,\frac{_1}{^2}\iota_{\bar A}^2H\ =\
\ \exp(-\iota_{\bar A})\big(H+v(F)\big)\,.&
{}
\qqq
\vskip -0.75cm\
\end{proof}

\begin{remark}
The minimal coupling operator $\,\exp(-\iota_{\bar A})\,$ may be
naturally interpreted within equivariant cohomology, see
\cite{Kalk}. \,Let us only mention here that it satisfies the
relation
\qq
\exp(\iota_{\bar A})\,d\,\exp(-\iota_{\bar A})\,=\,d\,-\,\iota_{\bar F}
\,+\,\CL_{\bar A}
{}
\qqq
for $\,\CL_{\bar A}=A^a\CL^{a}$.
\end{remark}

\subsection{More  equivariance properties}
\label{sec:lift}

\no We shall assume below that the 3-form $\,H\,$ extends to the
$\,\Gamma$-equivariantly closed 3-form $\,\hat H(X)=H+v(X)$. \,This
means, along with conditions (\ref{JJMO-HS}), \,that
\qq
\ell^*_\gamma H\ =\ H\qquad\quad{\rm and}\qquad\quad\ell_\gamma^*v(X)\
=\ v(Ad_{\gamma^{-1}}X)
\label{eqq}
\qqq
for all $\,\gamma\in \Gamma\,$ and all $\,X\in\Ng$, \,see
Eq.\,(\ref{relg}). \,In this section, we shall calculate the
pullback $\,\ell^*H\,$ of the 3-form $\,H\,$ along the action map
$\,\ell:\Gamma\times M\rightarrow M$. \,The result provides another way
to express equivariance properties of $\,H\,$ that will be used in the sequel.

More generally, \,we shall discuss below forms and gerbes over
the product spaces $\,\Gamma^{p-1}\times M\,$ that will be considered
as $\,\Gamma$-spaces with the adjoint action of $\,\Gamma\,$ on the factors in
$\,\Gamma^{p-1}\,$ and the original one on $\,M$. \,For a sequence of indices
$\,1\leq i_1<\dots i_{k_1}<i_{k_1+1}<\dots<i_{k_2}<\,\dots\,<i_{k_{q}}\leq p$,
\,we shall denote by $\,\ell_{i_1\dots i_{k_1},i_{k_1+1}\dotsi_{k_2},\,\dots\,,
i_{k_{q-1}+1}\dotsi_{k_{q}}}\,$ the maps
\qq
&&\Gamma^{p-1}\times M\ni(\gamma_1,\dots,\gamma_{p-1},m)\cr\cr
&&\hspace{0.3cm}\longmapsto\
\begin{cases}
\,\hbox to 10.6cm{$(\gamma_{i_1}\cdots \gamma_{i_{k_1}},\gamma_{i_{k_1+1}}
\cdots \gamma_{i_{k_2}},\dots,\gamma_{i_{k_{q-1}+1}}
\cdots \gamma_{i_{k_{q}}})\hspace{0.68cm}\in \Gamma^{q}
$\hfill}{\rm if}\quad i_{k_{q}}<p\,, \cr
\,\hbox to 10.6cm{$(\gamma_{i_1}\cdots \gamma_{i_{k_1}},\gamma_{i_{k_1+1}}
\cdots \gamma_{i_{k_2}},\dots,\gamma_{i_{k_{q-1}+1}}\cdots
\gamma_{i_{k_{q}-1}}m)\,\in\Gamma^{q-1}\times M$
\hfill}{\rm if}\quad i_{k_{q}}=p\,,\end{cases}\qquad
{}
\qqq
e.g., $\ell_2(\gamma,m)=m$, \,$\ell_{12}(\gamma,m)=\gamma m$,
\,$\ell_{12}(\gamma_1,\gamma_2,m)=\gamma_1\gamma_2$, \,or
$\,\ell_{2,3}(\gamma_1,\gamma_2,m)=(\gamma_2,m)$. \,All these maps
commute with the action of $\,\Gamma$. \,Finally, we shall
abbreviate $\,\ell^*_{i_1\dots i_{k}p}H\defn H_{i_1 \dots i_{k}p}$.
\,Similar self-explanatory shorthand notations will be employed for
other forms, gerbes and gerbe 1- and 2-morphisms, \,also living on
other product spaces.

Let us start by considering the pullback $\,H_{12}=\ell^*H\,$ of the
3-form $\,H\,$ to $\,\Gamma\times M$. \,The 1-forms $\,v(X)\,$ on $\,M$
define a 2-form
\qq
\rho\ \defn\ -v(\Theta)+\frac{_1}{^2}(\iota_{\bar\Theta}v)(\Theta)
\label{rho}
\qqq
on $\Gamma\times M$. \,Note the similarity to formula (\ref{HA1})
for the 2-form $\,\rho_A$.

\begin{lemma}
\label{lem:2formrho}
$H_{12}\ =\ d\rho\ +\ H_2$.
\label{prp3}
\end{lemma}

\begin{proof}
In order to find an explicit
expression for $\,H_{12}$, \,a useful tool is the observation that,
for a form $\,\omega\in\Omega(M)$,
\qq
(\ell^*\omega)(\gamma,m)\ =\ \big(\exp[-\iota_{\bar\Theta(\gamma)}]\,
\ell_\gamma^*\omega\big)(m)\,,
\label{pist}
\qqq
where $\,\Theta=t^a\Theta^a=\gamma^{-1}d\gamma\,$ is the $\,{\bf g}$-valued
Maurer-Cartan 1-form on $\,\Gamma$. \,As before, we use
the notations $\,\iota_{\bar X\otimes\alpha}\defn\alpha\hspace{0.02cm}
\iota_{\bar X}\,$ and $\,v(X\otimes\alpha)\defn v(X)\hspace{0.01cm}\alpha\,$
for $\,X\in\Ng\,$ and $\,\alpha\,$ a form, dropping the exterior product
sign. Eq.\,(\ref{pist}) makes explicit the contributions to $\,\ell^*\omega\,$
with differentials along $\,\Gamma\,$ and along $\,M$.
\,Application of identity (\ref{pist}) to $\,\omega=H\,$ gives
\qq
(\ell^*H)(\gamma,m)\,&=&\,\big(\exp[-\iota_{\bar\Theta(\gamma)}]\,
\ell_\gamma^*H\big)(m)
\ =\ \big(\exp[-\iota_{\bar\Theta(\gamma)}]\,H\big)(m)
{}\cr
&=&\,H(m)\,-\,\Theta^a(\gamma)(\iota^aH)(m)\,-\,\frac{_1}{^2}
(\Theta^a\Theta^b)(\gamma)(\iota^a\iota^bH)(m)\,+\,\frac{_1}{^6}(\Theta^a
\Theta^b\Theta^c)(\gamma)(\iota^a
\iota^b\iota^cH)(m)\cr
&=&\,H(m)\,-\,\Theta^a(g)(dv^a)(m)\,-\,
\frac{_1}{^2}f^{abc}(\Theta^a\Theta^b)(\gamma)v^c(m)
\,+\,\frac{_1}{^2}(\Theta^a\Theta^b)(\gamma)(d\iota^av^b)(m)\cr
&&\hspace{8cm}+\,\frac{_1}{^2}f^{bcd}
(\Theta^a\Theta^b\Theta^c)(\gamma)(\iota^av^d)(m)\cr
&=&H(m)\,+\,\big[d\big(\Theta^av^a+\frac{_1}{^2}\Theta^a
\Theta^b\iota^av^b\big)\big](\gamma,m)\,,
{}
\qqq
where the last but one equality was obtained by employing relations
(\ref{JJMO-HS}) and Lemma \ref{lem:implicationsofhullspence}, and
the last equality follows from the structure equation
$\,d\Theta^c=-\frac{_1}{^2}f^{abc}\Theta^a\Theta^b\,$ for the
Maurer-Cartan forms. \,The result is the claimed identity.
\end{proof}

\begin{remark}
\item[\ 1.\ ]
Similarly one may prove the relation
\qq
\hat H_{12}\ =\ \hat{d}\hspace{0.02cm}\rho\ +\ \hat H_2\,
{}
\qqq
which gives an equivariant extension of Lemma \ref{lem:2formrho}.

\item[\ 2.\ ]
Lemma \ref{lem:2formrho} implies that if $\,w(X)\,$ is a 1-form
linearly dependent on $\,X\,$ that is $\,\Gamma$-equivariantly
closed, then the 2-form
\qq
\sigma\ =\ -w(\Theta)+\frac{_1}{^2}\iota_{\bar\Theta}w(\Theta)
\label{sigmafrm}
\qqq
on $\,\Gamma\times M\,$ is closed, \,see Remark \ref{rem:frch}. \,This
is still true if $\,w(X)\,$ is only $\,\Ng$-equivariantly closed.
\end{remark}
\vskip 0.2cm


\begin{lemma}
\label{lem:rhorho}
The 2-form $\,\rho$ defined in Eq. (\ref{rho})
has the following properties:
\vskip -0.8cm\
\begin{enumerate}
\item
$\rho\,$ is a $\,\Gamma$-invariant
form on $\,\Gamma\times M$.
\item
As forms on $\,\Gamma^2\times M$,
\vskip -0.7cm
\qq
\rho_{12,3}\ =\ \rho_{1,23}\ +\ \rho_{2,3}\,.
\label{rrr}
\qqq
\end{enumerate}
\end{lemma}

\no A proof of Lemma \ref{lem:rhorho} may be found in Appendix
\ref{app:2}.

\nsection{Global gauge anomalies}
\label{sec:GGAnomal}
\subsection{General gauge transformations}
\label{sec:gengt}

\no As we have seen, conditions (\ref{JJMO-HS}) assure the
infinitesimal gauge invariance of the Feynman amplitudes
(\ref{FAg1}). In the present section, we shall examine the
behavior of the amplitudes under general gauge transformations
generated by $\,\Gamma$-valued smooth maps $\ \Ch:\Sigma
\rightarrow\Gamma$. \ Such maps act on the space $\,\Sigma\times
M\,$ by
\qq
(x,m)\ \mapstolabel{L_{\Ch}} \ (x,\Ch(x)m)\,,
\label{xmact}
\qqq
on the sigma-model fields $\ \varphi:\Sigma\rightarrow M\ $ by
\qq
\varphi\ \ \longmapsto\ \ \Ch\varphi\,,
\label{vptogvp}
\qqq
where $\,(\Ch\varphi)(x)=\Ch(x)\varphi(x)$, \ and on the gauge
fields according to the formulae
\qq
A\ \ \longmapsto\ \ \Ch A\,\defn\,Ad_{\Ch}(A)+(\Ch^{-1})^*\Theta\,,
\qquad\quad F\ \ \longmapsto\ \ \Ch F\,=\,Ad_{\Ch}(F)\,.
\label{gauact}
\qqq
The infinitesimal gauge transformations are then generated by taking
$\,\Ch=\ee^{-t\Lambda}\,$ for $\,\Lambda:\Sigma\rightarrow\Ng\,$ and
expanding to the 1$^{\rm st}$ order in $\,t$. \,Let us start by
establishing the transformation rule of the curvature 3-form
$\,H_A\,$ of gerbe $\,\CG_A\,$ over $\,\Sigma\times M\,$ under
maps (\ref{xmact}).

\begin{lemma}
\label{lem:curvcoincide} The \,3-form $\,H_A\,$ defined in
(\ref{HA}) transforms covariantly under the general gauge transformations
$\,\Ch:\Sigma\rightarrow\Gamma\,$:
\qq
L_\Ch^*H_{A}\ =\ H_{\Ch^{-1}\hspace{-0.06cm}A}\,.
{}
\qqq
\end{lemma}

\begin{proof}
By virtue of the formula (\ref{pist}), Lemma \ref{lem:HAandF}, \,the identity
\qq
\iota_{\ov{Ad_{\gamma^{-1}}X}}\,\ell_\gamma^*\ =\ \ell_\gamma^*\,
\iota_{\bar X}
\label{comiot}
\qqq
that holds on $\,M$, \,and relations (\ref{eqq}), \,we have:
\qq
\big(L_\Ch^*H_{A}\big)(x,m)\ &=&\
H_{A}(x,\Ch(x)m)\ =\
\big[\exp[-\iota_{\ov{(\Ch^*\Theta)(x)}}]\,\ell_{\Ch(x)}^*
H_{A}(x,\cdot)\big](m)\cr\cr
&=&\ \big[\exp[-\iota_{\ov{(\Ch^*\Theta)(x)}}]\,\ell_{\Ch(x)}^*
\exp[-\iota_{\bar A(x)}]\big(H+v(F(x))\big)\big](m)\cr\cr
&=&\ \big[\exp[-\iota_{\ov{(\Ch^*\Theta)(x)}}]\,
\exp[-\iota_{\ov{(Ad_{\Ch^{-1}}(A))(x))}}]
\,\big(H+v((Ad_{\Ch^{-1}}(F))(x))\big)\big](m)\cr\cr
&=&\ \big[\exp[-\iota_{\ov{(\Ch^*\Theta
+Ad_{\Ch^{-1}}(A))(x)}}]\,\big(H+v((Ad_{\Ch^{-1}}(F))(x))
\big)\big](m)\cr\cr &=&\ H_{\Ch^{-1}\hspace{-0.05cm}A}(x,m)\,,{}
\qqq
where the last equality follows from relations (\ref{gauact}).
\end{proof}
\vskip 0.1cm

We shall need below a few simple facts from the theory of gerbes.
\,First, the pullback and the tensor product of gerbes commute.
\,Second, the pullback of the gerbe $\,\CI_{B}\,$ associated to a
2-form $\,B\,$ is a similar gerbe associated to the pullback 2-form.
\,Third, the tensor product of gerbes
$\,\CI_{B_1}\otimes\CI_{B_2}\,$ for 2-forms $\,B_i\,$ on the same
space may be identified with the gerbe $\,\CI_{B_1+B_2}$. \,Fourth,
the tensor product $\,\CG\otimes\,\CG^*\,$ of a gerbe with its dual
is canonically isomorphic to the trivial gerbe $\,\CI_0\,$ which
provides the unity of the tensor product. \,Fifth, if two gerbes are
1-isomorphic then so are their tensor products by a third gerbe and
their pullbacks by the same map.

To find out the transformation rules of the Feynman amplitudes
under general gauge transformations, \,we have to compare
the amplitudes $\,{\bm A}_\WZ(\Ch\varphi,\Ch A)\,$ and
$\,{\bm A}_\WZ(\varphi,A)$. \ Since
\qq
{\bm A}_\WZ(\Ch\varphi,\Ch A)\ =\ \Hol_{\CG_{\Ch A}}(L_\Ch\circ\phi)\ =\
\Hol_{{L^*_\Ch}\CG_{_{\Ch A}}}(\phi)\qquad{\rm and}\qquad{\bm A}_\WZ(\varphi,A)\ =\
\Hol_{\CG_A}(\phi)
\label{bsar}
\qqq
for $\,\phi=(Id,\varphi)$, \ it will be enough to compare the
gerbes $\,L_\Ch^*\CG_{\Ch A}\,$ and $\,\CG_{A}\,$ whose curvatures,
equal to $\,L_\Ch^*H_{\Ch A}\,$ and $\,H_{A}$, \,respectively,
\,coincide by Lemma \ref{lem:curvcoincide}. \,From the latter
property, it follows that those two gerbes are related up to
1-isomorphism by tensoring with a flat gerbe which we shall identify
now. \,Consider the gerbe
\qq
\CF\ =\ \CG_{12}\otimes\CG_2^*\otimes\CI_{-\rho}
\label{CF}
\qqq
over $\,\Gamma\times M$. \,It follows from  Lemma \ref{lem:2formrho}
that $\,\CF\,$ is flat.

\begin{proposition}
\label{prop:isoflatgerbe}
The gerbes $\ L_\Ch^*\,\CG_{\Ch A}\ $ and $\ \CG_{A}\otimes
(\Ch\times\Id)^*\hspace{0.015cm}\CF\ $ over $\,\Sigma \times M\,$
are  1-isomorphic.
\end{proposition}

\no A proof of Proposition \ref{prop:isoflatgerbe} by a chain of relations,
\,based on the properties of gerbes listed above, \,may
be found in Appendix \ref{app:3}.
\vskip 0.2cm

Taking into account relations (\ref{bsar}) and the identities
$\ \Hol_{(\Ch\times\Id)^*\CF}(\phi)=\Hol_\CF((\Ch\times\Id)\circ\phi)
=\Hol_\CF((\Ch,\varphi))$, \,Proposition \ref{prop:isoflatgerbe}
implies immediately the following transformation property of the
Wess-Zumino amplitudes:

\begin{theorem}
\label{thm:ggtr} Under the gauge transformation induced by a
map $\,\Ch:\Sigma\rightarrow\Gamma$,
\qq
{\bm A}_\WZ(\Ch\varphi,\Ch A)\ =\ {\bm A}_\WZ(\varphi,A)\,\,\Hol_{\CF}((\Ch,\varphi))\,.
\label{gganom}
\qqq
\end{theorem}
\vskip -0.4cm
\hspace{13cm}$\blacksquare$

\no One can be more specific.
Note that from Eq.\,(\ref{CF}) it follows that
\qq
\Hol_{\CF}((\Ch,\varphi))\ =\ \Hol_{\CG}(\Ch\varphi)\
\Hol_{\CG}(\varphi)^{-1}\,\,\eee{-\si\int\limits_\Sigma(\Ch,\varphi)^*\rho}\,.
\label{prov}
\qqq
In particular, taking $\,\Ch=1$, \,we infer that
$\ \Hol_\CF((1,\varphi))=1$. \,Indeed, the 2-form
$\,(1,\varphi)^*\rho\,$ on $\,\Sigma\,$ vanishes because
the 2-form $\,\rho\,$ is composed of terms of degree $\,\leq1\,$ in
the direction of $\,M$.
\ More generally, since the flat-gerbe holonomies of homotopic fields coincide
in virtue of the holonomy property (\ref{ChS}),
$\ \Hol_\CF((\Ch,\varphi))=1\ $ if $\,\Ch\,$ is homotopic
to $\,1$.

\begin{corollary}
The Feynman amplitudes (\ref{FAg1}) are invariant
under gauge transformations homotopic to $\,1$.
\end{corollary}

The gauge transformations homotopic to $\,1\,$ are often
called {\it small}. The remaining issue is the invariance of the
amplitudes (\ref{FAg1}) under {\it large} gauge
transformations that are not homotopic to $\,1$. \,The holonomy of
the flat gerbe $\,\CF\,$ on $\,\Gamma\times M\,$ defines a
cohomology class $\,[\CF]\in H^2(\Gamma\times M,U(1))\,$ which is
trivial if and only if the flat gerbe $\,\CF\,$ is 1-isomorphic to
the trivial gerbe $\,\CI_0$. \,By virtue of definition (\ref{CF}),
the latter holds if and only if the gerbes $\,\CG_{12}\,$ and
$\,\CI_{\rho}\otimes\CG_2\,$ over $\,\Gamma\times M\,$ are
1-isomorphic. \,Consequently,

\begin{corollary}
\label{co:invariancecond} The amplitudes (\ref{FAg1}) are invariant
under all gauge transformations if and only if the gerbes
$\,\CG_{12}\,$ and $\,\CI_{\rho}\otimes\CG_2\,$ over $\,\Gamma\times
M\,$ are 1-isomorphic.
\end{corollary}

\no The class $\,[\CF]$, \,that will be more carefully studied in
Sec.\,\ref{sec:obstrclass}, is the obstruction to the invariance
of the Feynman amplitudes (\ref{FAg1}) under large gauge transformations.
\,In other words, a non-triviality of the class $\,[\CF]\,$
leads to a global gauge
anomaly in the two-dimensional sigma model with the Wess Zumino term
corresponding to the gerbe $\,\CG\,$ and coupled to topologically
trivial gauge fields.

In the above analysis, we kept fixed the $\,\Gamma$-equivariant
extension $\,\hat H+v(X)\,$ of the curvature $\,H\,$ of the gerbe
$\,\CG$. \,A natural question arises whether one may use the freedom
in the choice of $\,v(X)\,$ to annihilate the global gauge anomaly.
Clearly, the answer is that it may be done if and only if there
exists a 1-form $\,w(X)\,$ that is $\,\Gamma$-equivariantly closed
for which $\,[\CF]=[\sigma]$, \,where $\,[\sigma]\,$  denotes the
cohomology class in $\,H^2(\Gamma\times M,U(1))\,$ induced by the
closed 2-form $\,\sigma\,$ of Eq.\,(\ref{sigmafrm}). \,In many
contexts, however, \,e.g., in applications to WZW and coset
models of conformal field theory, that we shall discuss below,
$\,v(X)\,$ is a part of the structure tied to the symmetries of the
theories and should not be changed.

Similarly, one may ask whether it is possible to annihilate the global gauge
anomaly by an appropriate choice of gerbe $\,\CG$, \,keeping
the curvature form fixed. \,Since this involves tensoring $\,\CG\,$ with
flat gerbes whose 1-isomorphism classes belong to $\,H^2(M,U(1))$, \,the
answer  is that this is possible if and only if $\ [\CF]=[b]_{12}-[b]_2\ $
for some class $\ [b]\in H^2(M,U(1))$. \,A change of $\,\CG\,$ to
another non 1-isomorphic gerbe, however, implies a non-trivial change
of the Feynman amplitudes of the ungauged sigma model, i.e. of the model
itself.

\subsection{Global gauge  anomalies in WZW amplitudes}
\label{sec:expl1}

\noindent As an example, \,let us consider the case when $\,M=G$, \,where
$\,G\,$ is a connected compact semi-simple Lie group, not
necessarily simply connected. \,One has: $\,G=\tilde G/Z$, \,where
$\,\tilde G=\mathop{\times}\limits_{l}\tilde G_l\,$ is the covering
group of $\,G\,$ that decomposes into the product of simple factors,
and $\,Z\,$ is a subgroup of the center $\,\tilde Z
=\mathop{\times}\limits_{l}\tilde Z_{l}\,$ of $\,\tilde G$. \,The
factors $\,\tilde Z_{l}\,$ are cyclic except for those equal to
$\,{\mathbb Z}_2^2\,$ corresponding to $\,\tilde G_l
=Spin(4r)$. \,The Lie algebra $\,\Ng\,$ of $\,\tilde G\,$ decomposes as
$\,\oplus_{l}\Ng_{l}\,$ into the direct sum of simple factors.
Let $\,\Nh\,$ be a Lie subalgebra of $\,\Ng\,$ corresponding to
a connected but not necessarily simply connected closed subgroup
$\,\tilde H\subset\tilde G\,$ that maps onto
a closed connected subgroup $\,\Gamma\,$ of $\,\tilde G/\tilde Z$.
\,Clearly, $\,\Nh\,$ is also
the Lie algebra of $\,\Gamma\,$ and $\,\Gamma=\tilde H/Z_\Gamma\,$
with $\,Z_\Gamma=\tilde H\cap\tilde Z$. We shall consider $\,G\,$
with the adjoint action of $\,\Gamma$.

\begin{definition}
\label{def:context}
Below, we shall call  a $\,\Gamma$-space $\,M=G\,$ as above the one of the
coset-model context.
\end{definition}

\no In the simplest case, $\,\Nh=\Ng\,$ and $\,\Gamma=\tilde G/\tilde Z$.
\,In what follows, the reader may think about this example.

Over the group $\,G=\tilde G/Z$, \,we shall consider the gerbe
$\,\CG_k\,$ with the curvature 3-form
\qq
H_k\ =\ \frac{_1}{^{12\pi}}\,k\hspace{0.07cm}\tr\,\Theta^3,
{}
\qqq
where $\,\Theta=g^{-1}dg\,$ is the $\,\Ng$-valued Maurer-Cartan
1-form on $\,G\,$ and $\,k\,\tr\,XY\defn\sum k_l
\hspace{0.03cm}\tr_l\,X^lY^l\,$ stands for the $\,ad$-invariant
negative-definite bilinear form on $\,\Ng\,$ given by the sum of such
forms on $\,\Ng_l\,$ normalized so that, if $\,G=\tilde G$, \,then
the form $\,H_k$ has periods in $\,2\pi{\mathbb Z}\,$ if and only if
the {\it level} $\,k=(k_l)\,$ is composed of integers. \,For non-simply
connected groups $\,G$, $\,k\,$ has to satisfy more stringent
selection rules \cite{FGK,KS,GR1}. \,The holonomy of gerbes
$\,\CG_k\,$ provides the Wess-Zumino part of amplitudes for the
WZW sigma models of conformal field theory
\cite{Witt}, \,see the next section.

\begin{definition}
We shall call $\,\CG_k\,$ the WZW gerbes.
\end{definition}

There may be several non-1-isomorphic WZW gerbes $\,\CG_k\,$ over
$\,G\,$ (their 1-isomorphism classes are counted by elements of
$\,H^2(Z,U(1))\,$ in the discrete group $\,Z\,$ cohomology \cite{Brown}).
\,The adjoint action of group $\,\tilde G/\tilde Z\,$ leaves the
3-forms $\,H_k\,$ invariant. \,For $\,X\in\Ng$, \,the vector field
$\,\bar X\,$ on $\,G\,$ induced by the infinitesimal adjoint action:
$\ \bar X(g)=\frac{d}{dt}|_{_{t=0}} Ad_{\ee^{-tX}}(g)\,$ satisfies
the relation $\ \iota_{\bar X}\Theta(g) =X-Ad_{g^{-1}}(X)$. \ Hence,
\qq
\iota_{\bar X}H_k\ =\ \frac{_1}{^{8\pi}}\,k\,\tr\,X\,(1-Ad_g)([\Theta(g),
\Theta(g)])\ =\ -\,\frac{_1}{^{4\pi}}\,d\,k\,\tr\,X\,(1+Ad_g)(\Theta(g))
{}
\qqq
so that, upon setting
\qq
v_k(X)\ =\ -\,\frac{_1}{4\pi}\,k\,\tr\,X\,(1+Ad_g)(\Theta(g))\,,
{}
\qqq
the left one of conditions (\ref{JJMO-HS}) is satisfied.
The 1-forms $\,v_k(X)\,$ satisfy also the other conditions of (\ref{JJMO-HS}).
Indeed,
\qq
\iota_{\bar X}v_k(Y)&=&-\frac{_1}{^{4\pi}}\,k\,
\tr\hspace{0.08cm}Y\,\big(-Ad_{g{-1}}(X)+Ad_g(X)\big)\cr\cr
&=&\frac{_1}{^{4\pi}}\,
\,\tr\hspace{0.08cm}X\,\big(-Ad_{g{-1}}(Y)+Ad_g(Y)\big)\
=\ -\iota_{\bar Y}v_k(X)\,,\\ \cr
\CL_{\bar X}v_k(Y)&=&-\,\frac{_d}{^{dt}}\big|_{_{t=0}}\,
\frac{_1}{^{4\pi}}\,k\,\tr\hspace{0.08cm}Y\,Ad_{\ee^{-tX}}^{\,\,*}
\big((1+Ad_g)(\Theta(g))\big)\ =\ -\,\frac{_d}{^{dt}}\big|_{_{t=0}}\,
\frac{_1}{^{4\pi}}\,k\,\tr\hspace{0.08cm}Y\,Ad_{\ee^{-tX}}
\big((1+Ad_g)(\Theta(g))\big)\cr\cr
&=&-\,\frac{_d}{^{dt}}\big|_{_{t=0}}\,
\frac{_1}{^{4\pi}}\,k\,\tr\hspace{0.08cm}Y\,Ad_{\ee^{tX}}(Y)\,
(1+Ad_g)(\Theta(g))\ =\ v_k([X,Y])\,.
{}
\qqq
Of course, we may restrict $\,X,Y\,$ above to take values in the
subalgebra $\,\Nh\subset\Ng$. \,The 2-form $\,\rho_{k,A}\,$ on
$\,\Sigma\times\Gamma\,$ given by Eq.\,(\ref{HA1}) and the 2-form
$\,\rho_{k}\,$ on $\,\Gamma\times G\,$ given by Eq.\,(\ref{rho})
\,are given now by the formulae
\qq
\rho_{k,A}&=&\,\frac{_1}{^{4\pi}}\,k\,\tr\hspace{0.08cm}\big((1+Ad_g)(\Theta(g))
\,+\,Ad_{g^{-1}}(A)\big)\,A\,,\\ \cr
\rho_k\ \,\,\,&=&\,\frac{_1}{^{4\pi}}\,k\,\tr\hspace{0.08cm}\big((1+Ad_g)(\Theta(g))\,+\,Ad_{g^{-1}}(\Theta(\gamma))\big)\,\Theta(\gamma)\,,
{}
\qqq
where $\,\Theta(\gamma)=\gamma^{-1}d\gamma\,$ is the Maurer-Cartan form
on $\,\Gamma$. The 2-form $\,\rho_{k,A}\,$ enters the coupling, described
in Definition \ref{def:trivialamplitudes},  of the
Wess-Zumino action to the $\,\Nh$-valued 1-form $\,A\,$ on
$\,\Sigma$.

\begin{figure}[ht]
\begin{center}
\vskip 0.2cm
\leavevmode
\epsfxsize=8cm 
\epsfysize=2.5cm 
\epsfbox{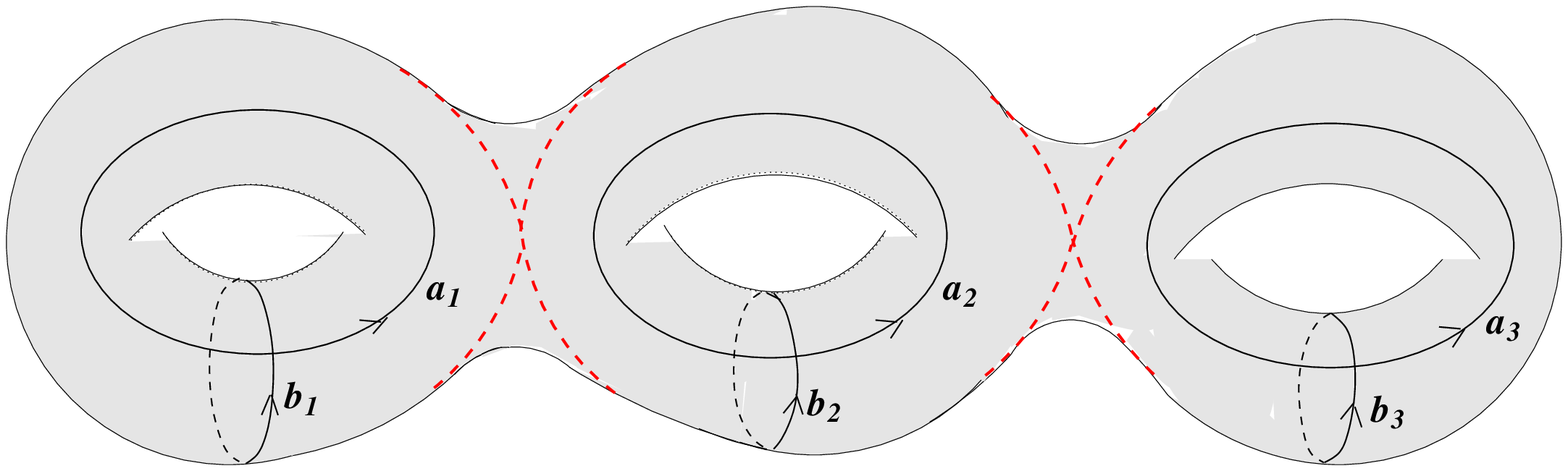}
\end{center}
\caption{Genus 3 surface with a marking; broken red lines indicate the
contours of its version with pinched handles}
\vskip 0.2cm
\label{fig:fig1}
\end{figure}

Let us compute the holonomy of the flat gerbe $\ \CF_k=(\CG_k)_{12}\otimes
(\CG_k)_2^*\otimes\CI_{-\rho_k}\ $ over
$\,\Gamma\times G$, \,see Eq.\,(\ref{CF}). \,Recall that the non-triviality
of such holonomy obstructs the
invariance of the Wess-Zumino amplitudes of Definition
\ref{def:trivialamplitudes} under large gauge transformations.
\,By Eq.\,(\ref{prov}),
\,for $\,\Ch:\Sigma\rightarrow\Gamma\,$ and $\,\varphi:\Sigma\rightarrow G$,
\qq
\Hol_{\CF_k}((\Ch,\varphi))\ =\ \Hol_{\CG_k}(Ad_{\Ch}(\varphi))
\ \Hol_{\CG_k}(\varphi)^{-1}\ \eee{-\si\int\limits_\Sigma(\Ch,\varphi)^*
\rho_k}\ =:\ c_{\Ch,\varphi}\,.
\label{HolFk}
\qqq
Since $\,\CF_k\,$ is flat, the above holonomy depends only on the
homotopy classes $\,[\Ch]\,$ and $\,[\varphi]\,$ of the maps
$\,\Ch\,$ and $\,\varphi$. \,Besides it does not depend on whether
we treat $\,h\,$ as a map with values in $\,\Gamma\,$ or in
$\tilde G/\tilde Z$. \,In the latter case, the homotopy classes of
the maps $\,\Ch\,$ are in one-to-one relation with
the elements of $\,Z_\Gamma^{2\omega}$, \,where $\,\omega\,$ is the
genus of $\Sigma$.
\,The element $\,(\tilde z_1,\tilde z_2,...,\tilde z_{2\omega-1},
\tilde z_{2\omega})\,$ corresponding to $\,[\Ch]\,$ is given by
the windings of $\,\Ch\,$ described by the holonomies
\qq
\tilde z_{2j-1}\,=\,\CP\,\eee{\,\int\limits_{a_j}\Ch^*\Theta},\qquad
\tilde z_{2j}\,=\,\CP\,\eee{\,\int\limits_{b_j}\Ch^*\Theta},
{}
\qqq
of the non-Abelian flat gauge field $\,\Ch^*(\Theta)$ on
$\,\Sigma$. \,Above,
$\,\CP\,$ stands for the path-ordering (from left to right) along
paths $\,a_j,b_j$, $\,j=1,\dots,\omega$, \,that generate a fixed
marking of the surface $\,\Sigma$, \,the latter assumed here to be
connected, see Fig.\,\ref{fig:fig1}.
Similarly for elements $\,(z_1,\dots,z_{2\omega})\,$ describing the
windings of $\,\varphi\,$ belonging to $\,Z^{2\omega}$.
\,By pinching off the handles of the surface
the same way as in Sec.\,III of \cite{GW}, \,one notes, using the
commutativity of the fundamental groups of $\,\tilde G/\tilde Z\,$ and
of $\,G$, \,that
\qq
c_{\Ch,\varphi}\,\equiv\,c_{(\tilde z_1,\dots,\tilde z_{2\omega}),(z_1,\dots,z_{2\omega})}
\ =\ \prod\limits_{j=1}^\omega c_{(\tilde z_{2j-1},\tilde z_{2j}),(z_{2j-1},z_{2j})}\,.
{}
\qqq
Hence, the calculation of $\ c_{\Ch,\varphi}\ $ reduces to the genus
1 case with $\,\Sigma=S^1\times S^1$. \,Let us choose the Cartan
subalgebras $\,\Nt_\Nh\subset\Nh\,$ and $\,\Nt_\Ng\subset\Ng\,$ so
that $\,\Nt_\Nh\subset \Nt_\Ng$. \,On $\,\Sigma=S^1\times S^1$, one
may take $\,\Ch=\Ch_{\tilde p^\vee_1,\tilde p^\vee_2}\,$ and
$\,\varphi=\varphi_{p^\vee_1.p^\vee_2}\,$ with
\qq
\Ch_{\tilde p^\vee_1,\tilde p^\vee_2}(e^{\,\si\sigma_1},\ee^{\,\si\sigma_2})\
=\ \ee^{\,\si(\sigma_1\tilde p^\vee_1+
\sigma_2\tilde p^\vee_2)}\,,\qquad\
\varphi_{{p^\vee_1,p^\vee_2}}(e^{\,\si\sigma_1},\ee^{\,\si\sigma_2})\ =\
\ee^{\,\si(\sigma_1 p^\vee_1+
\sigma_2 p^\vee_2)}\,,
\label{mnm'n'}
\qqq
where $\,\tilde p^\vee_i\in\si\Nt_\Nh\,\,$ and
$\,p^\vee_i\in\si\Nt_\Ng\,$ are such that the windings $\,\tilde
z_i=\ee^{\,2\pi\si\tilde p^\vee_i} \in Z_\Gamma\,$ and
$\,z_i=\ee^{\,2\pi\si p^\vee_i}\in Z$. \,Note that $\,\tilde
p^\vee_i\,$ and $\,p^\vee_i\,$ have to belong to the
coweight lattice $\,P^\vee_\Ng\,$ composed of elements $\,p^\vee\in\si\Nt_\Ng\,$
such that $\,\ee^{\,2\pi\si p^\vee}\in\tilde Z$.
\,Since $\,Ad_{\Ch_{\tilde p^\vee_1,\tilde p^\vee_2}}
(\varphi_{{p^\vee_1,p^\vee_2}})=\varphi_{p^\vee_1,p^\vee_2}$, \,the
formula (\ref{HolFk}) gives
\qq
c_{(\tilde z_1,\tilde z_2),(z_1,z_2)}\,&=&\,\exp\Big(-i\hspace{-0.1cm}
\int\limits_{S^1\times S^1}
(\Ch_{{\tilde p^\vee_1,\tilde p^\vee_2}},\varphi_{{p^\vee_1,p^\vee_2}})^*\rho_k\Big)
\cr\cr
&=&\,\exp\Big(\frac{i}{2\pi}\int\limits_0^{2\pi}
\int\limits_0^{2\pi}k\,\tr\,(d\sigma_1p^\vee_1+d\sigma_2p^\vee_2)
(d\sigma_1(\tilde p^\vee_1+d\sigma_2\tilde p^\vee_2)\Big)\cr\cr
&=&\,\exp\big(2\pi i\,k\,\tr\,(p^\vee_1\tilde p^\vee_2
-\tilde p^\vee_1p^\vee_2)\big)
\,.
\label{ugf}
\qqq
That the right hand side depends only on the windings is assured by
the restrictions on the level $\,k\,$ that guarantee the existence
of the WZW gerbe $\,\CG_k\,$ on $\,G$. \, \,The holonomy of the flat
gerbe $\,\CF_k\,$ is trivial if and only if the above expression is
always equal to 1 for the windings restricted as above \,(compare to a
similar discussion in \cite{GW}). \,From Corollary \ref{thm:ggtr},
\,we obtain

\begin{proposition}
\label{prop:secE}
For the $\,\Gamma$-space $\,M=G\,$ in the coset-model context, see
Definition \ref{def:context}, \,the WZ Feynman amplitudes (\ref{FAg1})
are invariant under all gauge transformations if and only if
the phases (\ref{ugf}) are trivial.
\end{proposition}
\vskip -0.35cm
\hspace{13cm}$\blacksquare$

The examples where the phases (\ref{ugf}) are non-trivial are
numerous. \,They include $\,G=\Gamma=\tilde G/\tilde Z\,$ for
$\,\tilde G=SU(r+1)\,$ with $\,r\,$ even and $\,k=1\,$ or with
$\,r\geq 3\,$ odd and $\,k=2$. Another example is
$\,G=\Gamma=Spin(2r)/{\mathbb Z}_2^2\,$ with $\,r\,$ divisible by
$\,4\,$ and $\,k=1$. \,In all those cases (and many others),
the amplitudes (\ref{FAg1}) of
Definition \ref{def:trivialamplitudes} exhibit a global gauge
anomaly. \,On the other hand, there is no global gauge anomaly for
the gauge fields $\,A\,$ given by global 1-forms on $\,\Sigma\,$ for
$\,G=\tilde G\,$ because in this case $\,p_i^\vee=0$.

The best known case with a non-simple group $\,\tilde G\,$ is
$\,\tilde G=SU(2)\times SU(2)$. \,The
restrictions on the level $\,k=(k_1,k_2)\,$ imposed by the existence
of the gerbe $\,\CG_k\,$ with curvature $\,H_k\,$ on $\,G=\tilde
G/Z\,$ depend on $\,Z\subset\tilde Z={\mathbb Z}_2\times{\mathbb
Z}_2$.
\qq
&&\hbox to 1.8cm{$k_1\in{\mathbb Z},$\hfill}\hbox to 1.8cm{$k_2\in{\mathbb Z}$
\hfill}\hbox to 2.5cm{$\ $\hfill}{\rm if}\qquad Z=0\,,\cr
&&\hbox to 1.8cm{$k_1\in2{\mathbb Z},$\hfill}\hbox to 1.8cm{$k_2\in{\mathbb Z}$
\hfill}\hbox to 2.5cm{$\ $\hfill}{\rm if}\qquad Z=\NZ_2\oplus 0\,,\cr
&&\hbox to 1.8cm{$k_1\in\NZ,$\hfill}\hbox to 1.8cm{$k_2\in2\NZ$
\hfill}\hbox to 2.5cm{$\ $\hfill}{\rm if}\qquad Z=0\oplus\NZ_2\,,\cr
&&\hbox to 1.8cm{$k_1\in\NZ,$\hfill}\hbox to 1.8cm{$k_2\in\NZ,$\hfill}
\hbox to 2.5cm{$k_1+k_2\in2\NZ$\hfill}{\rm if}\qquad Z={\rm diag}\,\NZ_2\,,\cr
&&\hbox to 1.8cm{$k_1\in2\NZ,$\hfill}\hbox to 1.8cm{$k_2\in2\NZ$
\hfill}\hbox to 2.5cm{$\ $\hfill}{\rm if}\qquad Z=\NZ_2\oplus\NZ_2\,.
\nonumber
\qqq
For $\,\Gamma=\tilde G/\tilde Z=SO(3)\times SO(3)$, \,with the adjoint
action on $\,G\,$ and $\,\tilde p_i,\tilde p'_i,p_i,p'_i\in\NZ$,
\qq
c_{_{(((-1)^{\tilde p_1},(-1)^{\tilde p'_1}),((-1)^{\tilde p_2},(-1)^{\tilde p'_2})),
(((-1)^{p_1},(-1)^{p'_1}),((-1)^{p_2},(-1)^{p'_2}))}}\ =\
(-1)^{k_1(p_1\tilde p_2-\tilde p_1p_2)+
k_2(p_1'\tilde p'_2-\tilde p'_1p'_2)}.\quad
{}
\qqq
We infer from this expression that the only case with a global anomaly of
the gauged WZ amplitudes (\ref{FAg1}) of Definition
\ref{def:trivialamplitudes} is the one with $\,G=(SU(2)\times SU(2))/{\rm
diag}\,\NZ_2\,$ with odd $\,k_1,k_2$. \,If one restricts, however,
the group $\,\Gamma\,$ to the diagonal $\,SO(3)\,$ subgroup of
$\,SO(3)\times SO(3)\,$ then the global gauge anomaly disappears.
\,Another anomalous example with a non-simple group is
$\ G=(SU(3)\times SU(3))/(\NZ_3\times\NZ_3)\ $ at level
$\,k=(1,1)\,$ with the adjoint action of $\,\Gamma={\rm diag}
(SU(3)/\NZ_3)$.
\vskip 0.1cm

The non-anomalous gauging of the adjoint action of the diagonal
$\,SO(3)\,$ subgroup in the WZW model with groups $\,(SU(2)\times SU(2))/Z\,$
is used in the coset model construction \cite{GKO} of the unitary minimal
models of conformal field theory \cite{GK,GB,Hori}. \,Other coset theories
involve other versions of gauged WZW amplitudes and may suffer from
global anomalies, as will be discussed below.

\subsection{Anomalies and WZW partition functions}
\label{sec:WZWextern}

\no The results of the calculation of the global-gauge-anomaly phases in the
last section are consistent with the exact solution for the toroidal partition
functions of the WZW models of conformal field theory in an external
gauge field.

Let us start by considering the level $\,k\,$ WZW sigma model on a
closed Riemann surface $\,\Sigma\,$ with the Lie group $\,G=\tilde
G/Z\,$ as the target manifold. \,The Feynman amplitude of a field $\
\varphi:\Sigma\rightarrow G\ $ in the background of the external
gauge field described by a $\,\Ng$-valued 1-form $\,A\,$ on $\Sigma\,$
is given by the formula
\qq
{\bm A}_{W\hspace{-0.03cm}ZW}(\varphi,A)\
=\ \exp\hspace{-0.05cm}\Big(\frac{_i}{^{4\pi}}\int\limits_\Sigma
k\,\tr\,\varphi^{-1}(\partial_A\varphi)(\varphi^{-1}\bar\partial_A\varphi)\Big)
\,\,{\bm \CA}_{W\hspace{-0.02cm}Z}(\varphi,A)\,,
\label{Awzw}
\qqq
where $\,\partial_A=\partial+ad_{A^{10}}\,$ and $\,\bar\partial_A
=\bar\partial+\ad_{A^{01}}\ $ are the minimally coupled Dolbeault
differentials relative to the complex structure of $\,\Sigma$, \,for
$\,A=A^{10}+A^{01}$. \,The WZ amplitude $\,{\bm
A}_{W\hspace{-0.02cm}Z}(\varphi,A)\,$ is related to the holonomy of
the WZW gerbe $\,\CG_k\,$ on $\,G$, \,with the adjoint action of the
group $\,\Gamma=\tilde G/\tilde Z\,$ gauged as described previously.

Let $\,\Sigma=\CT_\tau:=\NC/(2\pi\NZ+2\pi\tau\NZ)\ $ be the complex torus
with the modular parameter $\,\tau=\tau_1+i\tau_2$, \,where the imaginary
part $\,\tau_2>0$. \,The toroidal partition function is formally defined
by the functional integral over the space of maps
$\,\varphi:\CT_\tau\rightarrow G\,$
\qq
\CZ^{G}(\tau,A)\ \
=\mathop{\int}\limits_{Map(\CT_\tau,G)}\hspace{-0.2cm}
{\bm A}_{W\hspace{-0.03cm}ZW}(\varphi,A)\,\,D\varphi\,.
{}
\qqq
Its exact form may be found from (formal) symmetry properties of
the functional integral. \,The result has a specially
simple form for the gauge fields
\qq
A_u\,=\,\frac{_{\bar u\,dw-u\,d\bar w}}{^{2\tau_2}}
{}
\qqq
with $\,u\,$ in the complexified Cartan algebra $\,\Nt_\Ng^\NC\,$ of
$\,\Ng\,$ and $\,w\,$ the coordinate on the complex plane \,(for other
gauge fields, it is
then determined by chiral gauge transformations \cite{geom}). \,When
the group $\,G\,$ is simply connected, \,i.e.\ $G=\tilde G$, \,one
has
\qq
\CZ^{\tilde G}(\tau,A_u)\ =\ \sum\limits_{\Lambda\in P^+_k(\Ng)}\,|
\chi^{\hat\Ng}_{k,\Lambda}(\tau,u)|^2\,
\,\exp\hspace{-0.05cm}\big[\frac{_{\pi k\,\tr\,(u-\bar u)^2}}{^{2\tau_2}}\big]\,,
{}
\qqq
where $\,\chi^{\hat\Ng}_{k,\Lambda}\,$ are the affine characters,
\qq
\chi^{\hat\Ng}_{k,\Lambda}(\tau,u)\ =\ \tr_{V^{\hat\Ng}_{k,\Lambda}}\,\,
\exp\hspace{-0.05cm}\Big(2\pi i\big[\tau\,\big(L_0^{\hat\Ng}
-\frac{_{c_k^{\hat\Ng}}}{^{24}}\big)\,+\,u\big]\Big),
{}
\qqq
of the unitary highest-weight modules $\,V^{\hat\Ng}_{k,\Lambda}\,$
of level $\,k\,$ and highest weight $\,\Lambda\,$ of the affine
algebra $\,\hat\Ng\,$ associated to the Lie algebra $\,\Ng\,$ \cite{Kac,GepWit}.
$\,L_0^{\hat\Ng}\,$ stands for the corresponding Sugawara-Virasoro
generator and $\,c^{\hat\Ng}_k\,$ for the Virasoro central charge.
\,The admissible highest weights $\,\Lambda\,$ form
a finite set $\,P_k^+(\Ng)$. \,We consider
weights as elements of $\,i\Nt_\Ng$,
\,identifying the latter space with its dual by means of the bilinear
form $\,\tr$.

For non-simply connected groups $\,G=\tilde G/Z$, \,the toroidal
partition functions take a more complicated form \cite{FGK}.  \,The
space of (regular) maps from $\,\CT_\tau\,$ to $\,G\,$ has different
connected components that may be labeled by the windings:
\qq
Map(\CT_\tau,G)\
=\hspace{-0.1cm}\mathop{\sqcup}\limits_{(z_1,z_2)\in Z^2}
Map_{_{z_1,z_2}}(\CT_\tau,\tilde G)\,,
{}
\qqq
where for $\,z_i=\ee^{\,2\pi\si p^\vee_i}$, $\ Map_{_{z_1,z_2}}\ $
contains the maps homotopic to $\,\varphi_{p^\vee_1,p^\vee_2}\,$ of
\,Eq.\,(\ref{mnm'n'}) (viewed as a map on $\,\CT_\tau\,$ via the
parametrization of the complex plane by
$\,w=\sigma_1+\tau\sigma_2$). \,\,Let
\qq
\CZ^{G}_{_{z_1,z_2}}(\tau,A)\ =\mathop{\int}\limits_{Map_{_{z_1,z_2}}
(\CT_\tau,G)}\hspace{-0.3cm}
{\bm A}_{W\hspace{-0.03cm}ZW}(\varphi,A)\,\,D\varphi
\label{fitw}
\qqq
so that
\qq
\CZ^{G}(\tau,A)\ =\ \sum\limits_{(z_1,z_2)\in Z^2}
\CZ^{G}_{_{z_1,z_2}}(\tau,A)\,.
\label{dueto}
\qqq
By writing $\,\varphi=\varphi_{p^\vee_1,p^\vee_2}\tilde\varphi$,
\,where $\,\tilde\varphi\,$ has trivial windings and may be lifted
to a map from $\,\CT_\tau\,$ to $\,\tilde G$, \,one may relate the
functional integral for $\,\CZ^{G}_{_{z_1,z_2}}(\tau,A)\,$ to the
one for $\,Z^{\tilde G}(\tau,A)\,$ using the chiral Ward identities
\cite{geom}. \,One obtains this way the formula
\qq
\CZ^{G}_{_{z_1,z_2}}(\tau,A_u)\,&=&\,\frac{_1}{^{|Z|}}\,
\sum\limits_{\Lambda\in P^+_k(\Ng)}\,
H\hspace{-0.04cm}ol_{\CG_k}(\varphi_{p^\vee_1,p^\vee_2})\
\exp\hspace{-0.05cm}\big[-ik\,\tr\,p^\vee_1(p^\vee_2-\tau p^\vee_1)\,-\,
2\pi\si k\,\,tr\,u\hspace{0.025cm}p^\vee_1\big]\,\cr\cr
&&\hspace{3cm}\cdot\ \chi^{\hat\Ng}_{k,\Lambda}(\tau,u+p^\vee_2-\tau p^\vee_1)\
\overline{\chi^{\hat\Ng}_{k,\Lambda}(\tau,u)}\ \,
\exp\hspace{-0.05cm}\big[\frac{_{\pi k\,\tr\,(u-\bar u)^2}}{^{2\tau_2}}\big],
\label{trew}
\qqq
where $\,|Z|\,$ stands for the cardinality of $\,Z\,$ and the values
of $\,H\hspace{-0.04cm}ol_{\CG_k}(\varphi_{p^\vee_1,p^\vee_2})\,$
may be found in Sec.\,IV of \cite{GW}. \,There exists a {\it
spectral flow} $\ \Lambda\mapsto z\Lambda\ $ on $\,P_k^+(\Ng)\,$
(and on the set of the corresponding highest-weight modules of
$\,\hat\Ng$) \,induced by the elements $\,z\,$ of the center
$\,\tilde Z\,$ of $\,\tilde G\,$ \cite{FGK}. \,The highest weight
$\,z\Lambda\,$ is uniquely fixed by the property that
\qq
\ee^{\,2\pi\si\,k^{-1}z\Lambda}=z\,Ad_{w_z}(\ee^{\,2\pi\si\,k^{-1}\Lambda})
\label{flow}
\qqq
for some $\,w_z\,$ in the normalizer of the Cartan subgroup of
$\,\tilde G$. \,The characters of the $\,\hat\Ng$-modules with the
highest weights connected by the spectral flow satisfy the relation
\qq
&&\exp\hspace{-0.05cm}\big[-\pi\si\,k\,\tr\,(p^\vee_2
-\tau p^\vee_1)p^\vee_1\,
-\,2\pi\si k\,\tr\,u\hspace{0.025cm}
p^\vee_1\big]\ \,\chi^{\hat\Ng}_{k,\Lambda}(\tau,
u+p^\vee_2-\tau p^\vee_1)\cr\cr
&&\hspace{1.5cm}=\ \exp\hspace{-0.05cm}\big[2\pi\si\,\tr\,p^\vee_2
\Lambda\,-\,\pi\si\,k\,\tr\,p^\vee_1p^\vee_2\big]\ \,
\chi^{\hat\Ng}_{k,z_1^{-1}\Lambda}(\tau,u)
\label{flchar}
\qqq
for any $\,p^\vee_1\,$ and $\,p^\vee_2\,$ in the coweight lattice
$\,P^{^\vee}_\Ng$. \,As a result, \,Eq.\,(\ref{trew}) may be rewritten
in the form
\qq
\CZ^{G}_{_{z_1,z_2}}(\tau,A_u)\ =\
\frac{_1}{^{|Z|}}
\sum\limits_{\Lambda\in P_k^+(\Ng)}\hspace{-0.2cm}
\epsilon_{_{z_1,z_2}}(\Lambda)\ \,\chi^{\hat\Ng}_{k,z_1^{-1}\Lambda}
(\tau,u)\ \,\overline{\chi^{\hat\Ng}_{k,\Lambda}(\tau,u)}\ \,
\exp\hspace{-0.05cm}\big[\frac{_{\pi k\,\tr\,(u-\bar u)^2}}{^{2\tau_2}}\big]\,,
\label{twpf}
\qqq
where
\qq
\epsilon_{_{z_1,z_2}}(\Lambda)\
=\ H\hspace{-0.04cm}ol_{\CG_k}(\varphi_{p^\vee_1,p^\vee_2})\,\,
\exp\hspace{-0.05cm}\big[2\pi\si\,\tr\,p^\vee_2\Lambda\,
-\,\pi\si\,k\,\tr\,p^\vee_1p^\vee_2\big]
{}
\qqq
defines a character on $\,Z\,$ through its dependence on $\,z_2$.
\,Let, for $\,z\in Z$,
\qq
C_{z}\ :=\ \big\{\,\Lambda\in P^+_k(\Ng)\ \,\big|\,\
\epsilon_{{z,z_2}}(\Lambda)
=1\quad{\rm for\ all}\quad z_2\in Z\,\big\}\,.
\label{C_m}
\qqq
Summing both sides of Eq.\,(\ref{twpf}) over $\,z_1\,$ and $\,z_2$, \,
one obtains the following formula for the complete partition
function of the group $\,G\,$ WZW model at level $\,k\,$:
\qq
\CZ^G(\tau,A_u)\ =\ \sum\limits_{z\in Z}
\sum\limits_{\Lambda\in P^+_k(\Ng)\cap C_{z}}
\chi^{\hat\Ng}_{k,z^{-1}\Lambda}
(\tau,u)\ \,\overline{\chi^{\hat\Ng}_{k,\Lambda}(\tau,u)}\ \,
\exp\hspace{-0.05cm}\big[\frac{_{\pi k\,\tr\,(u-\bar u)^2}}{^{2\tau_2}}\big]\,.
{}
\qqq
The space of states of the model that can be read off from this formula
has the form \cite{FGK,KS}
\qq
\mathbb{H}^G\ =\ \mathop{\oplus}\limits_{z\in Z}
\Big(\mathop{\oplus}\limits_{\Lambda\in P^+_k(\Ng)\cap C_{z}}
V^{\hat\Ng}_{k,z^{-1}\Lambda}\otimes\,
\overline{V^{\hat\Ng}_{k,\Lambda}}\,\Big)\,.
{}
\qqq

The transformation properties of the WZW partition function (\ref{twpf})
under large gauge transformations $\,\Ch_{\tilde p^\vee_1,\tilde p^\vee_2}\,$
of \,Eq.\,(\ref{mnm'n'}) \,are determined by the equality
\qq
\Ch_{\tilde p^\vee_1,\tilde p^\vee_2}A_u\ =\ A_{u-\tilde p^\vee_2+\tau\tilde p^\vee_1}\,,
{}
\qqq
and by identity (\ref{flchar}) for the affine characters. \,With
help of these relations, \,one obtains
\qq
&&\CZ^{G}_{_{z_1,z_2}}(\tau,\Ch_{\tilde p^\vee_1,\tilde p^\vee_2}A_u)\
=\ \CZ^{G}_{_{z_1,z_2}}(\tau,A_{u-\tilde p^\vee_2+\tau\tilde p^\vee_1})\cr\cr
&&=\
\frac{_1}{^{|Z|}}\sum\limits_{\Lambda\in P_k^+(\Ng)}
\hspace{-0.2cm}\epsilon_{_{z_1,z_2}}(\Lambda)\
\exp\hspace{-0.05cm}\big[-2\pi i\,\tr\,
\tilde p^\vee_2(z_1^{-1}\Lambda-\Lambda)\big]\ \,
\chi^{\hat\Ng}_{k,\tilde z_1z_1^{-1}\Lambda}(\tau,u)\
\overline{\chi^{\hat\Ng}_{k,\tilde z_1\Lambda}(\tau,u)}\ \,
\exp\hspace{-0.05cm}\big[\frac{_{\pi k\,\tr\,(u-\bar u)^2}}{^{2\tau_2}}\big]\,,
\quad\qquad
\label{lst}
\qqq
where, as before, $\,\tilde z_i=\ee^{\,2\pi\si\tilde p^\vee_i}$.
\,It is easy to see, using Eq.\,(\ref{flow}), that
\qq
\exp\hspace{-0.05cm}\big[-2\pi i\,\tr\,\tilde p^\vee_2
(\tilde z_1^{-1}\Lambda-\Lambda)\big]
\,=\,\exp\hspace{-0.05cm}\big(2\pi i\,k\,\tr\,\tilde p^\vee_1
\tilde p^\vee_2\big)
\label{ltrn}
\qqq
for any $\,\Lambda\in P_k^+(\Ng)$. \,Replacing $\,\Lambda\,$ by
$\,\tilde z_1^{-1}\Lambda\,$ on the right-hand side of
\,Eq.\,(\ref{lst}) and using the relation
\qq
\epsilon_{_{z_1,z_2}}(\tilde z_1^{-1}\Lambda)\,
=\,\exp\hspace{-0.05cm}\big[-2\pi\si\,k\,\tr\,\tilde p^\vee_1p^\vee_2\big]
\ \epsilon_{_{z_1,z_2}}(\Lambda)
\label{epseps}
\qqq
that follows from Eq.\,(\ref{ltrn}), \,one obtains
\vskip 0.5cm

\begin{proposition}
\label{prop:anpf} The transformation law of the toroidal partition
function (\ref{twpf}) under large gauge transformations is described
by the identity
\qq
\CZ^{G}_{_{z_1,z_2}}(\tau,\Ch_{\tilde p^\vee_1,\tilde p^\vee_2}A_u)\
=\ c_{(\tilde z_1,\tilde z_2),
(z_1,z_2)}\
\CZ^{G}_{_{z_1,z_2}}(\tau,A_u)\,,
\label{anpf}
\qqq
where the phases $\,c_{(\tilde z_1,\tilde z_2),(z_1,z_2)}\,$ are given by
Eq.\,(\ref{ugf}).
\end{proposition}
\vskip -0.5cm
\hspace{13cm}$\blacksquare$
\vskip 0.2cm

If we assume the gauge invariance $\,D\varphi=D(\Ch\varphi)\,$ of
the formal functional integral measure, then the above anomalous
transformation property follows from the functional integral
expression (\ref{fitw}) and the relation
\qq
{\bm A}_{W\hspace{-0.03cm}ZW}(\Ch_{\tilde p^\vee_1,\tilde p^\vee_2}\varphi,
\Ch_{\tilde p^\vee_1,\tilde p^\vee_2}A)\ =\
c_{(\tilde z_1,\tilde z_2),(z_1,z_2)}\,\,{\bm A}_{W\hspace{-0.03cm}ZW}(\varphi,A)
{}
\qqq
for $\ \varphi\in Map_{_{z_1,z_2}}(\CT_\tau,G)$ which is a
consequence of Eq.\,(\ref{gganom}) (the minimally coupled term
of the WZW action (\ref{Awzw}) is invariant under all gauge
transformations).

As an example, let us consider the simplest gauged WZW model that
exhibits a global gauge anomaly, namely the one with the target
group $\,G=SU(3)/\NZ_3\,$ at level $\,k=1\,$ and the gauged adjoint
action of $\,\Gamma=G$. \,For the  simple coweights of $\,su(3)\,$
(identified with the simple weights), we may take
\qq
\lambda_1^\vee={\rm diag}[\frac{_2}{^3},-\frac{_1}{^3},-\frac{_1}{^3}]
=\lambda_1\,,\qquad\lambda_2^\vee={\rm diag}[\frac{_1}{^3},\frac{_1}{^3},
-\frac{_2}{^3}]=\lambda_2\,.
{}
\qqq
The element $\ z=\ee^{\,2\pi\si\lambda_1^\vee}={\rm diag}
[\ee^{\,\frac{4\pi\si}{3}},
\ee^{-\frac{2\pi\si}{3}},\ee^{-\frac{2\pi\si}{3}}]\ $ generates the
center $\,\NZ_3\,$ of $\,SU(3)$. \,The set $\,P^+_1(su(3))\,$
contains three weights $\ \Lambda=r_1\lambda_1+r_2\lambda_2\ $ with
$\,(r_1,r_2)=(0,0),(1,0),(0,1)$. \,We shall denote the corresponding
level 1 affine characters by $\,\hat\chi_{(r_1,r_2)}$. \,The toroidal
partition functions $\
\tilde\CZ^G(\tau,u):=\CZ^G(\tau,u)\,\exp\hspace{-0.05cm}
\big[-\frac{_{\pi k\,\tr\,(u-\bar u)^2}}{^{2\tau_2}}\big]\ $ with
fixed windings are, according to Eq.\,(\ref{twpf}),
\qq
&&\tilde\CZ^{G}_{1,1}\ =\ \frac{_1}{^3}\big(|\hat\chi_{(0,0)}|^2
+|\hat\chi_{(1,0)}|^2+|\hat\chi_{(0,1)}|^2\big)\,,\cr\cr
&&\tilde\CZ^{G}_{1,z}\ =\ \frac{_1}{^3}\big(|\hat\chi_{(0,0)}|^2
+\ee^{\,\frac{4\pi\si}{3}}|\hat\chi_{(1,0)}|^2
+\ee^{\,\frac{2\pi\si}{3}}|\hat\chi_{(0,1)}|^2\big)\,,
\cr\cr
&&\tilde\CZ^{G}_{1,z^2}\ =\ \frac{_1}{^3}\big(|\hat\chi_{(0,0)}|^2
+\ee^{\frac{\,2\pi\si}{3}}|\hat\chi_{(1,0)}|^2
+\ee^{\,\frac{\,4\pi\si}{3}}|\hat\chi_{(0,1)}|^2\big)\,,
\cr\cr
&&\tilde\CZ^{G}_{z,1}\ =\ \frac{_1}{^3}\big(\hat\chi_{(0,1)}\overline{\hat\chi_{(0,0)}}
+\hat\chi_{(0,0)}\overline{\hat\chi_{(1,0)}}
+\hat\chi_{(1,0)}\overline{\hat\chi_{(0,1)}}\big)\,,
\cr\cr
&&\tilde\CZ^{G}_{z,z}\ =\ \frac{_1}{^3}\big(\ee^{\,\frac{4\pi\si}{3}}
\hat\chi_{(0,1)}
\overline{\hat\chi_{(0,0)}}+\ee^{\,\frac{2\pi\si}{3}}\hat\chi_{(0,0)}\overline{\hat\chi_{(1,0)}}
+\hat\chi_{(1,0)}\overline{\hat\chi_{(0,1)}}\big)\,,
\cr\cr
&&\tilde\CZ^{G}_{z,z^2}\ =\ \frac{_1}{^3}\big(\ee^{\,\frac{2\pi\si}{3}}
\hat\chi_{(0,1)}\overline{\hat\chi_{(0,0)}}+\ee^{\,\frac{4\pi\si}{3}}\hat\chi_{(0,0)}
\overline{\hat\chi_{(1,0)}}+\hat\chi_{(1,0)}\overline{\hat\chi_{(0,1)}}\big)\,,
\cr\cr
&&\tilde\CZ^{G}_{z^2,1}\ =\ \frac{_1}{^3}\big(\hat\chi_{(1,0)}
\overline{\hat\chi_{(0,0)}}+\hat\chi_{(0,1)}\overline{\hat\chi_{(1,0)}}
+\hat\chi_{(0,0)}\overline{\hat\chi_{(0,1)}}\big)\,,
\cr\cr
&&\tilde\CZ^{G}_{z^2,z}\ =\ \frac{_1}{^3}\big(\ee^{\,\frac{2\pi\si}{3}}
\hat\chi_{(1,0)}
\overline{\hat\chi_{(0,0)}}+\hat\chi_{(0,1)}\overline{\hat\chi_{(1,0)}}
+\ee^{\,\frac{4\pi\si}{3}}\hat\chi_{(0,0)}\overline{\hat\chi_{(0,1)}}\big)\,,
\cr\cr
&&\tilde\CZ^{G}_{z^2,z^2}\ =\ \frac{_1}{^3}\big(\ee^{\,\frac{4\pi\si}{3}}
\hat\chi_{(1,0)}
\overline{\hat\chi_{(0,0)}}+\hat\chi_{(0,1)}\overline{\hat\chi_{(1,0)}}
+\ee^{\,\frac{2\pi\si}{3}}\hat\chi_{(0,0)}\overline{\hat\chi_{(0,1)}}\big)\,.
\nonumber
\qqq
Since
\qq
c_{(z^{\tilde p_1},z^{\tilde p_2}),(z^{p_1},z^{p_2})}\ =\ \exp\big(\frac{_{4\pi i}}{^3}
(p_1\tilde p_2-\tilde p_1 p_2)\big),
{}
\qqq
the transformation rule \ref{anpf} implies that all the sectors with non-trivial
windings suffer from global gauge anomalies. \,Summing over the windings,
one obtains the total partition function of the level 1
WZW theory for the target group $\,G=SU(3)/\NZ_3\hspace{0.02cm}$:
\qq
\tilde\CZ^G\ =\ |\hat\chi_{(0,0)}|^2+\hat\chi_{(1,0)}\overline{\hat\chi_{(0,1)}}
+\hat\chi_{(0,1)}\overline{\hat\chi_{(1,0)}}\,.
{}
\qqq
It should be contrasted with the anomaly-free level 1 partition
function for the covering group $\,\tilde G=SU(3)\hspace{0.02cm}$:
\qq
\tilde\CZ^{\tilde G}\ =\ |\hat\chi_{(0,0)}|^2+|\hat\chi_{(1,0)}|^2+
|\hat\chi_{(0,1)}|^2\,.
{}
\qqq

\subsection{Implications for coset models}
\label{sec:coset}

\no Consider now the group $\,\Gamma=\tilde H/Z_\Gamma$, \,where
$\,\tilde H\,$ is a connected closed subgroup of $\,\tilde G\,$ with
Lie algebra $\,\Nh\subset\Ng\,$ and $\,Z_\Gamma=\tilde H\cap\tilde Z$.
\,$\,\Gamma=\tilde\Gamma/\tilde Z_\Gamma\,$ where
$\,\tilde\Gamma\,$ is the covering group of $\,\Gamma\,$ (and of
$\,\tilde H$) \,and $\,\tilde Z_\Gamma\,$ is the subgroup of its
center composed of element that project to $\,Z_\Gamma\subset\tilde
H$. \,Of course, one has to distinguish between $\,\tilde
Z_\Gamma\,$ and $\,Z_\Gamma\,$ only if the subgroup $\,\tilde H\,$
is not simply connected. \,The so-called $\,G/\Gamma\,$ coset model
of the conformal field theory is obtained by gauging the adjoint
action of $\,\Gamma\,$ on $\,G=\tilde G/Z\,$ in the group $\,G\,$
level $\,k\,$ WZW model and by integrating out the
gauge fields in the functional integral \cite{BRS,GK,KPSY,Hori}.
\,In particular, the contribution of the topologically trivial
gauge fields to the toroidal partition function of the
$\,G/\Gamma\,$ coset model is formally given by
\qq
\CZ^{G/\Gamma}(\tau)\ =\ \int\,\CZ^{G}(\tau,A)\,\,DA\,,
\label{csfi}
\qqq
where $\,A\,$ are 1-forms on $\,\CT_\tau\,$ with values
in the Lie algebra $\,\Nh$. \,Clearly, due to the decomposition
(\ref{dueto}),
\qq
\CZ^{G/\Gamma}(\tau)\ =\ \sum\limits_{(z_1,z_2)\in Z^2}
\CZ^{G/\Gamma}_{_{z_1,z_2}}(\tau)\qquad\ {\rm with}\qquad\
\CZ^{G/\Gamma}_{_{z_1,z_2}}(\tau)\ =\ \int\,\CZ^{G}_{_{z_1,z_2}}(\tau,A)
\,\,DA\,.
{}
\qqq
The functional integral
(\ref{csfi}) may be computed by an appropriate parametrization of
the gauge fields $\,A\,$ \cite{GK}. \,In particular, when
$\,\Nh\,$ is semi-simple, the result is \cite{GK,Hori}
\qq
\CZ^{G/\Gamma}_{_{z_1,z_2}}(\tau)\ =\
\frac{_1}{^{|\tilde Z_\Gamma|\,|Z|}}\,\sum\limits_{\Lambda\in P_k^+(\Ng)}\,
\sum\limits_{\lambda\in P_{\tilde k}^+(\Nh)}
\hspace{-0.2cm}\epsilon_{_{z_1,z_2}}(\Lambda)\ \,
b^{\,\hat\Ng,\hat\Nh}_{k,z_1^{-1}\Lambda,\lambda}(\tau)\ \,
\overline{b^{\,\hat\Ng,\hat\Nh}_{k,\Lambda,\lambda}(\tau)}\,,
\label{twsectgh}
\qqq
where $\,b^{\,\hat\Ng,\hat\Nh}_{k,\Lambda,\lambda}(\tau)\,$ are the
{\it branching functions} that are the characters of the coset Virasoro
modules $\,V^{\hat\Ng,\hat\Nh}_{k,\Lambda,\lambda}$. \,The latter appear in
the decomposition \cite{Godd}
\qq
V^{\hat\Ng}_{k,\Lambda}\ =\ \mathop{\oplus}
\limits_{\lambda\in P^+_{\tilde k}(\hat\Nh)}
V^{\hat\Ng,\hat\Nh}_{k,\Lambda,\lambda}\otimes V^{\hat\Nh}_{\tilde k,\lambda}
\label{dcmgh}
\qqq
of the level $\,k\,$ unitary highest-weight modules of the affine
algebra $\,\hat\Ng\,$ into similar modules of the affine subalgebra
$\,\hat\Nh\subset\hat\Ng\,$ at the level $\,\tilde k\,$ induced by
restricting the bilinear form $\,k\,\,tr\,$ on $\,\Ng\,$ to $\,\Nh$.
\,By definition,
\qq
b^{\,\hat\Ng,\hat\Nh}_{k,\Lambda,\lambda}(\tau)\ =\
\tr_{V^{\hat\Ng,\hat\Nh}_{k,\Lambda,\lambda}}\,\,
\exp\hspace{-0.05cm}\Big(2\pi i\,\tau\,\big(L_0^{\hat\Ng}-L_0^{\hat\Nh}
-\frac{_{c_k^{\hat\Ng}-c_{\tilde k}^{\hat\Nh}}}{^{24}}\big)\Big)\,.
{}
\qqq
The decomposition (\ref{dcmgh}) implies the one for the characters:
\qq
\chi^{\hat\Ng}_{k,\Lambda}(\tau,u)\ =\ \sum\limits_{\lambda\in P_{\tilde k}^+(\Nh)}
b^{\,\hat\Ng,\hat\Nh}_{k,\Lambda,\lambda}(\tau)\
\,\chi^{\hat\Nh}_{\tilde k,\lambda}(\tau,u)
{}
\qqq
for $\,u\,$ in the complexified Cartan algebra $\,\Nt_\Nh^\NC\,$ of $\,\Nh$.

From the gauge transformation rule (\ref{anpf}), \,we should expect that
the sectors with fixed windings $\,(z_1,z_2)\,$ of the
group $\,G\,$ WZW theory which transform in the anomalous way under
the large gauge transformations $\,\Ch_{\tilde p^\vee_1,\tilde p^\vee_2}:
\CT_\tau\rightarrow\Gamma\,$ give vanishing contributions to the partition
function of the coset theory. This is, indeed, the case.
\vskip 0.5cm

\begin{proposition}
\label{prop:vanish}
If $\ c_{(\tilde z_1,\tilde z_2),(z_1,z_2)}\not=1\ $ for some
$\,(\tilde z_1,\tilde z_2)\in Z_\Gamma^2\,$ then the partition
function $\,\CZ^{G/\Gamma}_{_{z_1,z_2}}\,$ given by Eq.\,(\ref{twsectgh})
vanishes.
\end{proposition}

\begin{proof} Denote by $\,P^{^\vee}_\Gamma\,$ the subset of the set
$\,P^{^\vee}_\Nh\subset\si\Nt_\Nh\subset\si\Nt_\Ng\,$ of coweights
of $\,\Nh\,$ composed of such $\,\tilde p^\vee\,$ that $\,\tilde
z=\ee^{\,2\pi\si\tilde p^\vee}\in\tilde Z_\Gamma\,$ when viewed as
elements of $\,\tilde\Gamma\,$ (or that $\,\tilde z\in Z_\Gamma\,$
when viewed as elements of $\,\tilde H\subset \tilde G$). \,Clearly
$\,P^{^\vee}_\Gamma\subset P^{^\vee}_\Ng$. \,The vanishing result is
a consequence of the following properties of the branching
functions:
\qq
&&b^{\,\hat\Ng,\hat\Nh}_{k,\Lambda,\lambda}\ =\ 0\qquad{\rm if}\qquad
\exp[2\pi \si\,\tr\,\tilde p^\vee\Lambda]\not=\exp[2\pi \si\,
\tr\,\tilde p^\vee\lambda]
\qquad{\rm for\ some}\qquad\tilde p^\vee\in P^{^\vee}_\Gamma\,,
\quad\qquad\label{of1}
\\ \cr
&&b^{\,\hat\Ng,\hat\Nh}_{k,{\tilde z}\Lambda,{\tilde z}\lambda}\
=\ b^{\,\hat\Ng,\hat\Nh}_{k,\Lambda,\lambda}\qquad{\rm for}\,\qquad\,
\tilde z=\ee^{\,2\pi\si\tilde p^\vee}\qquad{\rm and}\qquad\tilde p^\vee
\in P^{^\vee}_\Gamma\,.
\label{of2}
\qqq
The first of these relations follows from the fact that the central
elements $\,\ee^{2\pi\si\tilde p^\vee}\,$ act by multiplication by
the same scalars in the modules $\,V^{\hat\Ng}_{k,\Lambda}\,$ and
$\,V^{\hat\Nh}_{\tilde k,\lambda}\,$ appearing in the decomposition
(\ref{dcmgh}). \,The second one is a consequence of the isomorphism
between the coset Virasoro modules $\,V^{\hat\Ng,\hat\Nh}_{k,\Lambda,
\lambda}\,$ with the weights related by the spectral flows under
elements $\,\ee^{\,2\pi\si\tilde p^\vee}$.
\,Note that both relations are consistent with the fact
that the identity (\ref{flchar}) is satisfied by the characters of
both affine algebras $\,\hat\Ng\,$ and $\,\hat\Nh$.

If $\,c_{(\tilde z_1,\tilde z_2),(z_1,z_2)}\not=1$, \,then either $\
\exp[2\pi\si k\,\tr\,p^\vee_1\tilde p^\vee_2]\not=1\ $ or $\
\exp[-2\pi\si k\,\tr\,\tilde p_1^\vee p^\vee_2]\not=1\ $ for some
$\,\tilde p^\vee_i\in P^{^\vee}_\Gamma$, \,see Eq.\,(\ref{ugf}).
\,Relation (\ref{of1}) implies that if $\ \exp[2\pi
\si\,k\,\tr\,p^\vee_1\tilde p^\vee_2] =\exp[-2\pi\si\,\tr\,\tilde
p^\vee_2(z_1^{-1}\Lambda-\Lambda)]\not=1\ $ for some $\,\tilde p^\vee_2\in
P^{^\vee}_\Gamma$, \,then, for each pair $\,(\Lambda,\lambda)$,
\,either $\ b^{\,\hat\Ng,\hat\Nh}_{k,z_1^{-1}\Lambda,\lambda}=0\ $ or
$\ b^{\,\hat\Ng,\hat\Nh}_{k,\Lambda,\lambda}=0\ $ so that
$\,\CZ^{G/\Gamma}_{_{z_1,z_2}}\,$ vanishes. \ Similarly, using
relation (\ref{of2}) and Eq.\,(\ref{epseps}), \,we infer that if
$\ \exp[-2\pi\si\,k\,\tr\,\tilde p^\vee_1p^\vee_2]\not=1\ $ for some
$\,\tilde p^\vee_1\in P^{^\vee}_\Gamma$, \,then
$\,\CZ^{G/\Gamma}_{_{z_1,z_2}}\,$ vanishes too.
\vskip -0.8cm
\
\end{proof}
\vskip 0.2cm

\no As we see, global gauge anomalies in the WZW model lead to
selection rules for the contributions to the partition functions of
the $\,G/\Gamma\,$ coset model.

Let $\,Z'\subset Z\,$ be the {\it non-anomalous} subgroup that is
composed of the elements $\,z=\ee^{\,2\pi\si p^\vee}\in Z\,$ such
that $\ \exp[2\pi\si\,k\,\tr\,p^\vee\tilde p^\vee]=1\ $ for all
$\,\tilde p^\vee\in P^{^\vee}_\Gamma$, \,and let $\,G'=\tilde
G/Z'\,$ be the corresponding quotient of $\,\tilde G$. \,Proposition
\ref{prop:vanish} and Eqs.\,(\ref{twsectgh}) imply that
\qq
\CZ^{G/\Gamma}(\tau)\ =\ \frac{_{|Z^{^\prime}|}}{^{|Z|}}\,\,\CZ^{G'/\Gamma}(\tau)\,.
\label{ZGG'}
\qqq
Upon summation over windings in $\,(Z')^2$, \,the partition
function on the right-hand side may be rewritten in the form
\qq
\CZ^{G'/\Gamma}(\tau)\ =\
\frac{_1}{^{|\tilde Z_\Gamma|}}\,\sum\limits_{z\in Z'}\,
\sum\limits_{\Lambda\in P_k^+(\Ng)\cap C^{^\prime}_{z}}\,
\sum\limits_{\lambda\in P_{\tilde k}^+(\Nh)}
b^{\,\hat\Ng,\hat\Nh}_{k,z^{-1}\Lambda,\lambda}(\tau)\ \,
\overline{b^{\,\hat\Ng,\hat\Nh}_{k,\Lambda,\lambda}(\tau)}\,,
\label{twsectgh1}
\qqq
where $\,C'_{z}\,$ is defined as in (\ref{C_m}) but with the
subgroup $\,Z\,$ replaced by $\,Z'$. \,Due to relation (\ref{of1}),
we may restrict the sum on the right-hand side to pairs
$\,(\Lambda,\lambda)\,$ such that the elements of
$\,\ee^{\,2\pi\si\tilde p^\vee}\,$ for $\,\tilde p^\vee\in
P^{^\vee}_\Gamma\,$ act by multiplication by the same scalar in
$\,V^{\hat\Ng}_{k,\Lambda}\,$ and in $\,V^{\hat\Nh}_{\tilde
k,\lambda}$. \,Then, also the pairs $\,(z^{-1}\Lambda,\lambda)\,$
for $\,z\in Z'\,$ and $\,(\tilde z\Lambda,\tilde z\lambda)\,$ for
$\,\tilde z =\ee^{\,2\pi\si\tilde p^\vee}\,$ will have this property
due to Eq.\,(\ref{ltrn}). \,Besides, it follows from
Eq.\,(\ref{epseps}) that if $\,\Lambda\in C'_{z}\,$ then $\,\tilde
z\Lambda\in C'_{z}\,$ for all $\,\tilde z\in Z_\Gamma\,$ (unlike for
$\,C_{z}\,$ if $\,Z'\,$ is strictly smaller than $\,Z$). \,As a
result of this observations and of relation (\ref{of2}), \,one may
rewrite the sum over weights on the right-hand side of
Eq.\,(\ref{twsectgh1}) as a sum over orbits $\,[\Lambda,\lambda]\,$
of the diagonal spectral flow of $\,\tilde Z_\Gamma$. \,Denoting by
$\,\CP_{z}\,$ the space of such orbits with $\,\Lambda\in C'_{z}$,
\,we infer that
\qq
\CZ^{G'/\Gamma}(\tau)\ =\ \,\sum\limits_{z\in Z'}\,
\sum\limits_{[\Lambda,\lambda]\in\CP_{z}}
\frac{_1}{^{|S_{[\Lambda,\lambda]}|}}\,\,
b^{\,\hat\Ng,\hat\Nh}_{k,z^{-1}\Lambda,\lambda}(\tau)\ \,
\overline{b^{\,\hat\Ng,\hat\Nh}_{k,\Lambda,\lambda}(\tau)}\,,
\label{twsectgh2}
\qqq
where $\,S_{[\Lambda,\lambda]}\subset\tilde Z_\Gamma\,$ denotes the
stabilizer subgroup of the elements of the orbit
$\,[\Lambda,\lambda]$. \,If $\,S_{[\Lambda,\lambda]}\,$ is trivial
for all orbits $\,[\Lambda,\lambda]\,$ then the last expression for
the partition function $\,\CZ^{G'/\Gamma}(\tau)\,$ is consistent
with the following form of the space of states:
\qq
{\mathbb{H}}^{G'/\Gamma}\ =\ \mathop{\oplus}\limits_{z\in Z'}
\Big(\mathop{\oplus}\limits_{[\Lambda,\lambda]\in\CP_{z}}
V^{\hat\Ng,\hat\Nh}_{k,z^{-1}\Lambda,\lambda}\otimes
\overline{V^{\hat\Ng,\hat\Nh}_{k,\Lambda,\lambda}}\,\Big)
{}
\qqq
Identity (\ref{ZGG'}) now implies that, on the contrary, barring
further identifications of the coset Virasoro representations
\cite{maverick}, the partition function $\,Z^{G/\Gamma}(\tau)\,$
lacks a Hilbert-space interpretation if the group $\,Z'\,$ is
strictly smaller than $\,Z$, \,i.e. if the group $\,G\,$ WZW model
suffers from global gauge anomalies relative to the adjoint action
of $\,\Gamma$. \,This points to the inconsistency of the
$\,G/\Gamma\,$ coset model in that case. On the level of the
partition function, this inconsistency is of a mild nature since one
may turn the inconsistent partition function $\,\CZ^{G/\Gamma}\,$
into the consistent one $\,\CZ^{G'/\Gamma}\,$ by changing the
normalization.

In the case when $\,G=SU(3)/\NZ_3=\Gamma$, \,the $\,G/\Gamma\,$
coset theory is topological and its partition function is
$\,\tau$-independent. The branching functions vanish if
$\,\Lambda\not=\lambda\,$ and are equal to 1 otherwise. At level 1,
all coset partition functions with non-trivial windings vanish and
\qq
\CZ^{G/\Gamma}\ =\ \CZ^{G/\Gamma}_{1,1}\ =\ \frac{_1}{^3}\,.
{}
\qqq
In a consistent two-dimensional  topological field theory, the
partition function is equal to the dimension of the space of states
and cannot take a fractional value, confirming the inconsistency of
the level 1 $\,G/\Gamma\,$ coset model for $\,G=SU(3)/\NZ_3=\Gamma$.
\,On the other hand, the non-anomalous subgroup $\,Z'\subset
Z=\NZ_3\,$ is trivial so that $\,G'=\tilde G=SU(3)\,$ in that case,
and for the anomaly-free level 1 $\,\tilde G/\Gamma\,$ coset theory,
\qq
\CZ^{\tilde G/\Gamma}\ =\ 1\,,
\qqq
corresponding to a 1-dimensional space of states.

It was pointed out in \cite{SY} (for the diagonal coset models
corresponding to simply connected groups $\,G=\tilde G=G'$) that, in
the presence of fixed points $\,(\Lambda_0,\lambda_0)\,$ of the
diagonal spectral flow of $\,\tilde Z_\Gamma$, \,there is a further
problem with the Hilbert space interpretation of the partition
function (\ref{twsectgh2}) because of the appearance of the fraction
$\,\frac{1}{|S_{[\Lambda_0,\lambda_0]}|}$. It was shown in
\cite{FSS} within an algebraic approach how to resolve such fixed
points to repair this defect. \,Somewhat earlier, in \cite{Hori}, it
was argued that the problem may be resolved on the Lagrangian level
by adding to the partition function (\ref{twsectgh2}) contributions
from the sectors with gauge fields in the topologically non-trivial
principal $\,\Gamma$-bundles $\,P\,$ over $\,\CT_\tau\,$ (it was
also shown that such contributions vanish if there are no fixed
points $\,(\Lambda_0,\lambda_0)\,$ of the diagonal spectral flow of
$\,\tilde Z_\Gamma$). \,For the sectors with topologically
non-trivial gauge fields, the WZW sigma model fields are sections of
the associated bundle $\,P\times_\Gamma \hspace{-0.05cm}G\,$ with
respect to the adjoint action of $\,\Gamma\,$ and the globally gauge
invariant WZW amplitudes in the gauge field background may be
defined with the help of a $\,\Gamma$-equivariant structure on the
WZW gerbe $\,\CG_k\,$ over $\,G$, \,as will be explained in the
following section.
\vskip -0.9cm

\

\nsection{Coupling to general gauge fields}
\label{sec:generalgaugefields}
\subsection{Equivariant gerbes}
\label{sec:equivgerbes}

\

\vskip -0.9cm

\no We showed in Sec.\,\ref{sec:trivialgaugefields} that the
invariance of the Feynman amplitudes (\ref{FAg1}) under all
gauge transformations requires the existence of
a 1-isomorphism between the gerbes $\,\CG_{12}\equiv\ell^*\CG\,$ and
$\,\CI_{\rho}\otimes\CG_2\,$ over $\,\Gamma\times M$. \,Here, we shall
strengthen this property by introducing the notion of $\,\Gamma$-equivariant
gerbes in the way that will subsequently assure the gauge invariance
of the Feynman amplitudes coupled to topologically non-trivial gauge fields.

\begin{definition}
\label{def:equivgerbe}A gerbe $\CG$ with the curvature $H$ possessing
a $\,\Gamma$-equivariantly closed extension $\,\hat H(X)=H+v(X)\,$
will be called $\,\Gamma$-equivariant relative to the 2-form
$\,\rho\,$ given by Eq.\,(\ref{rho}) if it is equipped with a pair
$(\alpha,\beta)$, \,called a $\,\Gamma$-equivariant structure,
\,such that
\begin{enumerate}
\item[(i)]
$\alpha:\ \CG_{12}\,\rightarrow\ \CI_{\rho}\otimes\CG_2$ \quad is a
1-isomorphism of gerbes over $\,\Gamma \times M$;

\item[(ii)]
$\beta:\,(\Id\otimes\alpha_{2,3})\circ\alpha_{1,23}$\,
{\large$\Rightarrow$}\,\,$\alpha_{12,3}$\quad is a 2-isomorphism of
1-isomorphisms of gerbes over $\,\Gamma^2\times M$;

\item[(iii)]
the following diagram of 2-isomorphisms between 1-isomorphisms of
gerbes over $\,\Gamma^3\times M\,$ is commutative:
\end{enumerate}
\vskip -0.2cm
\qq
\xymatrix@C=-2cm@R=1.4cm{&(\Id\otimes\alpha_{3,4})\circ(\Id\otimes\alpha_{2,34})
\circ\alpha_{1,234} \ar@{=>}[dl]_>>>>>>>{(\Id\otimes\beta_{2,3,4})\circ\Id}  \ar@{=>}[dr]^>>>>>>>{\Id\circ\beta_{1,2,34}} &\cr
(\Id\otimes\alpha_{23,4})\circ\alpha_{1,234} \ar@{=>}[dr]_{\beta_{1,23,4}} &&
(\Id\otimes\alpha_{3,4})\circ\alpha_{12,34} \ar@{=>}[dl]^{\beta_{12,3,4}}
\label{axiom}\\&\alpha_{123,4}&}
{}
\qqq
\end{definition}

$\Gamma$-equivariant gerbes over $\,M\,$
form a 2-category $\Ggrb MG$.
\,A 1-isomorphism between two $\,\Gamma$-equivariant gerbes,
\qq
(\chi,\eta)\,:\ (\CG^a,\alpha^a,\beta^a)\,\rightarrow\,
(\CG^b,\alpha^b,\beta^b),
{}
\qqq
is a  1-isomorphism $\,\chi:\CG^a\rightarrow\CG^b\,$
and a 2-isomorphism $\ \eta:\,(\Id\otimes\chi_2)\circ \,\alpha^a\,\Rightarrow
\,\alpha^b\circ\chi_{12}\ $ between 1-isomorphisms of gerbes over
$\,\Gamma\times M$, \,such that the diagram
\qq
\xymatrix@C=-2.4cm@R=1.4cm{&&(\Id\otimes\chi_3)\circ(\Id\otimes\alpha^a_{2,3})
\circ\alpha^a_{1,23} \ar@{=>}[drr]^-{\Id\circ\beta^a} \ar@{=>}[dll]_-{(\Id\otimes\eta_{2,3})\circ\Id~~~}&&
\\(\Id\otimes(\alpha^b_{2,3}\circ\chi_{23}))\circ\alpha^a_{1,23} \ar@{=>}[dr]_{\Id\circ\eta_{1,23}}&\hspace{4cm}&&\hspace{4cm}&(\Id\otimes\chi_3)\circ\alpha^a_{12,3} \ar@{=>}[dl]^{\eta_{12,3}}
\\\hspace{4cm}&(\Id\otimes\alpha^b_{2,3})\circ\alpha^b_{1,23}\circ\chi_{123} \ar@{=>}[rr]_-{\beta^b\circ\Id}&&\alpha^b_{12,3}\circ\chi_{123}&\hspace{4cm}}
\label{commdiag1iso}
\qqq
 of 2-isomorphisms between 1-isomorphisms
of gerbes over $\,\Gamma^2\times M\,$  is commutative.
\,1-isomorphic $\,\Gamma$-equivariant gerbes necessarily correspond
to the same curvature $\,H\,$ and to the same 2-form $\,\rho\,$ and,
consequently, to the same $\,\Gamma$-equivariantly closed extension
$\,\hat H$. \,The
identity 1-isomorphism of $\,\Gamma$-equivariant gerbes is given by
the pair $\,(\chi,\eta)=(\Id,\Id)\,$ for which the diagram
(\ref{commdiag1iso}) \,reduces to a trivially commutative one.
\ Finally, a $\,\Gamma$-equivariant 2-isomorphism
\qq
\epsilon\,:\ (\chi,\eta)\quad\hbox to 0.7cm{\Large$\Rightarrow
$\hfill}\ (\chi',\eta')
{}
\qqq
is a 2-isomorphism $\,\ \epsilon\,:\,\chi$\,{\Large$\Rightarrow$}$\chi'\ \,$
such that the diagram
\qq
\xymatrix{(\Id\otimes\chi_2)\circ\alpha^a \ar@{=>}[d]_{(\Id\otimes\epsilon_2)\circ\Id} \ar@{=>}[r]^-{\eta} & \alpha^b\circ\chi_{12} \ar@{=>}[d]^{\Id\circ
\epsilon_{12}} \\ (\Id\otimes\chi'_2)\circ\alpha^a \ar@{=>}[r]_-{\eta'} & \alpha^b\circ\chi'_{12} }
{}
\qqq
is commutative, \,which is trivially the case for the identity
2-isomorphism $\ \Id:\,\chi$\,{\Large$\Rightarrow$}$\chi\,$ when
also $\,\eta'=\eta$.

\

\begin{remark}
\label{rem:ind}
\item[\ 1.\ ]
We shall say that two $\,\Gamma$-equivariant structures
$\,(\alpha^a,\beta^a)\,$ and $\,(\alpha^b,\beta^b)\,$ on the gerbe $\,\CG\,$
are isomorphic if the $\,\Gamma$-equivariant gerbes
$\,(\CG,\alpha^a,\beta^a)\,$ and $\,(\CG,\alpha^b,\beta^b)\,$
are 1-isomorphic.

\item[\ 2\ ]
If $\,(\CG^a,\alpha^a,\,\beta^a)\,$ is a $\,\Gamma$-equivariant gerbe, then
for each 1-isomorphism of gerbes $\,\delta:\CG^a\rightarrow\CG^b\,$
there exists a $\,(\Gamma$-equivariant structure $\,(\alpha^b,\beta^b)\,$
such that the $\,\Gamma$-equivariant gerbe $\,(\CG^a,\alpha^a,\beta^a)\,$
and $\,(\CG^b,\alpha^b,\beta^b)\,$ are 1-isomorphic.

\item[\ 3.\ ]
$\Gamma$-equivariant gerbes $\,(\CG,\alpha,\beta)\,$ over a
$\,\Gamma$-space $\,M\,$ may be pulled back to
$\,\Gamma$-equivariant gerbes $\,(f^*\CG,f_2^*\alpha,f_3^*\beta)\,$
over another $\,\Gamma$-space $\,N\,$ along $\,\Gamma$-equivariant
maps $\,f:N\rightarrow M$. \,Similarly, their 1- and 2-isomorphisms
may be pulled back.

\item[\ 4.\ ]
For any subgroup $\,\Gamma'\subset\Gamma$, the restriction induces a
$\,\Gamma'$-equivariant gerbe from a $\Gamma$-equivariant gerbe
$\,(\CG,\alpha,\beta)$.

\item[\ 5.\ ]
For discrete groups $\,\Gamma$, the above definition of
$\,\Gamma$-equivariant gerbes, 1-isomorphisms and 2-isomorphisms is
equivalent to those  introduced in \cite{GSW} (where the actions of
$\,\Gamma\,$ that change the sign of the curvature 3-form $\,H\,$
were also considered).
\end{remark}

There is a sub-2-category $\Ggrbz MG$ composed of those
$\,\Gamma$-equivariant gerbes $\,\mathcal{G}\,$ whose curvature
$\,H\,$ is $\,\Gamma$-equivariantly closed and the
2-form $\,\rho=0$. \,Below, we shall need the following result, a
particular consequence of the general descent theory for gerbes:

\begin{theorem}
\label{thm:descent} Suppose that $\,\Gamma\,$ acts on $\,M\,$ in
such a way that $\,M'=M/\Gamma\,$ is a smooth manifold and $\,M\,$
forms a smooth (left) principal $\,\Gamma$-bundle
$\,\omega:M\rightarrow M'$. \,Then, there exists a canonical
equivalence
\qq
\Ggrbz M \Gamma\,\cong\ \grb {M'}\text{.}
{}
\qqq
\end{theorem}

\no In particular, a gerbe $\,\CG\,$ over $\,M\,$ that is
$\,\Gamma$-equivariant
relative to the zero 2-form descends to a gerbe $\,\CG'\,$ over $\,M'\,$
whose pullback by $\,\omega\,$ is 1-isomorphic to $\,\CG$. \,The equivalence
of Theorem \ref{thm:descent} commutes with the pullback functors: $\,f^*\,$ of
$\Ggrbz M \Gamma$ induced by a $\,\Gamma$-equivariant map
$\,f:N\rightarrow M\,$ and $\,f'^*\,$ of $\grb{M'}$ induced by the projected
map $\,f':N'\rightarrow M'$.

\no We give a proof of Theorem \ref{thm:descent} in Appendix \ref{app:4},
employing results of \cite{Stev}.

\subsection{WZ amplitudes with topologically non-trivial gauge fields}
\label{sec:topntr}

\no In Sec.\,\ref{sec:trivialgaugefields}, we discussed only
topologically trivial two-dimensional gauge fields, i.e. connections
in the trivial principal $\,\Gamma$-bundle over the worldsheet
$\,\Sigma$. \,Here, we shall consider connections in a general
principal $\,\Gamma$-bundle $\,\pi:P\to\Sigma$. \,Such connections
are $\Ng$-valued 1-forms $\,\CA\,$ on $\,P\,$ with the following
defining property:
\qq
(r^*\CA)(p,\gamma^{-1})\ =\ Ad_{\gamma}\big(\CA(p)\,-\,\Theta(\gamma)\big)\,,
\label{rstar}
\qqq
where $\ r:P\times\Gamma\rightarrow P\ $ is the right action of $\,\Gamma\,$
on $\,P$. \,For a $\,\Gamma$-equivariantly closed 3-form $\,\hat H(X)=H+v(X)$,
\,consider the 2-form $\,\tilde\rho_\CA\,$ on
$\,\tilde M \defn P\times M\,$
given by the formula
\qq
\tilde\rho_{\CA}\,\defn\,
-v(\CA)+\frac{_1}{^2}\iota_{\bar\CA}v(\CA)\,,
\label{HAgl}
\qqq
compare to the left one of Eqs.\,(\ref{HA1}). \,Below, the map
$\,\tilde\ell:\Gamma\times\tilde M\rightarrow\tilde M\,$ will denote
the left action of $\,\Gamma\,$ on $\,\tilde M\,$:
\qq
\tilde\ell\big(\gamma,(p,m)\big)\ \defn\ \big(r(p,\gamma^{-1}),\,
\ell(\gamma,m)\big)
\ =\ (p\gamma^{-1},\,\gamma m)\,.
\label{tell}
\qqq
For maps and forms on the product spaces $\,\Gamma^n\times\tilde M$,
\,we shall use the notation from the beginning of
Sec.\,\ref{sec:lift}, marking the subscript indices with a {\it
tilde}. \,The subscript indices without a {\it tilde} will be
reserved for the factors in the expanded expression
$\,\Gamma^n\times P\times M\,$ for the same spaces. One has the
following counterpart of Eq.\,(\ref{rrr}):

\begin{lemma}
\label{lem:tilderho}
As forms on $\,\Gamma\times\tilde M=\Gamma\times P\times M$,
\qq
(\tilde\rho_\CA)_{\tilde 1\tilde 2}\ =\ (\tilde\rho_\CA)_{2,3}\,
-\,\rho_{1,3}\ =\ (\tilde\rho_\CA)_{\tilde 2}\,-\,\rho_{1,3}\,.
\label{elltr}
\qqq
\end{lemma}

\no A proof of Lemma \ref{lem:tilderho} is given in Appendix
\ref{app:5}.
\vskip 0.2cm

Let $\,\CG\,$ be a gerbe over $\,M$ \,with the curvature $\,H\,$ which
extends to the $\,\Gamma$-equivariantly closed form $\,\hat H=H+v(X)$.
\,Define a gerbe $\,\tilde\CG_\CA\,$ over $\,\tilde M=P\times M\,$ by setting
\qq
\tilde\CG_\CA\ \defn \ \CI_{\tilde\rho_\CA}\otimes\CG_2\,.
\label{tGA}
\qqq
Note that the curvature of $\,\tilde\CG_\CA\,$ is given by the
closed 3-form
\qq
\tilde H_\CA\ \defn \ d\tilde\rho_\CA\,+\,H_2\,.
{}
\qqq
For the pullback of $\,\tilde H_\CA\,$ under the action
$\,\tilde\ell\,$ of $\,\Gamma\,$ on $\,\tilde M$, \,we obtain from
Lemmas \ref{lem:tilderho} and \ref{lem:2formrho}:
\qq
(\tilde H_\CA)_{\tilde 1\tilde 2}\ =\ d(\tilde\rho_\CA)_{\tilde 1\tilde 2}\ +\
(\ell^*H)_{1,3}\ =\
d(\tilde\rho_\CA)_{\tilde 2}\,-\,d\rho_{1,3}\,+\,d\rho_{1,3}\,+\,H_3\,
\ =\ (\tilde H_\CA)_{\tilde 2}\,.
{}
\qqq
It follows that $\,\tilde H_\CA\,$ (without any further extension) is
a $\,\Gamma$-equivariantly closed form on $\,\tilde M$.

\begin{proposition}
\label{prop:equiv} Let $\,(\mathcal{G},\alpha,\beta)\,$ be a
$\,\Gamma$-equivariant gerbe over $\,M\,$ in the sense of Definition
\ref{def:equivgerbe} and let $\,P\,$ be a principal
$\,\Gamma\,$-bundle over the surface $\,\Sigma\,$ with connection
$\,\mathcal{A}$. \,Then the gerbe ${\tilde\CG}_\CA$ over $\,\tilde M
= P\times M\,$ may be canonically equipped with the structure of a
$\,\Gamma$-equivariant gerbe relative to the zero 2-form.
\end{proposition}

\begin{proof}
First, \,we have to construct a 1-isomorphism
$\,\tilde\alpha_\CA\,$ of gerbes over $\,\Gamma\times\tilde M\,$:
\qq
\tilde\alpha_\CA:\,(\tilde\CG_\CA)_{\tilde 1\tilde 2}\ \rightarrow\
(\tilde\CG_\CA)_{\tilde 2}\,.
{}
\qqq
\vskip -0.3cm
\no It is obtained as the composition
\qq
(\tilde\CG_\CA)_{\tilde 1\tilde 2}\ =\ \CI_{(\tilde\rho_\CA)_{\tilde 1\tilde 2}}
\otimes\CG_{13}
\tolabel{1.6cm}{Id\otimes\alpha_{1,3}}
\CI_{(\tilde\rho_\CA)_{\tilde 1\tilde 2}}\otimes\,\CI_{\rho_{1,3}}\otimes\,\CG_3
\ =\ \CI_{(\tilde\rho_\CA)_{\tilde 2}}\otimes\,\CG_3\ =\
(\tilde\CG_\CA)_{\tilde 2}\,,
{}
\qqq
where we used Lemma \ref{lem:tilderho}. \,Hence,
$\,\tilde\alpha_\CA\,$ is the tensor product of the identity
1-isomorphism of the gerbe $\,\CI_{(\tilde\rho_\CA)_{\tilde 1\tilde
2}}\,$ with the 1-isomorphism $\,\alpha_{1,3}$.

Next, \,we have to construct a 2-isomorphism
$\,\tilde\beta_\CA\,$ between 1-isomorphisms of gerbes
$\,(\tilde\CG_\CA)_{\tilde 1\tilde 2\tilde 3}\,$ and
$\,(\tilde\CG_\CA)_{\tilde 3}\,$ over $\,\Gamma^2\times\tilde M\,$
\vskip -0.65cm
\qq
\tilde\beta_\CA\,:\ (\tilde\alpha_\CA)_{\tilde 2,\tilde 3}\circ(\tilde
\alpha_\CA)_{\tilde 1,\tilde 2\tilde 3}\hbox to 1cm{\,\,{\Large$\
\Rightarrow\ $}\hfill}
(\tilde\alpha_\CA)_{\tilde 1\tilde 2,\tilde 3}\,.
{}
\qqq
Note that  $\,(\tilde\alpha_\CA)_{\tilde 1,\tilde 2\tilde 3}\,$ is
the 1-isomorphism
\qq
(\tilde\CG_\CA)_{\tilde 1\tilde 2\tilde 3}\ =\
\CI_{(\tilde\rho_\CA)_{\tilde 1\tilde 2\tilde 3}}
\otimes\CG_{124}\ \tolabel{1.5cm}{\Id\otimes\alpha_{1,24}}
\CI_{(\tilde\rho_\CA)_{\tilde 1\tilde 2\tilde 3}}
\otimes\CI_{\rho_{1,24}}\otimes\,\CG_{24}\
=\ \CI_{(\tilde\rho_\CA)_{\tilde 2\tilde 3}}\otimes\,\CG_{24}\ =\
(\tilde\CG_\CA)_{\tilde 2\tilde 3}
{}
\qqq
since Lemma \ref{lem:tilderho} implies that
$\,(\tilde\rho_\CA)_{\tilde 1\tilde 2\tilde 3}+\rho_{1,24}
=(\tilde\rho_\CA)_{\tilde 2
\tilde 3}$.
\ Similarly, $\,(\tilde\alpha_\CA)_{\tilde 2,\tilde 3}\,$ is the 1-isomorphism
\qq
(\tilde\CG_\CA)_{\tilde 2\tilde 3}\ =\
\CI_{(\tilde\rho_\CA)_{\tilde 2\tilde 3}}\otimes\,\CG_{24}\
\tolabel{1.6cm}{\Id\otimes\alpha_{2,4}}
\CI_{(\tilde\rho_\CA)_{\tilde 2\tilde 3}}\otimes\CI_{\rho_{2,4}}\otimes\,\CG_4
\ =\ \CI_{(\tilde\rho_\CA)_{\tilde 3}}\otimes\CG_4\
=\ (\tilde\CG_\CA)_{\tilde 3}\,,
{}
\qqq
where we used the relation $\,(\tilde\rho_\CA)_{\tilde 2\tilde
3}+\rho_{2,4} =(\tilde\rho_\CA)_{\tilde 3}$, \,again following from
Lemma \ref{lem:tilderho}.
\ Hence, $\ (\tilde\alpha_A)_{\tilde 2,\tilde 3}
\circ(\tilde\alpha_\CA)_{\tilde 1,\tilde 2\tilde 3}\ $ is the
1-isomorphism
\qq
\CI_{(\tilde\rho_\CA)_{\tilde 1\tilde 2\tilde 3}}
\otimes\CG_{124}\, \tolabel{1cm}{\Id\otimes\alpha_{1,24}}
\CI_{(\tilde\rho_\CA)_{\tilde 1\tilde 2\tilde 3}}
\otimes\CI_{\rho_{1,24}}\otimes\,\CG_{24}\,\tolabel{1cm}{\Id\otimes\alpha_{2,4}}\,\CI_{(\tilde\rho_\CA)_{\tilde 1\tilde 2
\tilde 3}}\otimes\,\CI_{\rho_{1,24}}\otimes\,\CI_{\rho_{2,4}}\otimes\,\CG_4
\,=\,\CI_{(\tilde\rho_{\CA})_3}\otimes\,\CG_4\qquad
\label{1is}
\qqq
that is the tensor product of the identity 1-isomorphism of the
gerbe $\,\CI_{(\tilde\rho_\CA)_{\tilde 1\tilde 2\tilde 3}}\,$ with
the 1-isomorphism $\ (\Id\otimes\alpha_{2,4})\circ\alpha_{1,24}$.
\ On the other hand, $\,(\tilde\alpha_\CA)_{\tilde 1\tilde 2,\tilde
3}\,$ is the 1-isomorphism given by
\qq
(\tilde\CG_\CA)_{\tilde 1\tilde 2\tilde 3}\ =\ \CI_{(\tilde\rho_\CA)_{\tilde 1
\tilde 2\tilde 3}}
\otimes\CG_{124}\ \tolabel{1.3cm}{\Id\otimes\alpha_{12,4}}\
\CI_{(\tilde\rho_\CA)_{\tilde 1\tilde 2\tilde 3}}\otimes\CI_{\rho_{12,4}}
\otimes\,\CG_4\ =\ \CI_{(\rho_\CA)_{\tilde 3}}\otimes\,\CG_4\
=\ (\tilde\CG_\CA)_{\tilde 3}
\label{2is}
\qqq
because $\ (\tilde\rho_\CA)_{\tilde 1\tilde 2\tilde 3}+\rho_{12,4}=
(\tilde\rho_\CA)_{\tilde 3}$, \,once again in virtue of Lemma
\ref{lem:tilderho}. \,Comparison between (\ref{1is}) and
(\ref{2is}), and Definition \ref{def:equivgerbe} (ii) show that we
may take for $\,\tilde\beta_\CA\,$ the 2-isomorphism obtained by
tensoring the identity 2-isomorphism between the identity
1-isomorphisms of the gerbe $\,\CI_{(\tilde\rho_\CA)_{\tilde 1\tilde
2\tilde 3}}\,$ with the 2-isomorphism $\,\beta_{1,2,4}\,$:
\vskip -0.6cm
\qq
\tilde\beta_\CA\ \defn\ \Id\otimes\beta_{1,2,4}\,.
\label{tilbet}
\qqq
We have to check that the 1-isomorphism $\,\tilde\alpha_\CA\,$
and 2-isomorphism $\,\tilde\beta_\CA\,$ make the diagram
\qq
\xymatrix@C=-1.2cm@R=1.4cm{&(\tilde\alpha_\CA)_{\tilde 3,\tilde
4}\circ(\tilde\alpha_\CA)_{\tilde 2, \tilde 3\tilde 4}
\circ(\tilde\alpha_\CA)_{\tilde 1,\tilde 2\tilde 3\tilde 4}
\ar@{=>}[dl]_>>>>>>>{(\tilde\beta_\CA)_{\tilde 2,\tilde 3,\tilde
4}\circ\Id}  \ar@{=>}[dr]^>>>>>>>{\Id\circ(\tilde
\beta_\CA)_{\tilde 1,\tilde 2,\tilde 3\tilde 4}} &\\
(\tilde\alpha_\CA)_{\tilde 2\tilde 3,\tilde
4}\circ(\tilde\alpha_\CA)_{\tilde 1,\tilde 2\tilde 3 \tilde 4}
\ar@{=>}[dr]_{(\tilde\beta_\CA)_{\tilde 1,\tilde 2\tilde 3,\tilde
4}} && (\tilde\alpha_\CA)_{\tilde 3,\tilde
4}\circ(\tilde\alpha_\CA)_{\tilde 1\tilde 2, \tilde 3 \tilde 4}
\ar@{=>}[dl]^{(\tilde\beta_\CA)_{\tilde 1 \tilde 2,\tilde 3,\tilde
4}}
\\&(\tilde\alpha_\CA)_{\tilde 1\tilde 2\tilde 3,\tilde 4}&}
\qqq
commutative. \,It is easy to see that the above diagram may be
identified with the tensor product of the identity 2-isomorphism
between the identity 1-isomorphisms of the gerbe $\
\CI_{(\tilde\rho_\CA)_{\tilde 1\tilde 2\tilde 3\tilde 4}}\ $ by
the pullback of diagram (\ref{axiom}) along the projection from
$\ \Gamma^3\times P\times M\ $ to $\ \Gamma^3\times M$.
\,This assures its commutativity, \,completing the proof of Proposition
\ref{prop:equiv}. \end{proof}

The action (\ref{tell}) of $\,\Gamma\,$ on $\,\tilde M\,$ is free
and the quotient space $\,\tilde
M/\Gamma=P\times_\Gamma\hspace{-0.05cm}M=:P_M\,$ is
the associated bundle
over $\,\Sigma\,$ with the typical fiber $\,M$. \ The space
$\,\tilde M\,$ may be viewed as a (left) principal $\,\Gamma$-bundle
$\,\tilde\omega:\tilde M\rightarrow P_M$.
\,Theorem \ref{thm:descent} and Proposition
\ref{prop:equiv} have as the immediate consequence

\begin{corollary}\label{cor:5.6}
The gerbe $\,\tilde\CG_\CA\,$ on $\,\tilde M\,$ descends to a gerbe
$\,\CG_\CA\,$ on $\,P_M\,$ whose pullback along $\,\tilde\omega\,$
is 1-isomorphic to $\,\tilde\CG_\CA$. \ In particular, \,the curvature
of $\,\CG_\CA\,$ is equal to the closed 3-form $\,H_\CA\,$ on
$\,P_M\,$ whose pullback to $\,\tilde M\,$ coincides with $\,\tilde H_\CA$.
\end{corollary}

In order to couple the sigma model with target $\,M\,$ to a gauge
field $\,\CA\,$ in the principal $\,\Gamma$-bundle
$\,\pi:P\rightarrow\Sigma$, \,one has to modify also the sigma-model
fields. In the gauged model, they become global sections $\
\Phi:\Sigma\rightarrow P_M\ $ of the
associated bundle rather than maps from $\,\Sigma\,$ to $\,M$.

\begin{definition}
\label{def:nontrivialamplitudes} Let
$\,(\mathcal{G},\alpha,\beta)\,$ be a $\,\Gamma$-equivariant gerbe
over $\,M\,$ and $\,P\,$ a principal $\,\Gamma$-bundle with
connection $\,\CA\,$ over a closed oriented surface $\,\Sigma$.
\,The Wess-Zumino contribution of a field $\ \Phi:\Sigma\rightarrow
P_M\ $ to the gauged Feynman amplitude is defined  by
\qq
{\bm A}_\WZ(\Phi,\CA)\ \defn\ \Hol_{\CG_{\CA}}(\Phi)\,.
\label{FAg2}
{}
\qqq
\end{definition}

\begin{remark}\label{rem:bundtrivbund}
The above constructions are functorial with respect to isomorphisms
of principal bundles $\,P$. \,If $\,P\,$ is trivial,
\,i.e. $P=\Sigma\times\Gamma$, \,then the gauge fields $\,\CA\,$ may be
related to $\,\Ng$-valued 1-forms $\,A\,$ on $\,M\,$ by the formula
$\,\CA(x,\gamma^{-1})=Ad_{\gamma}\big(A(x)-\Theta(\gamma)\big)$.
\,In this case, the associated bundle $\,P_M\,$ may be naturally identified
with $\,\Sigma\times M$, \,and the gerbe $\,\CG_{\CA}\,$ with the
gerbe $\,\CG_A\,$ defined by relation (\ref{HA1}). One recovers
this way the coupling to the topologically trivial gauge fields
discussed previously, see Definition \ref{def:trivialamplitudes}.
\end{remark}

\subsection{General  gauge invariance}
\label{sec:GGInv}

\no For the general case of gauge fields $\,\CA\,$ corresponding to
connections in a principal $\,\Gamma$-bundle $\,\pi:P\rightarrow\Sigma$,
\,the general gauge transformations $\,{\Ch}\,$ are defined as sections
of the associated bundle $\,Ad(P)=P\times_{Ad}\Gamma$. \,The latter is
composed of the orbits $\ \{\,(p\gamma'^{-1},Ad_\gamma'(\gamma))\ |\
\gamma'\in\Gamma\,\}\defn [(p,\gamma)]\ $ of the action of
$\,\Gamma\,$ on $\,P\times\Gamma$. \,Orbits $\,[(p,\gamma_1)]\,$ and
$\,[(p,\gamma_2)]\,$ may be multiplied to $\,[(p,\gamma_1\gamma_2)]\,$
so that $\,Ad(P)\,$ is a bundle of groups. \,Consequently,
sections of $\,Ad(P)\,$ may be multiplied point-wise, forming the
group of gauge transformations. \,An orbit $\,[(p,\gamma)]\,$
acts (from the left) on the fiber $\,\pi^{-1}(\pi(p)) \subset P\,$
by the mapping
\qq
p\gamma'\ \longmapsto\ p\gamma\gamma'\,=:\,[(p,\gamma)]\cdot p\gamma'\,.
{}
\qqq
This action induces a left action of gauge transformations $\,\Ch\,$
on $\,P\,$ by principal $\,\Gamma$-bundle automorphisms $\,\Cl_\Ch\,$
given by
\qq
P\ni p\ \mapstolabel{\Cl_{\Ch}}\ \Ch(x)\cdot p\,.
{}
\qqq
Gauge transformations of the gauge field $\,\CA\,$ are
defined as
\qq
\CA\ \longmapsto\ \Ch\CA \defn \Cl_{\Ch^{-1}}^{\,*}\CA\,.
\label{glgtrA}
\qqq
\vskip -0.3cm
\no Note that the maps
\vskip -0.4cm
\qq
\tilde L_\Ch\,:=\,
\Cl_\Ch\times\Id
\label{TL}
\qqq
from $\,\tilde M=P\times M\,$ into itself are
$\,\Gamma$-equivariant, \,i.e.\ they commute with the action
(\ref{tell}) of $\,\Gamma\,$ on $\,\tilde M$. \,Consequently, they
descend to automorphisms $\,L_\Ch\,$ of the associated bundle
$\,P_M=P\times_\Gamma\hspace{-0.05cm}M$. \,Gauge transformations
of sections $\,\Phi\,$ of $\,P_M\,$ are defined by the formula
\qq
\Phi\ \longmapsto\ L_\Ch\circ\Phi\,=:\,\Ch\Phi\,.
{}
\qqq

In the case of the trivial bundle $\,P$, \,the associated bundle
$\,Ad(P)\,$ is also trivial and the sections $\,\Ch\,$ of $\,Ad(P)\,$
reduce to maps from $\,\Sigma\,$ to $\,\Gamma$. \,Their
action on gauge fields $\,\CA\,$ agrees with the action
(\ref{gauact}) on the 1-forms $\,A\,$
related to $\,\CA\,$ as in Remark \ref{rem:bundtrivbund}.
\,Similarly, their action on sections $\,\Phi\,$ of the trivial
associated bundle agrees with the one considered in Eq.\,(\ref{vptogvp}).
\,The invariance of the amplitudes ${\bm A}_\WZ(\Phi,\CA)$ from Definition
\ref{def:nontrivialamplitudes} in the case of the trivial bundle
$\,P\,$ is assured by the assumption of the $\,\Gamma$-equivariance
of the gerbe $\,\CG$. \,Indeed, as follows from Corollary
\ref{co:invariancecond}, only property $(i)$ of Definition
\ref{def:equivgerbe} is needed in that case to guarantee the
gauge invariance under general gauge transformations. Here, we shall
prove for a general principal $\,\Gamma$-bundles $\,P$,

\begin{theorem}
\label{thm:globalgaugeinv}
The amplitudes ${\bm A}_\WZ(\Phi,\CA)$ of Definition
\ref{def:nontrivialamplitudes} are invariant under
all gauge transformations, i.e.
\qq
{\bm A}_\WZ({\Ch}\Phi,{\Ch}\CA)\ =\ {\bm A}_\WZ(\Phi,\CA)
{}
\qqq
for all sections $\,{\Ch}\,$ of the bundle $\,Ad(P)$.
\end{theorem}

\begin{proof}
We have to show that
\qq
\Hol_{\CG_{\Ch\CA}}(L_\Ch\circ\Phi)\ =\ \Hol_{L_\Ch^*\CG_{\Ch\CA}}(\Phi)
\ =\ \Hol_{\CG_\CA}(\Phi)
{}
\qqq
for every $\,\Ch,\ \,\Phi\,$ and $\,\CA$. \,This follows if there
exists a 1-isomorphism between gerbes $\,L_\Ch^*\CG_{\Ch\CA}\,$
and $\,\CG_{\CA}$. \,Recall that gerbe $\,\CG_{\CA}\,$ over
$\,P_M\,$ descended from the $\,\Gamma$-equivariant gerbe
$\,(\tilde\CG_\CA,\tilde\alpha_\CA,\tilde\beta_\CA)\,$ over
$\,\tilde M$, \,see Proposition \ref{prop:equiv} and Corollary
\ref{cor:5.6}. \,Since maps $\,\tilde
L_\Ch\,$ of $\,\tilde M\,$ are $\,\Gamma$-equivariant, \,gerbe
$\,L_\Ch^*\CG_A\,$ descends, in turn, from the
$\,\Gamma$-equivariant gerbe $\ (\tilde
L_\Ch^*\,\tilde\CG_\CA,\,(\tilde L_\Ch)_{\tilde 2}^*
\,\tilde\alpha_\CA,\,(\tilde L_\Ch)_{\tilde 3}^*\,
\tilde\beta_\CA)$, \ see Theorem \ref{thm:descent}. \,We claim that
the two gerbes
$$\ (\tilde L_\Ch^*\,\tilde\CG_\CA,\,
(\tilde L_\Ch)_{\tilde 2}^*\,\tilde\alpha_\CA, \,(\tilde
L_\Ch)_{\tilde 3}^*\,\tilde\beta_\CA)\ \qquad\text{ and }\qquad\
(\tilde\CG_{\Ch^{-1}\hspace{-0.05cm}\CA},\,
\tilde\alpha_{\Ch^{-1}\hspace{-0.05cm}\CA},\,
\tilde\beta_{\Ch^{-1}\hspace{-0.05cm}\CA})$$coincide. The claim
implies, \,in virtue of Theorem \ref{thm:descent}, \,that the descended
gerbes $\ L_\Ch^*\CG_\CA\ $ and $\ \CG_{\Ch^{-1}\hspace{-0.05cm}\CA}\ $
over $\ P_M\ $ coincide as well and, hence, so do
$\ L_\Ch^*\CG_{\Ch\CA}\,$ and $\ \CG_{\CA}\,$.

It remains to prove the above claim. From definitions (\ref{HAgl})
of the form $\,\tilde\rho_A$, \,(\ref{glgtrA}) of $\Ch\CA\,$ and
\,(\ref{TL}) of the action $\,\tilde L_\Ch\,$ on $\,\tilde M$, \,
using, in particular, the fact that $\,\tilde L_\Ch\,$ acts trivially
on the factor $\,M\,$ in $\,P\times M$, \ it follows immediately that
\qq
\tilde L_\Ch^*\,\tilde\rho_{\CA}\
=\ \tilde\rho_{\Ch^{-1}\hspace{-0.05cm}\CA}\,.
{}
\qqq
This, in conjunction with definition (\ref{tGA}), implies, \,in turn,
\,the equality of gerbes
\qq
\tilde L^*_\Ch\,\tilde\CG_{\CA}\
=\ \tilde\CG_{\Ch^{-1}\hspace{-0.05cm}\CA}\,.
\label{LhtGA}
\qqq
Recall from the proof of Proposition \ref{prop:equiv} that
$\,\tilde\alpha_\CA\,$ is the tensor product of the identity
1-isomorphism of the gerbe $\ \CI_{(\tilde\rho_\CA)_{\tilde 1\tilde
2}}\ $ with the 1-isomorphism $\,\alpha_{1,3}$. \,Now, \,the map
$\,(\tilde L_\Ch)_{\tilde 2}\,$ of $\,\Gamma\times\tilde
M=\Gamma\times P\times M\,$ acts only on the factor $\,P$.
\,Besides,
\qq
(\tilde L_\Ch)_{\tilde 2}^*\,(\tilde\rho_\CA)_{\tilde 1\tilde 2}\ =\
(\tilde L_\Ch^*\,\tilde\rho_\CA)_{\tilde 1\tilde 2}\
=\ (\tilde\rho_{\Ch^{-1}\hspace{-0.05cm}\CA})_{\tilde 1\tilde 2}\,.
{}
\qqq
We infer this way that the 1-isomorphism $\ (\tilde L_\Ch)^*_{\tilde
2}\,\tilde\alpha_\CA\ $ is the tensor product of the identity
1-isomorphism of the gerbe
$\ \CI_{\tilde\rho_{\Ch^{-1}\hspace{-0.05cm}\CA}}\ $ with the
1-isomorphism $\,\,\alpha_{1,3}\,\,$ so that
\qq
(\tilde L_\Ch)^*_{\tilde 2}\,\tilde\alpha_\CA\
=\ \tilde\alpha_{\Ch^{-1}\hspace{-0.05cm}\CA}\,.
\label{ddon}
\qqq
Additionally, \,equalities (\ref{LhtGA}) and (\ref{ddon}) allow to
relate the 2-isomorphisms $\ (\tilde L_\Ch)_{\tilde
3}^*\,\tilde\beta_\CA\,$ and
$\,\tilde\beta_{\Ch^{-1}\hspace{-0.05cm}\CA}$. \,Indeed, both are
tensor products of the identity 2-isomorphism between the identity
1-isomorphisms of the gerbe $\ (\tilde L_\Ch)_{\tilde
3}^*\CI_{(\tilde\rho_\CA)_{\tilde 1\tilde 2\tilde 3}}=
\CI_{(\tilde\rho_{\Ch^{-1}\hspace{-0.05cm}\CA})_{\tilde 1\tilde
2\tilde 3}}\ $ with the 2-isomorphism $\,\beta_{1,2,4}$, \,see
Eq.\,(\ref{tilbet}). \,Hence,
\qq
(\tilde L_\Ch)_{\tilde 3}^*\,\tilde\beta_\CA\
=\ \tilde\beta_{\Ch^{-1}\hspace{-0.05cm}\CA}
{}
\qqq
and the claim is established.
\vskip -0.6cm\
\end{proof}

\nsection{Obstructions and classification of equivariant structures}
\label{sec:obstrclass}

\no In this section, \,we shall treat the obstructions to the
existence and the classification of equivariant structures on a gerbe
$\,\CG\,$ over a $\,\Gamma$-space, \,see Definition \ref{def:equivgerbe}.
\,We shall start by discussing subsequently the obstructions to the three
parts of the structure: 1-isomorphism $\,\alpha$,\ \,2-isomorphism
$\,\beta$, \,and the commutative diagram \,(\ref{axiom}).

\subsection{Obstructions to 1-isomorphisms $\,\alpha$}
\label{sec:alphaobstr}

\no The first obstruction concerns the existence of 1-isomorphism
$\ \alpha:\CG_{12}\rightarrow\CI_\rho\otimes\CG_2\ $ or,
equivalently, the triviality of 1-isomorphism class $\ [\CF]\in
H^2(\Gamma\times M,U(1))\ $ of flat gerbe
$\,\CF=\CG_{12}\otimes\CG_2^*\otimes\CI_{-\rho}\,$ over
$\,\Gamma\times M$. \,It coincides with the obstruction to the
general gauge invariance of the WZ amplitudes (\ref{FAg1}) coupled to
topologically trivial gauge fields, \,see Corollary
\ref{co:invariancecond}. \,By the Universal Coefficient Theorem, $\
H^2(\Gamma\times M,U(1))=Hom(H_2(\Gamma\times M),U(1))$. \ In the
latter presentation, \,class $\,[\CF]\,$ is given by the
holonomy of flat gerbe $\,\CF\,$ along maps $\
(\Ch,\varphi):\Sigma\rightarrow\Gamma \times M\ $ defining
singular 2-cycles, and its triviality is equivalent to the
triviality of the holonomy. \,By the K\"unneth Theorem,
\qq
H_2(\Gamma\times M)\,=\,H_2(\Gamma)\otimes H_0(M)\oplus H_1(\Gamma)
\otimes H_1(M)\oplus H_0(\Gamma)\otimes H_2(M)\,.
{}
\qqq
Subgroup $\ H_2(\Gamma)\otimes H_0(M)\cong H_2(\Gamma)^{\pi_0(M)}\ $
is generated by the singular 2-cycles corresponding to maps
$\,(\Ch,\varphi)\,$ with $\,\varphi\,$ taking a constant value in one
of the connected
components of $\,M\,$ ($\pi_0(M)\,$ is the set of such components).
\,Similarly for $\,H_0(\Gamma)\otimes H_2(M)\cong
H_2(M)^{\pi_0(\Gamma)}$. \,Subgroup $\
H_1(\Gamma)\otimes H_1(M)\ $ is generated by the maps
\qq
S^1\times S^1\,\ni\,(\ee^{\,i\sigma_1},\ee^{\,i\sigma_2})\ \longmapsto\
\big(\Ch(\ee^{\,i\sigma_1}),\varphi(\ee^{\,i\sigma_2})\big)\,\in\,\Gamma\times
M \label{s1s1}
\qqq
with $\,\Ch\,$ and $\,\varphi\,$ giving rise to singular 1-cycles
in $\,\Gamma\,$ and $\,M$, \,respectively. \,Thus,
\qq
&&H^2(\Gamma\times M,U(1))\cr\cr
&&=\ Hom(H_2(\Gamma)^{\pi_0(M)},U(1))\oplus
Hom(H_1(\Gamma)\otimes H_1(M),U(1))\oplus Hom(H_2(M)^{\pi_0(\Gamma)},U(1))\cr\cr
&&=\
H^2(\Gamma,U(1))^{\pi_0(M)}\,\oplus\,Hom(H_1(\Gamma)\otimes H_1(M),U(1))\,\oplus\,
H^2(M,U(1))^{\pi_0(\Gamma)}\,.
{}
\qqq
Accordingly, \,we obtain
\vskip 0.5cm

\begin{proposition}
Class
$\,[\CF]\in H^2(\Gamma\times M,U(1))\,$ that obstructs the existence
of 1-isomorphism $\,\alpha\,$ of \,Definition \ref{def:equivgerbe}
decomposes as
\qq
[\CF]\ =\ [\CF]_{20}\,+\,[\CF]_{11}\,+\,[\CF]_{02}\,,\qquad
\label{201102}
\qqq
with the summands $\ [\CF]_{20}\in H^2(\Gamma,U(1))^{\pi_0(M)}$, $\
[\CF]_{11}\in Hom(H_1(\Gamma)\otimes H_1(M),U(1))\ $ and $\
[\CF]_{02}\in H^2(M,U(1))^{\pi_0(\Gamma)}$.
\end{proposition}

\hspace{13cm}$\blacksquare$
\vskip 0.5cm

\no Components of $\ [\CF]_{20}\,$ are the 1-isomorphism classes
of flat gerbes $\ r_m^*\CG\otimes\CI_{-\rho_m}\ $ over
$\,\Gamma\,$ for fixed points $\,m\,$ in different connected
components of $\,M\,$ with $\,r_m(\gamma)=\gamma m=\ell_\gamma(m)\,$
and $\,\rho_m =\frac{1}{2}(\iota^av^b)(m) \Theta^a\Theta^b$. \,Components
of $\,[\CF]_{02}\,$ are the 1-isomorphism classes of flat
gerbes $\ \ell_\gamma^*\CG\otimes\CG^*\ $ for fixed points
$\,\gamma\,$ in different connected components of $\,\Gamma$.
Finally, the bihomomorphism $\ [\CF]_{11}\in Hom(H_1(\Gamma)\otimes
H_1(M),U(1))\,$ is given by the gerbe $\,\CF\,$ holonomy of the maps
(\ref{s1s1}).

\begin{corollary}
If the connected components of $\,M\,$ and $\,\Gamma\,$ are
2-connected, then there is no obstruction to the existence of
1-isomorphism $\,\alpha\,$ of \,Definition \ref{def:equivgerbe}.
\end{corollary}

\no This applies to the case, studied in \cite{GSW,GSW1}, of
$\,\Gamma$-equivariant structures on the WZW gerbe $\,\CG_k\,$ over
$\,\tilde G\,$ for $\,\Gamma=Z\subset\tilde Z\,$ acting on $\,\tilde
G\,$ by multiplication.

For the $\,\Gamma$-space $\,M=G\,$ in the coset-model context, \,see
Definition \ref{def:context}, \,and with a WZW gerbe $\,\CG_k\,$
over $\,G$, \,the flat gerbe $\,\CF=\CF_k$, \,see
Sec.\,\ref{sec:expl1}. \,In decomposition (\ref{201102}) of
cohomology class $\,[\CF_k]\in H^2(\Gamma\times G,U(1))$, \,terms
$\,[\CF_k]_{20}\,$ and $\,[\CF_k]_{02}\,$ are trivial as
determined by the $\,\CF_k$-holonomy of the maps $\,(\Ch_{\tilde
p^\vee_1,\tilde p^\vee_2},\varphi_{p^\vee_1,p^\vee_2})\,$ of
Eqs.\,(\ref{mnm'n'}) with $\,p^\vee_1=p^\vee_2=0\,$ or $\,\tilde
p^\vee_1 =\tilde p^\vee_2=0$, \,respectively, \,whereas the
bihomomorphism $\,[\CF_k]_{11}\in Hom(\tilde Z_\Gamma\otimes
Z,U(1))\,$ is determined by the holonomy with $\,\tilde
p^\vee_2=p^\vee_1=0$, \,i.e. by
\qq
b_{\tilde z_1,z_2}\ =\ \exp\hspace{-0.05cm}\big[-2\pi i\,k\,
\tr\,\tilde p^\vee_1p^\vee_2\big],
\label{ugf1}
\qqq
see Eq.\,(\ref{ugf}), \,and may be non-trivial.

\subsection{Local description of gerbes}
\label{sec:localdata}

\no In order to discuss further obstructions to the existence of
a $\,\Gamma$-equivariant structure on a gerbe
$\,\CG\,$ over a $\,\Gamma$-space $\,M$, \,it will be convenient to
use local data for gerbes and their 1- and 2-isomorphisms. \,We
shall follow the discussion in the first part of Sec.\,VII of
\cite{GW}. \,The local data live in the Deligne complex $\,\mathcal{D}(2)\,$
\qq
0\ \longrightarrow\ A^0(\mathcal{O})\ \mathop{\longrightarrow}\limits^{D_0}
\ A^1(\mathcal{O})\ \mathop{\longrightarrow}^{D_1}\ A^2(\mathcal{O})
\ \mathop{\longrightarrow}^{D_2}\ A^3(\mathcal{O})
\label{complx}
\qqq
associated to an open covering $\mathcal{O}\,$ of $\,M$.
\,With $\,\mathcal{U}\,$ standing for the
sheaf of smooth $U(1)$-valued functions and $\,\Lambda^{n}\,$
for the sheaf of $n$-forms, \,the groups of the Deligne complex are
\qq
\hbox to 7.4cm{$A^0(\mathcal{O})\ =\ C^0(\mathcal{O},
\mathcal{U})\,\textrm{,}$\hfill}&&\hbox to 7.4cm{$A^1(\mathcal{O})\ =\
C^0(\mathcal{O},\Lambda^1)
\oplus C^1(\mathcal{O},\mathcal{U})\,\textrm{,}$\hfill}\\
\hbox to 7.4cm{$A^2(\mathcal{O})\ =\ C^0(\mathcal{O},\Lambda^2)
\oplus C^1(\mathcal{O},\Lambda^1) \oplus C^2(\mathcal{O},
\mathcal{U})\,\textrm{,}$\hfill}&&\hbox to 7.4cm{$
A^3(\mathcal{O})\ =\ C^1(\mathcal{O},\Lambda^2) \oplus
C^2(\mathcal{O},\Lambda^1) \oplus
C^3(\mathcal{O},\mathcal{U})\,\textrm{,}$\hfill}\quad
\qqq
where $C^\ell(\mathcal{O},\mathcal{S})$ denotes the $\ell^{\rm th}$\v Cech
cochain group of the open cover $\mathcal{O}$, \,with values
in a sheaf $\mathcal{S}$ of Abelian groups.  \,The differentials are
\qq
&&D_0(f_i)\,=\,(-\mathrm{i}f_i^{-1}df_i,
\; f_j^{-1}f_i)\textrm{,}\quad
D_1(\Pi_i,\chi_{ij})\,=\,(d\Pi_i,\;
-\mathrm{i}\chi_{ij}^{-1}d\chi_{ij}+\Pi_{j}
- \Pi_{i},\;\chi_{jk}^{-1}\chi_{ik}\chi_{ij}^{-1})\textrm{,}\\
&&D_2(B_i,A_{ij},g_{ijk})\ =\ (dA_{ij}-B_{j} + B_i,\;
-\mathrm{i}g_{ijk}^{-1}dg_{ijk}+A_{jk}-A_{ik}+A_{ij},\;
 g_{jkl}^{-1}g_{ikl}g_{ijl}^{-1}g_{ijk})\textrm{.}
{}
\qqq
A refinement $r: \mathcal{O}' \to \mathcal{O}$ of the covering
induces a restriction map on complexes (\ref{complx}). \,Local
data for gerbe $\,\mathcal{G}\,$ over $\,M\,$ form a cocycle $\,c
\in A^2(\mathcal{O})$, $\,D_2c=0$, \,for a sufficiently fine
covering $\mathcal{O}$ of $\,M$. \,Local data for 1-isomorphism
$\,\alpha:\CG_1\rightarrow\CG_2\,$ of gerbes with the respective local
data $\,c_i\in A^2(\mathcal{O}_i)\,$ are given by a cochain
$\,b \in A^1(\mathcal{O})\,$ for $\,\mathcal{O}\,$ a common refinement of
$\,\mathcal{O}_1\,$ and $\,\mathcal{O}_2\,$ such that, \,upon
restricting the $\,c_i\,$ to it, $\,c_2 = c_1 + D_1b\,$ (we use the additive
notation for the Abelian group law in all $\,A^n(\mathcal{O})$).
\,Finally, local data
for 2-isomorphism $\,\beta:\alpha_1\Rightarrow\alpha_2\,$ are
given by a cochain $\,a\in A^0(\mathcal{O})\,$ for a sufficiently
fine covering $\mathcal{O}\,$ such that, given the local data
$\,b_i\,$ for 1-isomorphisms $\,\alpha_i\,$ restricted to
$\mathcal{O}$, $\,b_2=b_1+D_0a$. \,For sufficiently fine
$\,\mathcal{O}$, \,the cohomology of the complex (\ref{complx}) is
\qq
&\displaystyle{\mathbb{H}^2(\mathcal{O},\mathcal{D}(2))\ =\
\frac{{\rm ker}\,D_2}{{\rm Im}
\,D_1},\ \ \qquad
\mathbb{H}^1(\mathcal{O},\mathcal{D}(2))\ =\ \frac{{\rm ker}\,D_1}{{\rm Im}
\,D_0}\ \cong\ H^1(M,U(1))}\,,&\\
&\displaystyle{\mathbb{H}^0(\mathcal{O},\mathcal{D}(2))
\ =\ {\rm ker}\,D_0\ \cong\ H^0(M,U(1))\,.}&
\qqq
These groups may be identified, respectively, with the group of
1-isomorphism classes of gerbes, \, the group of isomorphism classes
of flat line bundles, and the group of locally constant
$\,U(1)$-valued functions on $\,M$.
\smallskip

In the following, we want to consider local data for gerbes and
their 1- and 2-isomorphisms over the spaces $\,\Gamma^p\times M\,$
that form a simplicial manifold with {\it face maps} $\,\Delta_q^p:
\Gamma^{p}\times M \to\Gamma^{p-1}\times M\,$ for all $p\geq 1$ and
$0\leq q\leq p\,$ given by
\qq
{}
\Delta_q^{p}(\gamma_1,...,\gamma_p,m) := \left\lbrace\begin{array}{ll}
(\gamma_2,...,\gamma_p,m) & \text{ for } q=0, \\
(\gamma_1,...,\gamma_{q}\gamma_{q+1},...,\gamma_p,m) & \text{ for }
1 \leq q < p, \\
(\gamma_1,...,\gamma_{p-1},\gamma_pm) & \text{ for }q=p\text{.}
\end{array} \right.
\qqq
They satisfy satisfy the simplicial relations
\qq
\label{simprel}
\Delta^{p-1}_r\hspace{-0.06cm}
\circ \Delta^p_q = \Delta^{p-1}_{q-1} \circ \Delta^{p}_r
\qqq
for all $r<q$. \,We shall use simplicial sequences $\,\left \lbrace
\mathcal{O}^p \right \rbrace\,$ of open coverings
$\,\mathcal{O}^p=\left \lbrace O^p_i\right \rbrace{}_{i\in
I^{^p}}\,$ of the spaces $\,\Gamma^p\times M\,$ such that there are
face maps $\,\Delta_q^p: I^{p}\to I^{p-1}\,$ of the index sets
satisfying (\ref{simprel}), and such that
\qq
\Delta_q^p(O^{p}_i)\,\subset\,O^{p-1}_{\Delta^p_q(i)}
{}
\qqq
for all $p\geq 1$, all $0 \leq q \leq p$ and all $i\in I^p$. \,A
construction of ref.\,\cite{Tu}, reviewed in the Appendix of
\cite{GW}, permits to build a simplicial sequence $\,\left \lbrace
\mathcal{O}^p \right \rbrace\,$ whose coverings $\,\mathcal{O}^p\,$
refine the coverings of any given sequence of coverings of
$\,\Gamma^p\times M$. \ Given a simplicial sequence $\,\left \lbrace
\mathcal{O}^p \right \rbrace\,$ of coverings of $\,\Gamma^p\times
M$, \,one has induced cochain maps
\qq
(\Delta_q^p)^{*}: C^{\ell}(\mathcal{O}^{p-1}\hspace{-0.08cm},\mathcal{S})
\to C^{\ell}(\mathcal{O}^p,\mathcal{S})
\qquad\text{ defined \,by }\qquad \left ( (\Delta_q^p)^{*}f \right )_i
= (\Delta_q^p)^{*}(f_{\Delta_q^p(i)})\text{,}
\label{liftfacemaps}
\qqq
satisfying the co-simplicial relations
\qq
(\Delta_q^p)^{*} \circ (\Delta_r^{p-1})^{*} = (\Delta_r^p)^{*} \circ
(\Delta_{q-1}^{p-1})^{*}
\label{cosimplicial}
\qqq
for $r<q$. \ On the groups $\ A^n(\mathcal{O}^p)$,
\ besides the Deligne differentials
\qq
D_{n,p}: A^n(\mathcal{O}^p) \to A^{n+1}(\mathcal{O}^p)\,,
{}
\qqq
one has the simplicial operators
\qq
\Delta_{n,p}: A^n(\mathcal{O}^p) \to A^n(\mathcal{O}^{p+1})
\quad\text{\quad with\quad}\quad
\Delta_{n,p} :=\sum_{q=0}^{p+1} (-1)^{q}
(\Delta_q^{p+1})^{*}
{}
\qqq
whose definition uses the lift (\ref{liftfacemaps}) of the face maps
to groups $\,A^n(\mathcal{O}^p)$. \,Due to the co-simplicial
relations (\ref{cosimplicial}), \,we have $\,\Delta_{n,p+1} \circ
\Delta_{n,p}=0$. \,The differentials $\,D_{n,p}\,$ commute with pullbacks,
and thus also with operators $\,\Delta_{n,p}$. \,This endows the family
$\,\CK=\left(A^n(\mathcal{O}^p) \right)\,$ of Abelian groups with
the structure of a double complex.

\subsection{Obstructions to 2-isomorphism $\,\beta$}
\label{sec:betaobstr}

\noindent If cocycle $\,c\in A^2(\mathcal{O}^0)\,$ describes
local data for gerbe $\,\CG\,$ on $\,M\,$ then $\
-(\Delta_{2,0}c+\rho)\in A^2(\mathcal{O}^1)$, \ where $\,\rho\,$ is
identified with the cochain $\ (\rho|_{O^1_i},0,1)\ $ for $\,i\in I^1$,
\ represents local data for
flat gerbe $\,\CF=\CG_{12}\otimes\CG_2^*\otimes\CI_{-\rho}$.
\ The triviality of 1-isomorphism class $\,[\CF]$, \,discussed in
\,Sec.\,\ref{sec:alphaobstr}, \,means that, for a sufficiently fine
simplicial sequence of coverings $\,\left \lbrace \mathcal{O}^p
\right \rbrace$,
\qq
\Delta_{2,0}c\,+\,\rho\ =\ D_{1,1}b
{}
\qqq
for some $\,b\in A^1(\mathcal{O}^1)$. \,The cochain $\ b\,$ provides
local data for a 1-isomorphism $\
\alpha:\CG_{12}\rightarrow\CI_\rho\otimes\CG_2$, \,see Definition
\ref{def:equivgerbe}. \,It is defined modulo the addition $\
b\mapsto b+b'$, \ where $\,D_{1,1}b'=0$. \,This freedom corresponds
to the freedom of choice of $\,\alpha\,$ and of local data for
it. \,The cochains $\,(\Delta^2_0)^*b$, $\,(\Delta^2_1)^*b\,$ and
$\,(\Delta^2_2)^*b\,$ provide, in turn, local data for
1-isomorphisms $\,\alpha_{2,3}$, $\,\alpha_{12,3}\,$ and
$\,\alpha_{1,23}$, \,respectively. \,The existence of
2-isomorphism $\ \beta:\,(\Id\otimes\alpha_{2,3})\circ\alpha_{1,23}$\,
{\large$\Rightarrow$}\,\,$\alpha_{12,3}\ $ is equivalent to the
requirement that, for sufficiently fine $\,\left \lbrace
\mathcal{O}^p\right\rbrace$,
\qq
\Delta_{1,1}b\ =\ -D_{0,2}a
\label{ba}
\qqq
with $\ a\in A^0(\mathcal{O}^2)\ $ representing local data for
$\,\beta$. \,Let us first note that
\qq
D_{1,2}\Delta_{1,1}b\ =\ \Delta_{2,1}D_{1,1}b\ =\ \Delta_{2,1}\Delta_{2,0}c\,
+\,\Delta_{2,1}\rho\ =\ 0\,,
{}
\qqq
where the last equality is a consequence of relations $\
\Delta_{2,1}\circ \Delta_{2,0}=0\ $ and $\
\Delta_{2,1}\rho=\big((\rho_{2,3}-\rho_{12,3}
+\rho_{1,23})|_{O^2_i},0,0\big)$, \,and of
Eq.\,(\ref{rrr}) of Lemma \ref{lem:rhorho}. \,It follows that $\
\Delta_{1,1}b\ $ defines a cohomology class
\qq
[\Delta_{1,1}b]\,\in\,\frac{{\rm ker}\,D_{1,2}}{{\rm Im}\,
D_{0,2}}\,\cong\,H^1(\Gamma^2\times M,U(1))
{}
\qqq
that obstructs the solution of Eq.\,(\ref{ba}). \,However, since
$\,b\,$ was defined up to $\,D_{1,1}$-cocycles $\,b'\in A^1(\mathcal{O}^1)$,
\ the class $\,[\Delta_{1,1}b]\,$ is defined modulo the image
$\,{\mathcal H}^{1,2}\,$ of the map $\ [\Delta_{1,1}]:H^1(\Gamma\times
M,U(1))\rightarrow H^1(\Gamma^2\times M,U(1))\ $ that sends class
$\,[b']\,$ to class $\,[\Delta_{1,1}b']$. \ We obtain this way

\begin{proposition}
\label{prop:2obstrc}
Let $\ \alpha:\CG_{12}\rightarrow\CI_{\rho}\otimes\CG_2\ $ be
a 1-isomorphism with local data $\,b\in A^1(\CO^1)\,$ for a
sufficiently fine family of coverings $\{\CO^p\}$. \,Then there
exists 2-isomorphism $\,\beta\,$
for a, possibly modified, choice of 1-isomorphism $\,\alpha\,$
if and only if the obstruction class
\qq
[\Delta_{1,1}b]+{\mathcal H}^{1,2}\ \in\
H^1(\Gamma^2\times M,U(1))\hspace{0.02cm}\big/\hspace{0.02cm}
{\mathcal H}^{1,2}\,.\qquad
\label{betaod}
\qqq
\vskip -0.3cm
\no vanishes.
\vskip -0.3cm
\hspace{13cm}$\blacksquare$
\end{proposition}
\vskip 0.3cm

\no In the particular case with simply connected components of
 $\,\Gamma\,$ and $\,M$, \,groups
$\,H^1(\Gamma^p\times M,U(1))\,$ are trivial and we obtain
\vskip 0.3cm

\begin{corollary}
If the connected components of $\,\Gamma\,$ and $\,M\,$ are simply
connected then the class (\ref{betaod}) obstructing the existence of
2-isomorphism $\,\beta\,$ is zero.
\end{corollary}
\

\vskip -0.8cm

\no This applies to the case of $\,Z$-equivariant structures on
gerbes $\,\CG_k\,$ over groups $\,\tilde G\,$ discussed in
\cite{GSW,GSW1}.
\vskip 0.2cm

In the general situation, a more precise description of spaces
$\ H^1(\Gamma^2\times M,U(1))\supset{\mathcal H}^{1,2}\,$ may be
provided with the help of the Universal Coefficient and K\"unneth Theorems.
\,One has
\qq
H^1(\Gamma\times M,U(1))\ \cong\ H^1(\Gamma,U(1))^{\pi_0(M)}
\oplus H^1(M,U(1))^{\pi_0(\Gamma)}\,.
\label{1dec0}
\qqq
The element $\,[b']\in H^1(\Gamma\times M,U(1))\,$ is represented
by the sequences with elements
\qq
[b']_1([m])\,:=\,(\iota^1_m)^*[b']\,\in\,H^1(\Gamma,U(1))\,,\qquad\quad
[b']_2([\gamma])\,:=\,(\iota^2_\gamma)^*[b']\,\in\,H^1(M,U(1))\,,
\label{1dec1}
\qqq
where $\,m$, \,resp. $\gamma$, \,are chosen
points in the connected components $\,[m]\in\pi_0(M)$, \,resp.
$[\gamma]\in\pi_0(\Gamma)$,
\,and $\,\iota^1_m:\Gamma\rightarrow \Gamma\times M$,
\,resp. $\iota^2_\gamma:M\rightarrow\Gamma\times M$, \,are the injections
\,with $\,\iota^1_m(\gamma)=\iota^2_\gamma(m)=(\gamma,m)$.
\ Similarly,
\qq
H^1(\Gamma^2\times M,U(1))\ \cong\ H^1(\Gamma,U(1))^{\pi_0(\Gamma)\times\pi_0(M)}
\oplus H^1(\Gamma,U(1))^{\pi_0(\Gamma)\times\pi_0(M)}
\oplus H^1(M,U(1))^{\pi_0(\Gamma)^2}.
{}
\qqq
An element $\ [d]\in H^1(\Gamma^2\times M,U(1))\ $ is represented by
the sequences with elements
\qq
&&[d]_1([\gamma_2],[m])\,:=\,(\iota^1_{\gamma_2,m})^*[d]\,
\in\,H^1(\Gamma,U(1))\,,\\
&&[d]_2([\gamma_1],[m])\,:=\,
(\iota^2_{\gamma_1,m})^*[d]\,\in\,H^1(\Gamma,U(1))\,,\\
&&[d]_3([\gamma_1],[\gamma_2])\,:=\,(\iota^3_{\gamma_1,\gamma_2})^*[d]
\hspace{0.02cm}\in\,H^1(M,U(1))\,,
{}
\qqq
where $\ \iota^1_{\gamma_2,m},\,\iota^2_{\gamma_1,m}:\Gamma\rightarrow
\Gamma^2\times M\ $ and $\ \iota^3_{\gamma_1,\gamma_2}:M\rightarrow \Gamma^2
\times M\ $ are the injections with $\ \iota^1_{\gamma_2,m}(\gamma_1)
=\iota^2_{\gamma_1,m}(\gamma_2)=\iota^3_{\gamma_1,\gamma_2}(m)
=(\gamma_1,\gamma_2,m)$. \ Compositions of the
above injections with simplicial maps $\,\Delta^2_q\,$ are
\qq
&&\Delta^2_0\circ\iota^1_{\gamma_2,m}(\gamma_1)\,=\,\Delta^2_0\circ
\iota^2_{\gamma_1,m}(\gamma_2)\,=\,\Delta^2_0\circ\iota^3_{\gamma_1,
\gamma_2}(m)\,=\,(\gamma_2,m)\,,\label{1of3}\\
&&\Delta^2_1\circ\iota^1_{\gamma_2,m}(\gamma_1)\,=\,
\Delta^2_1\circ\iota^2_{\gamma_1,m}
(\gamma_2)\,=\,\Delta^2_1\circ\iota^3_{\gamma_1,\gamma_2}(m)
\,=\,(\gamma_1\gamma_2,m)\,,\\
&&\Delta^2_2\circ\iota^1_{\gamma_2,m}(\gamma_1)\,=\,
\Delta^2_2\circ\iota^2_{\gamma_1,m}
(\gamma_2)\,=\,\Delta^2_2\circ\iota^3_{\gamma_1,\gamma_2}(m)\,=\,
(\gamma_1,\gamma_2m)\,.
{}
\qqq
Since $\ \Delta_{1,1}=(\Delta^2_0)^*-(\Delta^2_1)^*+(\Delta^2_2)^*$, \ it
follows that
\qq
&&[\Delta_{1,1}b]_1([\gamma_2],[m])\,=\,[-R_{\gamma_2}^*(\iota^1_m)^*b\,+\,
(\iota^1_{\gamma_2m})^*b]\,,\\
&&[\Delta_{1,1}b]_2([\gamma_1],[m])\,=\,[(\iota^1_m)^*b\,
-\,L_{\gamma_1}^*(\iota^1_m)^*b\,+\,(\iota^2_{\gamma_1}\circ r_m)^*b]\,,\\
&&[\Delta_{1,1}b]_3([\gamma_1],[\gamma_2])\,=\,[(\iota^2_{\gamma_2})^*b\,
-\,(\iota^2_{\gamma_1\gamma_2})^*b\,+\,\ell_{\gamma_2}^*(\iota^2_{\gamma_1})^*b]\,,
\qqq
where $\,L_{\gamma},R_{\gamma}:\Gamma\rightarrow\Gamma\,$ denote,
\,respectively, \,the left and the right multiplication by
$\,\gamma$, $\,r_m(\gamma)=\gamma m\,$ (as before), \,and we used
the fact that the class in $\,H^1(\Gamma,U(1))\,$ of the pullback of
$\,A^1({\mathcal O}^1)\,$ along a constant map is trivial.
\,When the group $\,\Gamma\,$
is connected, we may choose its identity element as its special
point and the above equations reduce to
\qq
\label{cal3}
[\Delta_{1,1}b]_1([1],[m])\,=\,0\,,\qquad
[\Delta_{1,1}b]_2([1],[m])\,=\,[\,(\iota^2_{1}\circ r_m)^*b]\,,\qquad
[\Delta_{1,1}b]_3([1],[1])\,=\,[(\iota^2_{1})^*b]\,.\quad
\qqq

In the case of a $\,\Gamma$-space $\,M=G\,$ in the coset-model
context of Definition \ref{def:context}, \,we may take $\,m=1\in
G\,$ in the last formulae which reduce then further to the relations
\begin{eqnarray}
\label{cal4}
[\Delta_{1,1}b]_1([1],[1])\,=\,0\,,\qquad
[\Delta_{1,1}b]_2([1],[1])\,=\,0\,,\qquad
[\Delta_{1,1}b]_3([1],[1])\,=\,[(\iota^2_{1})^*b]\,,
\end{eqnarray}
because $\,\iota^2_{1}\circ r_1\,$ is a constant map. \,In particular,
for $\,b=b'\,$ with $\,D_{1,1}b'=0$,
\begin{eqnarray}
\label{cal5}
[\Delta_{1,1}b']_1([1],[1])\,=\,0\,,\qquad
[\Delta_{1,1}b']_2([1],[1])\,=\,0\,,\qquad
[\Delta_{1,1}b']_3([1],[1])\,=\,[b']_2([1])\,.
\end{eqnarray}
Since $\,[b']_2([1])\,$ runs through arbitrary element of $\,H^1(M,U(1))$,
\,it follows that the obstruction class (\ref{betaod}) vanishes and,
for an appropriate choice of $\,b'\,$ with $\,D_{1,1}b'=0$,
$\,[\Delta_{1,1}(b+b')]=0\,$ so that $\ \Delta_{1,1}(b+b')=-D_{0,2}a\ $
for some $\,a\in A^0(\CO^2)$.
\,We obtain this way

\begin{corollary}
\label{cor:betaE}
For the $\,\Gamma$-space $\,M=G\,$ in the coset-model context, \,an appropriate
choice of 1-isomorphism $\,\alpha\,$ of Definition
\ref{def:equivgerbe} assures the existence of 2-isomorphism $\,\beta$.
\end{corollary}

\subsection{Obstructions to the commutativity of \,diagram
(\ref{axiom})}
\label{sec:diagrobstr}

\no By Proposition \ref{prop:2obstrc}, \,the vanishing of obstruction
(\ref{betaod}) guarantees in the general case that 2-isomorphism
$\,\beta\,$ exists for a suitable choice of \,1-isomorphism $\,\alpha$.
\,In terms of local data, \,the condition
$\,[\Delta_{1,1}b]\in\CH^{1,2}\,$  assures that after the modification
of local data $\,b\,$ by an appropriate $\,D_{1,1}$-cocycle $\,b'$,
\,determined up to the change $\,b'\mapsto b'-D_{0,1}a''$, \,there exists
$\,a\in A^0(\mathcal{O}^2)\,$ such that
\qq
\Delta_{1,1}(b+b')\ =\ -D_{0,2}a\,.
\qqq
In view of the freedom of choice of $\,b'$, \,the cochain $\,a\,$
is determined up to the replacement $\ a\mapsto
a+\Delta_{0,1}a''+a'''\ $ for $\,a''\in A^0(\mathcal{O}^1)\,$
and $\,a'''\in{\rm ker}\,D_{0,2}\cong H^0(\Gamma^2\times M,U(1))$.
Cocycle $\,a'''\,$ describes the possible choices of
2-isomorphism $\,\beta$. \,The commutativity of the diagram (\ref{axiom})
of 2-isomorphisms of gerbes
over $\,\Gamma^3\times M\,$ is now equivalent to the condition that, after
the restriction to a sufficiently fine simplicial sequence of coverings,
\qq
\Delta_{0,2}a\ =\ 0\,.
{}
\qqq
Note that, in any case,
\qq
D_{0,3}\Delta_{0,2}a\ =\ \Delta_{1,2}D_{0,2}a\ =\ -\,
\Delta_{1,2}\Delta_{1,1}(b+b')\ =\ 0
{}
\qqq
so that $\ \Delta_{0,2}a\,\in\,{\rm ker}\,D_{0,3}\,\cong
\,H^0(\Gamma^3\times M,U(1))$.
\,Let us denote by $\,{\mathcal H}^{0,3}\,$ the image of the map
$\ [\Delta_{0,2}]:H^0(\Gamma^2\times M,U(1))\rightarrow
H^0(\Gamma^3\times M,U(1))\ $ that sends $\,a'''\,$ to $\,\Delta_{0,2}a'''$.
\,Using the freedom in the choice of the cochain $\,a\,$ and the relation
\qq
\Delta_{0,2}(a+\Delta_{0,1}a''+a''')\ =\ \Delta_{0,2}a\,+\,\Delta_{0,2}a'''\,,
{}
\qqq
we infer

\begin{proposition}
2-isomorphism $\,\beta\,$ may be chosen so that the diagram
\,(\ref{axiom}) \,of \,Definition \ref{def:equivgerbe}
is commutative if and only if the obstruction class
\qq
\Delta_{0,2}a\,+\,{\mathcal H}^{0,3}\
\in\ H^0(\Gamma^3\times M,U(1))\hspace{0.02cm}\big/
\hspace{0.02cm}{\mathcal H}^{0,3}
\qquad\label{3obstr}
\qqq
vanishes.
\end{proposition}
\vskip -0.4cm
\hspace{13cm}$\blacksquare$
\vskip 0.3cm

Elements $\,f^p\in H^0(\Gamma^p\times M,U(1))\,$ are locally
constant $\,U(1)$-valued functions on $\,\Gamma^p\times M$. \,One
may identify them with $p$-chains $\,v^p\,$ on the group
$\,\pi_0(\Gamma)\,$ with values in the $\,\pi_0(\Gamma)$-module
$\,U(1)^{\pi_0(M)}\cong H^0(M,U(1))\,$ of $\,U(1)$-valued functions
on $\,\pi_0(M)$, \,where the action of $\,\pi_0(\Gamma)\,$ on
$\,U(1)^{\pi_0(M)}\,$ is induced from the action of $\,\Gamma\,$ on
$\,M$. \,If the identification is done by the formula
\qq
f^p(\gamma_1,\dots,\gamma_p,m)\ =\ v^p_{[\gamma_p^{-1}],\dots,
[\gamma_1^{-1}]}([m])\,,
\label{fpvp}
\qqq
then the induced maps $\ [\Delta_{0,p}]:H^0(\Gamma^p\times
M,U(1))\rightarrow H^0(\Gamma^{p+1} \times M,U(1))\ $ become the
coboundary operators $\,\delta^p\,$ of the group $\,\pi_0(\Gamma)\,$
cohomology:
\qq
(\Delta_{0,p}f^p)(\gamma_1,\dots,\gamma_p,\gamma_{p+1},m)\
=\ (-1)^{p+1}(\delta^pv^p)_{[\gamma_{p+1}^{-1}],[\gamma_p^{-1}],
\dots,[\gamma_1^{-1}]}([m])\,.
{}
\qqq

\begin{corollary}
\label{cor:3.8}
Under identification (\ref{fpvp}),
the cochain $\,\Delta_{0,2}a\,$ generates a 3-cocycle $\,v^3\,$ of the group
$\,\pi_0(\Gamma)\,$ taking values in $\,U(1)^{\pi_0(M)}\,$ and the obstruction
coset (\ref{3obstr}) is the cohomology class
$\ [v^3]\,\in\,H^3(\pi_0(\Gamma),\hspace{0.02cm}U(1)^{\pi_0(M)})$.
\end{corollary}

\no In particular, when $\,\Gamma\,$ is discrete and $\,M\,$ is
connected, \,then $\,[v^3]\in H^3(\Gamma,U(1))$. \,That is the
situation for the $\,Z$-equivariant structures on gerbes $\,\CG_k\,$
over groups $\,\tilde G\,$ discussed in \cite{GSW,GSW1} and
mentioned already above. \,The obstruction cohomology classes
$\,[v^3]\in H^3(Z,U(1))\,$ were computed for these cases and simple
$\,\tilde G\,$ in \cite{GR1}.

Since the cohomology groups
$\ H^p(\pi_0(\Gamma),\hspace{0.02cm}U(1)^{\pi_0(M)})\ $ for $\,p>1\,$
are trivial if $\,\pi_0(\Gamma)\,$ is a trivial group, \,we obtain

\begin{corollary}\label{cor:6.8}
If the symmetry group $\,\Gamma\,$ is connected and
2-isomorphism $\,\beta\,$ of Definition \ref{def:equivgerbe} exists,
then it may always be chosen so that the diagram (\ref{axiom})
commutes.
\end{corollary}

Putting together Proposition \ref{prop:secE} and Corollaries
\ref{cor:betaE} and \ref{cor:6.8}, we summarize the results for the
situation discussed in Sec.\,\ref{sec:expl1}:

\begin{theorem}
\label{thm:obstrsecE}
For a $\,\Gamma$-space $\,M=G\,$ in the coset-model context of
Definition \ref{def:context}, a $\,\Gamma$-equivariant structure on
the WZW gerbe $\,\CG_k\,$ over $\,G\,$ exists if and only if the
global-anomaly phases (\ref{ugf}) are trivial, \,as, for example,
for $\ G=\tilde G$.
\end{theorem}
\vskip -0.3cm
\hspace{13cm}$\blacksquare$

\subsection{Classification of equivariant structures}
\label{sec:classif}

\no Suppose now that we are given two equivariant structures $\,(\alpha_i,
\beta_i)$, $i=1,2$, \,on gerbe $\,\CG\,$ with local data
$\,c\in A^2(\CO^0)$, $\,D_{2,0}c=0$, \, for a sufficiently fine
simplicial sequence of coverings $\,\left \lbrace \mathcal{O}^p
\right \rbrace$. \ Their local data are $\,(b_i,a_i)$, \,with
$\,b_i\in A^1(\CO^1)\,$ and $\,a_i\in A^0(\CO^2)$, \,that satisfy
\qq
\Delta_{2,0}c+\rho\ =\ D_{1,1}b_i\,,\qquad \Delta_{1,1}b_i\
=\ -D_{0,2}a_i\,,\qquad \Delta_{0,2}a_i\ =\ 0\,.
{}
\qqq
The difference $\ (b',a')=(b_2-b_1,a_2-a_1)\,$ gives local data
for a $\,\Gamma$-equivariant structure on the trivial gerbe
$\,\CI_0\,$ (relative to $\,\rho=0$).  \,It
satisfies the homogeneous equations
\qq
D_{1,1}b'\ =\ 0\,,\qquad \Delta_{1,1}b'\ =\ -D_{0,2}a'\,,\qquad \Delta_{0,2}a'\
=\ 0\,.
\label{homeq}
\qqq
There is an isomorphism $\,(\chi,\eta)\,$ between the equivariant
structures $\,(\alpha_i,\beta_i)\,$ if there exist: a cocycle
$\,e\in A^1(\CO^0)$, $\,D_{1,0}e=0\,$ (providing local data for
1-isomorphism $\,\chi:\CG\rightarrow\CG$) \,and a cochain
$\,f\in A^0(\CO^1)\,$ (giving local data for 2-isomorphism
$\,\eta$) \,such that
\qq
b'\ =\ \Delta_{1,0}e\,+\,D_{0,1}f\,,\qquad a'\ =\ -\Delta_{0,1}f\,.
\label{imhom}
\qqq
These identities represent the definition of $\,\eta\,$ and the
commutativity of diagram (\ref{commdiag1iso}), respectively.
They imply Eqs.\,(\ref{homeq}). Classes of
solutions to Eqs.\,(\ref{homeq}) modulo solutions to Eqs.\,(\ref{imhom})
form the $\,2^{\rm nd}\,$ hypercohomology group $\,\mathbb H^2(\CJ)\,$
of the double complex $\,\CJ\,$
\qq
\xymatrix{0 \ar[r] & A^0(\CO^0) \ar[d]_{\Delta_{0,0}} \ar[r]^-{D_{0,0}}
& \mathrm{ker}D_{1,0} \ar[d]^{\Delta_{1,0}} \\ 0 \ar[r] & A^0(\CO^1)
\ar[d]_{\Delta_{0,1}} \ar[r]^-{D_{0,1}} & \mathrm{ker}D_{1,1}
\ar[d]^{\Delta_{1,1}} \\
0 \ar[r] & A^0(\CO^2) \ar[d]_{\Delta_{0,2}} \ar[r]^-{D_{0,2}}
& \mathrm{ker}D_{1,2} \ar[d]^{\Delta_{1,2}}
\\ 0 \ar[r] & A^0(\CO^3) \ar[r]^-{D_{0,3}} & \mathrm{ker}D_{1,3}}
{}
\qqq
obtained from the double complex $\,\CK=\big(A^n(\CO^p)\big)$.
$\,\mathbb H^2(\CJ)\,$ is the group of isomorphism classes of
$\,\Gamma$-equivariant structures on the trivial gerbe $\,\CI_0$.
It acts freely and transitively on the set of isomorphism classes
of $\,\Gamma$-equivariant structures on gerbe $\,\CG$. \,In other words,

\begin{proposition}
The set of isomorphisms classes of $\,\Gamma$-equivariant structures
on gerbe $\,\CG\,$ is a torsor for the Abelian group
$\,\mathbb H^2(\CJ)$.
\end{proposition}
\vskip -0.5cm
\hspace{13cm}$\blacksquare$
\vskip 0.2cm

Denote by $\,{\mathcal H}^{1,1}\,$ the image of the map
$\ [\Delta_{1,0}]:H^1(M,U(1))\rightarrow H^1(\Gamma\times M,U(1))\ $
that sends class $\,[e]\,$ to class $\,[\Delta_{1,0}e]$.
\,In terms of the decomposition (\ref{1dec0}) and (\ref{1dec1}),
\qq
[\Delta_{1,0}e]_1([m])\,=\,-[r_m^*e]\,,\qquad[\Delta_{1,0}e]_2([\gamma])
\,=\,[e]\,-\,[\ell_\gamma^*e]\,.
\label{1dec2}
\qqq
Since $\,b'\,$ is a $\,D_{1,1}$-cocycle, \,one may consider the map
\qq
(b',a')\ \longmapsto\ [b']+{\mathcal H}^{1,1}\,\in\,H^1(\Gamma
\times M,U(1))\big/{\mathcal H}^{1,1}\,.
\label{map1}
\qqq
Since $\,[b']\in{\mathcal H}^{1,1}\,$ for $\,(b',a')\,$ of the form
(\ref{imhom}), \,the map (\ref{map1}) induces a homomorphism
\qq
\kappa:\,\mathbb{H}^2(\CJ)\,\rightarrow\,H^1(\Gamma\times M,U(1))\big/
{\mathcal H}^{1,1}
\label{kappa}
\qqq
of Abelian groups. \,To describe the image and the kernel of $\,\kappa$,
\,we shall do some tracing of diagrams.

If $\,[b'']+{\mathcal H}^{1,1}\,$ is in the image of $\,\kappa$,
\,then $\,b''=b'+\Delta_{1,0}e+D_{0,1}f\,$ for some $\,(b',a')\,$ as
above, \,some $\,e\in A^1(\CO^0)$ with $\,D_{1,0}e=0$, \,and some
$\,f\in A^0(\CO^1)$. \,Consequently,
$\,\Delta_{1,1}b''=-D_{0,2}a'+\Delta_{1,1}D_{0,1}f=
-D_{0,2}(a'-\Delta_{0,1}f)\,$ so that $\ [\Delta_{1,1}][b'']=0$.
\,For any $\,[b'']\,$ that satisfies the latter equation, \,i.e. such
that $\,\Delta_{1,1}b''=-D_{0,2}a''$ for some $\,a''\in
A^0({\mathcal O}^2)$, \,we have
$\,D_{0,3}\Delta_{0,2}a''=\Delta_{1,2}D_{0,2}a''=0$, \,hence
$\ \Delta_{0,2}a''\in H^0(\Gamma^3\times M,U(1))\ $ and it generates, via
Eq.\,(\ref{fpvp}), a 3-cocycle $\,v^3\,$ on group
$\,\pi_0(\Gamma)\,$ with values in $\,U(1)^{\pi_0(M)}$. \,If
$\,[b'']+{\mathcal H}^{1,1}\,$ is in the image of $\,\kappa$,
\,then, \,for $\,a''':=a'-\Delta_{0,1}f-a''$, \,we have
$\,D_{0,2}a'''=0\,$ so that $\ a'''\in H^0(\Gamma^2\times M,U(1))\ $
generates, again via Eq.\,(\ref{fpvp}), a 2-cochain $\,u^2\,$
on $\,\pi_0(\Gamma)\,$ with values in $\,U(1)^{\pi_0(M)}$. \,The relation
$\,\Delta_{0,2}a'''=-\Delta_{0,2}a''\,$ implies then that
$\,\delta^2u^2=v^3\,$ so that the cohomology class of
$\ [v^3]\in H^3(\pi_0(\Gamma),U(1)^{\pi_0(M)})\ $
vanishes. \,Conversely, if this is the case, then $\,\Delta_{0,2}a''
=-\Delta_{0,2}a'''\,$ for some $\,a'''\in H^0(\Gamma^2\times M,U(1))\,$
so that, for $\,a'=a''+a'''$, \,one has $\,\Delta_{1,1}b''=-D_{0,2}a'\,$
and $\,\Delta_{0,2}a'=0$. \,We have proven this way

\begin{lemma}
\label{lem:image}
$\,[b'']+{\mathcal H}^{1,1}\,$ is in the image of $\,\kappa\,$ if and only if
\item[\ 1.\ ]
$[\Delta_{1,1}][b'']\,=\,0\ $ so\ that\ $\ \Delta_{1,1}b''\,=\,-D_{0,2}a''$\,,
\item[\ 2.\ ]
\parbox[t]{15cm}{the\ cohomology\ class\ $\ [v^3]\,\in\,
H^3(\pi_0(\Gamma),U(1)^{\pi_0(M)})\ $\ of\ the\ 3-cocycle\ $\ v^3\ $
corresponding, via Eq.\,(\ref{fpvp}), to\ $\ \Delta_{0,2}a''\,
\in\,H^0(\Gamma^3\times M,U(1))\ $ vanishes.}
\end{lemma}
\vskip -0.5cm
\hspace{13 cm}$\blacksquare$

Now, \,let us study the kernel of $\,\kappa$. \,If
$\,[b']\in{\mathcal H}^{1,1}$, \,i.e. $b'=\Delta_{1,0}e+D_{0,1}f'\,$
for $\,e\in A^1({\mathcal O}^0)\,$ with $\,D_{1,0}e=0\,$ and
$\,f'\in A^0(\CO^1)$, \,then $\ \Delta_{1,1}b'=\Delta_{1,1}D_{0,1}f'
=D_{0,2}\Delta_{0,1}f'\ $ so that
\qq
a'+\Delta_{0,1}f'\,\in\,{\rm ker}\,D_{0,2}\,\cong\,H^0(\Gamma^2\times M,U(1))\,.
{}
\qqq
Since $\ \Delta_{0,2}(a'+\Delta_{0,1}f')=0$, \ the cochain
$\,a'+\Delta_{0,1}f'\,$ may be identified, by means of
Eq.\,(\ref{fpvp}), with a 2-cocycle $\,v^2\,$ on group
$\,\pi_0(\Gamma)\,$ with values in $\,U(1)^{\pi_0(M)}$. \,All
2-cocycles $\,v^2\,$ may be obtained this way by changing $\,a'\,$
to $\,a'+a''\,$ with $\,D_{0,2}a''=0=\Delta_{0,2}a''$. \,Since
$\,f'\,$ is defined modulo $\ f''\in {\rm ker}\,D_{0,1}\cong
H^0(\Gamma\times M,U(1))$, \,2-cocycle $\,v^2\,$ is defined modulo
coboundaries of the 1-cochains $\,u^1\,$ corresponding to $\,f''\,$
so that the cohomology class $\ [v^2]\in
H^2(\pi_0(\Gamma),U(1)^{\pi_0(M)})\ $ is well defined by the pair
$\,(b',a')\,$ with $\,[b']\in{\mathcal H}^{1,1}$. \ The class
$\,[v^2]\,$ vanishes if and only if $\,(b',a')\,$ is of the form
(\ref{imhom}). \,This shows

\begin{lemma}
\label{lem:kernel}
The kernel of the map $\,\kappa\,$ of (\ref{kappa})
may be identified with the cohomology group $\ H^2(\pi_0(\Gamma),
U(1)^{\pi_0(M)})$.
\end{lemma}
\vskip -0.5cm
\hspace{13 cm}$\blacksquare$

Let us look at some special cases. \,First, if
$\ H^1(\Gamma,U(1))=\{0\}=H^1(M,U(1))$, \,then the homomorphism
$\,\kappa\,$ vanishes and we obtain from Lemma \ref{lem:kernel}:

\begin{corollary}
In the case when the connected components of $\,\Gamma\,$ and $\,M\,$ are
simply connected, $\ \mathbb{H}^2(\CJ)\,\cong\,H^2(\pi_0(\Gamma),U(1)^{\pi_0(M)})$.
\end{corollary}

\no This is the result that gives, e.g., the classification of
$\,Z$-equivariant structures on gerbe $\,\CG_k\,$ over $\,\tilde
G\,$ for $\,Z\subset\tilde Z\,$ acting by multiplication, \,see
\cite{GR1,GSW,GSW1}.

Suppose now that $\,\Gamma\,$ is connected so that $\,\Gamma
=\tilde\Gamma/\tilde Z_\Gamma$, \,where $\,\tilde\Gamma\,$ is a simply
connected Lie group and $\,\tilde Z_\Gamma\,$ is a subgroup of the center
of $\,\tilde\Gamma$. \,One has $\,H_1(\Gamma)\cong\pi_1(\Gamma)\cong\tilde
Z_\Gamma$. \,Lemma \ref{lem:kernel}
implies in that case that $\,\kappa\,$ is injective onto its image
which, by Lemma \ref{lem:image} and Eq.\,(\ref{cal3}), is composed
of the cosets $\,[b'']+{\mathcal H}^{1,1}\,$ such that
$\,[b'']_2([1])=0\,$ in the decomposition (\ref{1dec0}) and
(\ref{1dec1}). \,From the explicit form (\ref{1dec2}) of
$\,{\mathcal H}^{1,1}$, \,we then infer

\begin{corollary}
\label{cor:gconn} If the group $\,\Gamma\,$ and manifold $\,M\,$ are
connected, then
\qq
{\mathbb H}^2(\CJ)\ \cong\ H^1(\Gamma,U(1))\big/[r_m^*](H^1(M,U(1)))\
\cong\ Z_M^*\,,
\qqq
\no where $\,Z_M^*\,$ is the group of characters of the kernel $\,Z_M\,$
of the homomorphism from $\,H_1(\Gamma)\,$ to $\,H_1(M)\,$  induced by the
map $\,\displaystyle{\gamma\mathop{\longmapsto}\limits^{r_m}\gamma m}$.
\end{corollary}

\no In particular, we have

\begin{corollary}
\label{cor:class}
For the $\,\Gamma$-space $\,M=G\,$ in the coset-model context, see Definition
\ref{def:context}, $\,Z_M=\tilde Z_\Gamma\,$ so that
\qq
\mathbb{H}^2(\CJ)\ \cong\ H^1(\Gamma,U(1))
\ \cong\ \tilde Z_\Gamma^{^*}
\qqq
and the $\,\Gamma$-equivariant structures on the WZW gerbes
$\,\CG_k\,$ over $\,G\,$ are classified by the group of characters
of $\,H_1(\Gamma)\cong\pi_1(\Gamma)\cong\tilde Z_\Gamma$.
\end{corollary}

Let us analyze closer the case when the
$\,\Gamma$-space $\,M\,$ is a (left) principal $\,\Gamma$-bundle
$\,\omega:M\rightarrow M'$. \,By the descent Theorem \ref{thm:descent},
\,each $\,\Gamma$-equivariant
structure on the trivial gerbe $\,\CI_0\,$ relative to the
vanishing 2-form descends to a flat gerbe on $\,M'\,$ whose
pullback to $\,M\,$ is 1-isomorphic to $\,\CI_0$. \,Passing
to isomorphism classes, one obtains the canonical injective homomorphism
\qq
\nu:\,{\mathbb H}^2(\CJ)\,\rightarrow\,H^2(M',U(1))
{}
\qqq
that maps into the kernel of the pullback map $\,[\omega^*]:
H^2(M',U(1))\rightarrow H^2(M,U(1))$. \,Now, suppose that we are
given a flat gerbe on $\,M'\,$ whose class is in the kernel
of $\,[\omega^*]$. \,It is easy to see, using Theorem \ref{thm:descent}
and Remark \ref{rem:ind}(2), that such a gerbe is 1-isomorphic
to a gerbe that descends from the trivial gerbe $\,\CI_0\,$ equipped with
a $\,\Gamma$-equivariant structure (relative to $\,\rho=0$).
\,This shows that $\,\nu\,$ maps onto the kernel of $\,[\omega^*]$.
\,We obtain this way

\begin{corollary}
\label{cor:ambig1}
In the case when $\,M\,$ is a principal $\,\Gamma$-bundle, \,there
is an exact sequence of Abelian groups
\qq
0\ \longrightarrow\ {\mathbb H}^2(\CJ)\ \mathop{\longrightarrow}\limits^{\nu}
\ H^2(M',U(1))\ \mathop{\longrightarrow}\limits^{[\omega^*]}\ H^2(M,U(1))
\label{exseq1}
\qqq
that induces an isomorphism between $\,{\mathbb H}^2(\CJ)\,$ and
the kernel of $\,[\omega^*]\,$ in $\,H^2(M',U(1))$.
\end{corollary}

\no If $\,\Gamma\,$ and $\,M\,$ are connected, then the exact
sequence (\ref{exseq1}) is induced, in virtue of Corollary \ref{cor:gconn},
by the cohomology exact sequence for the $\,\Gamma$-bundle
$\,\omega:M\rightarrow M'\,$ \cite{Serre,DFN}
\qq
H^1(M,U(1))\ \mathop{\longrightarrow}^{[r_m^*]}\ H^1(\Gamma,U(1))\
\mathop{\longrightarrow}\limits^\tau
\ H^2(M',U(1))\ \mathop{\longrightarrow}^{[\omega^*]}
\ H^2(M,U(1))\,.
\label{exseq2}
\qqq
\,The middle arrow $\tau\,$ may be easily described in terms of the
classifying space $\,B\Gamma\,$ of group $\,\Gamma$. \,The
transgression map $\,H^2(B\Gamma,U(1))\rightarrow H^1(\Gamma,U(1))\,$
is an isomorphism for connected $\,\Gamma$. \,Its composition
with $\,\tau\,$ is given by the pullback map from $\,H^2(B\Gamma,U(1))\,$
to $\,H^2(M',U(1))\,$ along the classifying map
$\,f_\omega:M'\rightarrow B\Gamma\,$ for the principal bundle
$\,\omega:M\rightarrow M'$. \,In Appendix
\ref{app:6}, we describe an equivalent construction of homomorphism
$\,\tau$. \,That construction, carried out in terms
of line bundles and gerbes, will be used below.

\subsection{Ambiguity of gauged amplitudes}
\label{sec:ambig}

\no Let us recall from Sec.\,\ref{sec:topntr} how the WZ amplitudes
coupled to a topologically non-trivial gauge field $\,\CA\,$ in the
principal $\,\Gamma$-bundle $\,P\,$ over the worldsheet $\,\Sigma\,$
were defined. They were given by the holonomy of gerbe $\,\CG_\CA\,$
over the associated bundle $\ P_M=P\times_\Gamma\hspace{-0.05cm}M$,
\,see Definition \ref{def:nontrivialamplitudes}.
\ That gerbe was obtained via Theorem \ref{thm:descent}
from gerbe $\ \tilde\CG_\CA= \CI_{\tilde\rho_\CA}\otimes\CG_2\ $
over $\,\tilde M=P\times M\,$ equipped with a $\,\Gamma$-equivariant
structure (relative to $\,\rho=0$) \,induced from that of
$\,\CG$. \,Let us use the subscript $\,M\,$ or $\,\tilde M\,$ to
distinguish between the two cases of $\,\Gamma$-spaces. \,If we
change the isomorphism class of a $\,\Gamma$-equivariant structure
on $\,\CG\,$ by a class $\,K\in{\mathbb H}^2(\CJ_M)$, \,then a quick
inspection of the proof of Proposition \ref{prop:equiv} shows
that the isomorphism class of the induced $\,\Gamma$-equivariant
structure on $\,\tilde\CG_\CA\,$ changes by the class
$\,K_2\in{\mathbb H}^2(\CJ_{\tilde M})\,$ obtained by the pullbacks
along the projection $\,pr_2\,$ of $\,P\times M$ on the second factor.
\,The isomorphism class of the descended gerbe
$\,\CG_\CA\,$ changes then, according to the discussion from
Sec.\,\ref{sec:classif}, by
\qq
\nu_{\tilde M}(K_2)\,\in\, H^2(P_M,U(1))\ \cong\ Hom(H_2(P_M),U(1))\,.
\qqq
Viewed as a character of $\,H_2(P_M)$, \,class $\,\nu_{\tilde
M}(K_2)\,$ describes the change of the holonomy of the gerbe
$\,\CG_{\CA}$. \,We obtain this way

\begin{corollary}
\label{cor:ambig2}
Under the change of the isomorphism class of a $\,\Gamma$-equivariant
structure on gerbe $\,\CG\,$ over $\,M\,$ by a class
$\,K\in{\mathbb H}^2(\CJ_M)$, \,the WZ amplitude (\ref{FAg2}) of a section
$\,\Phi:\Sigma\rightarrow P_M\,$
of the associated bundle is multiplied by the $\,U(1)\,$ phase
$\ \big\langle[\Phi],\nu_{\tilde M}(K_2)\big\rangle$, \ where
$\,[\Phi]\,$ denotes the homology class of $\,\,\Phi$.
\end{corollary}

\begin{remark}
The dependence of the gauged WZ amplitudes on the choice of an
equivariant structure is another manifestation of the phenomenon of
``discrete torsion'' \cite{Vafa}
\end{remark}

In the particular situation where manifolds $\,\,\Gamma$,
$\,M\,$ and $\,\Sigma\,$ are connected, \,Corollary \ref{cor:gconn}
implies that
\qq
{\mathbb H}^2(\CJ_M)\ \cong\ Z_M^*\,,\qquad{\mathbb H}^2(\CJ_{\tilde M})
\ \cong\ Z_{\tilde M}^*\,.
{}
\qqq
We shall denote by $\,\chi_K\,$ the character of $\,Z_M\,$ corresponding to
$\,K\in{\mathbb H}^2(\CJ_M)\,$ and by $\,\chi_{\tilde K}\,$ the one of
$\,Z_{\tilde M}\,$ corresponding to $\,\tilde K\in{\mathbb H}^2(\CJ_{\tilde M})$.
\,The relation $\,pr_2\circ r_{(p,m)}=r_m\,$ for $\,(p,m)\in\tilde M\,$
implies the inclusion $\,Z_{\tilde M}\subset Z_M$. \,
The map $\,Z_M^*\ni\chi_K\mapsto\chi_{K_2}\in Z_{\tilde M}^*\,$ is now given
by the restriction of the characters, whereas the homomorphism
$\,\nu_{\tilde M}\,$ is induced by the map $\ \tau_{\tilde M}:
H^1(\Gamma,U(1))\rightarrow H^2(P_M,U(1))\,$ of the exact sequence
(\ref{exseq2}). \,The problem of ambiguities of the gauged WZ amplitudes
may be completely settled in this case
 employing a construction of homomorphism $\,\tau_{\tilde M}\,$ along
the lines of Appendix \ref{app:6} and an explicit description of
principal $\,\Gamma$-bundles over $\,\Sigma\,$ \cite{Hori}.
\vskip 0.1cm

Up to isomorphism, such bundles may be obtained
by gluing $\,D\times\Gamma\,$ and
$\,(\Sigma\setminus {\dot D})\times\Gamma$, \,where $\,D\,$ is a closed
unit disc embedded into $\,\Sigma\,$ and $\,\dot{D}\,$ its interior,
via the identification
\qq
D\times\Gamma\,\ni\,(\ee^{\,\si\sigma},\gamma(\ee^{\,\si\sigma})\gamma)
\ =\ (\ee^{\,\si\sigma},\gamma)\,\in\,
(\Sigma\setminus{\dot D})\times\Gamma
\qqq
for a transition loop $\ S^1\ni
e^{\,\si\sigma}\mapsto\gamma(\ee^{\,\si\sigma}) \in\Gamma\,$ that we
assume based at the unit element: $\,\gamma(1)=1$. The
$\,\Gamma$-bundle $\,P\,$ depends, up to isomorphism, only on the
element $\,z_P\in\tilde Z_\Gamma\cong\pi_1(\Gamma)\,$ corresponding
to the homotopy class of the transition loop. The associated bundle
$\,P_M\,$ is then obtained by gluing $\,D\times M\,$ and
$\,(\Sigma\setminus{\dot D})\times M\,$ via the identification
\qq
D\times M\,\ni\,(\ee^{\,\si\sigma},\gamma(\ee^{\,\si\sigma})m)\ =\
(\ee^{\,\si\sigma},m)\,\in\,(\Sigma\setminus{\dot D})
\times M\,.
\qqq
A global section $\,\Phi:\Sigma\rightarrow P_M\,$ is given by two maps
\qq
D\ni x\,\mapsto\,\phi_1(x)\in M\qquad{\rm and}\qquad(\Sigma\setminus{\dot D})
\ni x\,\mapsto\,\phi_2(x)\in M
\qqq
such that
\qq
\phi_1(\ee^{\,\si\sigma})\ =\ \gamma(\ee^{\,\si\sigma})
\phi_2(\ee^{\,\si\sigma})\,.
\qqq

According to Appendix \ref{app:6}, the homomorphism
$\,\tau_{\tilde M}$, \,mapping $\,H^1(\Gamma,U(1))
\cong{\tilde Z}^{^*}_\Gamma\,$ to $\,H^2(P_M,U(1))$, \,associates to
a character $\,\chi\in{\tilde Z}^{^*}_\Gamma\,$ a 1-isomorphism class
of a flat gerbe $\,\CG_\chi\,$ on $\,P_M$. \,Consequently,
the phase $\,\big\langle[\Phi],\nu_{\tilde M}(K_2)\big\rangle\,$
is equal to the holonomy $\,\Hol_{\CG_\chi}(\Phi)\,$
for a character $\,\chi\,$ of $\,\tilde Z_\Gamma\,$ extending $\,\chi_K$.
The flat gerbe $\,\CG_\chi\,$ may be trivialized over
$\,D\times M\,$ and $\,(\Sigma\setminus{\dot D})\times M$. \,It is
then given by a transition line bundle \cite{Hitchin}
over $\,S^1\times M\,$
obtained by pulling back the flat line bundle $\,L_\chi\,$ over
$\,\Gamma$, \,described in Appendix \ref{app:6}, along the map
\qq
S^1\times M\ni(\ee^{\,\si\sigma},m)\
\longmapsto\ \gamma(\ee^{\,\si\sigma})\in\Gamma\,.
\qqq
Using such a presentation of gerbe $\,\CG_\chi$, \,it is easy to see
from the geometric definition of the holonomy of gerbes, see, e.g.,
\cite{GR}, \,that the phase $\,\Hol_{\CG_\chi}(\Phi)\,$ is given by
the holonomy of the loop
$\,\ee^{\,\si\sigma}\mapsto\gamma(\ee^{\,\si\sigma})\,$ in the line
bundle $\,L_\chi\,$ over $\,\Gamma$. \,The latter is equal to the
value of the character $\,\chi\,$ on the element $\,z_P\in\tilde
Z_\Gamma$. \vskip 0.1cm

The above phase should be independent of the extension
$\,\chi\,$ of the character $\,\chi_K\,$ from the subgroup $\,Z_M\,$ to
$\,\tilde Z_\Gamma$. \,This does not seem evident. \,Here is the
resolution of the puzzle. \,Let $\,\phi_i\,$ be the maps representing
section $\,\Phi\,$ of $\,P_M$. \,As a boundary value of
a map from the disc to $\,M$, \,the 1-cycle
\qq
S^1\ni\ee^{\,\si\sigma}\,\mapsto\,\phi_2(\ee^{\,\si\sigma})\in M
\qqq
is homologous to a constant 1-cycle. \,Hence the 1-cycle
\qq
S^1\ni\ee^{\,\si\sigma}\,\mapsto\,\phi_1(\ee^{\,\si\sigma})\,=\,
\gamma(\ee^{\,\si\sigma})\phi_2(\ee^{\,\si\sigma})\,,
\qqq
which is a boundary value of a map from $\,\Sigma\setminus{\dot
D}\,$ to $\,M\,$ and, as such, has a trivial class in $\,H_1(M)$,
\,is homologous to
\qq
S^1\ni\ee^{\,\si\sigma}\,\mapsto\,\gamma(\ee^{\,\si\sigma})m
\qqq
for any point $\,m\in M$. \,But the triviality of the class in $\,H_1(M)\,$
of the latter 1-cycle is just the condition that $\,z_P\in Z_M$. \,Note that
in the coset context, there always exists a section $\,\Phi\equiv1\,$
of the associated bundle given by $\,\phi_i\equiv1$. \,In that case
$\,z_P\,$ always belongs to $\,Z_M=\tilde Z_\Gamma$. \ We may summarize
the above discussion in

\begin{theorem}
Let $\,\Gamma$, $\,M$, and $\,\Sigma$ be connected and $\,P\,$
be the principal $\,\Gamma$-bundle over $\,\Sigma\,$ associated to
$\,z_P\in\pi_1(\Gamma)$. \,Then
\item[\ 1.\ ]
if $\,z_P\not\in Z_M\,$ then there are no global sections $\,\Phi\,$
of the associated bundle $\,P_M$;
\item[\ 2.\ ]
for any global section $\,\Phi:\Sigma\rightarrow P_M$,
\qq
\big\langle[\Phi],\nu_{\tilde M}(K_2)\big\rangle\ =\ \chi_K(z_P)\,\in\,U(1)\,.
\qqq
\end{theorem}
\vskip -0.5cm
\hspace{13 cm}$\blacksquare$

\no In particular, if $\,P\,$ is trivializable then $\,z_P=1\,$ and
the gauged WZ amplitudes are independent of the choice of a
$\,\Gamma$-equivariant structure (and may be defined in more general
circumstances discussed in the first part of the paper).
\vskip 0.2cm

In \cite{Hori}, \,K. Hori studied an example of the coset theory based
on the WZW model with simply connected group $\,\tilde G=SU(2)\times SU(2)\,$
at level $\,(k,2)\,$ and with gauged adjoint
action of $\,\Gamma={\rm diag}(SU(2))/{\rm diag}({\bf Z_2})$. \,He argued,
using the supersymmetry present in the coset model, that
the contribution of the sector with gauge fields in the non-trivial
$\,\Gamma$-bundle $\,P\,$ to the coset toroidal partition function was equal
to $\,\frac{1}{2}$, \,but that a choice of a different $\,\theta$-vacuum
should lead to the contribution equal to $\,-\frac{1}{2}$.
\,Since $\,\tilde Z_\Gamma={\bf Z_2}\,$ in this case,
the phase $\pm1$ describes indeed the ambiguity of the topologically
non-trivial sector due to the freedom of choice of an $\,SO(3)$-equivariant
structure on gerbe $\,{\cal G}_{(k,2)}$ over $SU(2)\times SU(2)$.
\,For $\,k\,$ odd, \,such ambiguity is not visible in the
partition function. Indeed, in this case there are no fixed points of
the joint spectral flow of weights under the non-trivial element
of $\,\tilde Z_\Gamma$,
\qq
(j_1,j_2,j')\ \longmapsto\
(\frac{_{k_1}}{^2}-j_1,\frac{_{k_2}}{^2}-j_2,\frac{_{k_1+k_2}}{^2}-j')\,,
\qqq
and, as a result, the contribution of the topologically non-trivial sector
to the toroidal partition function vanishes \cite{Hori}.

\nsection{$\,Ad$-equivariant WZW gerbes
over simply connected groups}
\label{sec:expl2}

\no In order to illustrate the concept of $\,\Gamma$-equivariant gerbes,
\,we shall return to the situation discussed in Sec.\,\ref{sec:expl1}
\,involving the WZW gerbes $\,\CG_k\,$ over connected compact simple
groups $\,G=\tilde G/Z\,$ viewed as $\,\Gamma$-spaces
for $\,\Gamma=\tilde G/\tilde Z\,$ acting by the adjoint action. \,Recall
that Theorem \ref{thm:obstrsecE} states that gerbes $\,\CG_k\,$ possess
$\,\Gamma$-equivariant structures whenever the phases (\ref{ugf})
are trivial, so always for $\,G=\tilde G$. \,Such structures are composed
of 1-isomorphism $\,\alpha\,$ and 2-isomorphism $\,\beta$,
\,see Definition \ref{def:equivgerbe}. \,They are classified
by the dual group of $\,\tilde Z$, \,see Corollary
\ref{cor:class}.  \,What follows is devoted to an explicit construction
of $\,\Gamma$-equivariant structures on gerbes $\,\CG_k\,$
over simply connected groups $\,\tilde G$.

Instead of the local data formalism used in Sec.\,\ref{sec:obstrclass},
we shall employ below a geometric presentation of gerbes
and their 1- and 2-isomorphisms, see, e.g., \cite{GSW1}.
In such a  presentation, a gerbe $\,\CG\,$ over $\,M\,$
with curvature $\,H\,$ is a quadruple $\,(Y,B,L,\mu)\,$ \,where
$\,\pi:Y\rightarrow M\,$ is a surjective submersion, $\,B\,$ is a
2-form on $\,Y$, \,called {\it curving}, \,such that $\,dB=\pi^*H$,
$\,L\,$ is a line bundle over the
fiber-product$\,Y^{[2]}=Y\times_MY\,$ with curvature
$\,F=B_{2}-B_{1}$, \,and $\,\mu:L_{12}\otimes L_{23}\rightarrow
L_{13}\,$ is an isomorphism of line bundles over
$\,Y^{[3]}=Y\times_MY\times_MY\,$ that defines a groupoid structure
on $\,L\rightrightarrows Y\,$ (the subscripts denote here the pullbacks
along projections from $\,Y^{[p]}\rightarrow Y^{[q]}$). \,An explicit
geometric construction of
gerbes $\,\CG_k\,$ over $\,M=\tilde G\,$ with $\,k\in\NZ\,$ was given
in \cite{Meinr} and is somewhat involved. \,We shall use here its
description from \cite{GR1}, \,see also Sec.\,4.1 of \cite{G}.

\subsection{WZW gerbes over compact simply connected simple Lie groups}
\label{sec:simpleconn}

\no As before, coroots, coweights, roots and coroots will be
considered as elements of the imaginary Cartan subalgebra
$\,\si\Nt\subset\si\Ng\,$ identified with its dual with the help of
the bilinear form $\,\tr$. \,The normalization of $\,\tr\,$ makes
the length squared of long roots equal to $\,2$. $\,\alpha_i,\
\alpha_i^\vee,\ \lambda_i\ \lambda_i^\vee $, \,where
$\,i=1,\dots,r$, \,will denote the simple roots, coroots, weights
and coweights, respectively, with $\,r\,$ the rank of $\,\Ng$.
\,The highest root
$\,\phi=\sum_ik_i\alpha_i$, \,where the positive integers $\,k_i\,$
are the Kac labels. Denote by $\,\CA_{_W}\subset i\Nt\,$ the
positive Weyl alcove. $\,\CA_{_W}\,$ is a simplex with vertices
$\,\tau_i=\frac{1}{k_i}\lambda_i^\vee$, $\,i=1,\dots,r$, \,and
$\,\tau_0=0$. \,For $\,i\in R\defn\{0,1,\dots,r\}$, \,let
\qq
\CA_i=\{\,\tau\in\CA_{_W}\ |\ \tau=\sum\limits_{j}s_j\tau_j\ {\rm with}
\ s_i>0\,\}\,,\quad\,
O_i=\{\,g=Ad_{h_g}(\ee^{\,2\pi\si\tau})\,|\,\,h_g\in\tilde G,\
\tau\in\CA_i\,\}\,,
{}
\qqq
and, for $\,I\subset R$, \,let
$\,\CA_I=\mathop{\cap}\limits_{i\in I}\CA_i\,$
and $\,\,O_I=\mathop{\cap}\limits_{i\in I}O_{i}$. \,Subsets $\,O_I\,$ of
$\,\tilde G\,$ are open and $\,Ad$-invariant. \,They are composed of elements
$\,g=Ad_{h_g}(\ee^{\,2\pi\si\tau})\,$ with $\,h_g\in G\,$ and $\,\tau
\in\CA_I$. \,The expressions
\qq
B_i\ =\ \frac{_{k}}{^{4\pi}}\,\tr\,\big(\Theta(h_g)
\,Ad_{\ee^{\,2\pi\si\tau}}(\Theta(h_g))\,+\ 2\pi\si\,(\tau-\tau_i)[\Theta(h_g),
\Theta(h_g)]\big),
\label{Bi}
{}
\qqq
where $\,\Theta(h_g)=h_g^{-1}dh_g$, \,define smooth 2-forms on
$\,O_i\,$ such that $\,dB_i=H_k|_{O_i}$. \,For groups $\,SU(n)$,
\,it is enough to take $\,Y=\mathop{\sqcup}\limits_iO_i$, \,see
\cite{Chatt,GR}. \,In order to have a unique construction of gerbes
$\,\CG_k\,$ for all compact simply connected simple Lie groups, one
makes a more involved choice \cite{Meinr}.

\,Consider the stabilizer subgroups,
\qq
G_I\ =\ \{\,\gamma\in\tilde G\,\,|\,\,\gamma\,\ee^{\,2\pi\si\tau}\,\gamma^{-1}
=\ee^{\,2\pi\si\tau}\ \,{\rm for\ (any)}\ \,
\tau\in\CA_I\setminus\mathop{\cup}\limits_{i\not\in I}\CA_{i}\,\}\,.
{}
\qqq
In particular, $\,G_i\,$ is composed of the elements of $\tilde G\,$
that commute with $\,\ee^{\,2\pi\si\tau_i}$. \,The Cartan subgroup
$\,T\subset\tilde G\,$ is contained in all $\,G_I$. \,The maps
\qq
\CO_I\,\ni\,g\,=\,Ad_{h_g}(\ee^{\,2\pi\si\tau})\ \mathop{\longmapsto}
\limits^{\eta_I}\ \ h_gG_I\,\in\,G/G_I
{}
\qqq
are well-defined because the adjoint-action stabilizers of
$\,\ee^{\,2\pi\si\tau}\,$ for $\,\tau\in\CA_I\,$ are contained in
$\,G_I$. \,They are smooth, \,see Sec.\,5.1 of \cite{Meinr}. \,One
introduces principal $\,G_I$-bundles $\,\pi_I:P_I\to O_I\,$
\qq
P_I\ =\ \{\,(g,h)\in O_I\times\tilde G\,\,|\,\,\eta_I(g)=hG_I\,\}\,.
{}
\qqq
For the gerbes $\,\CG_k=(Y,B,L,\mu)\m$, \,one sets
\qq
Y\ =\ \mathop{\sqcup}\limits_{i\in R}P_i
{}
\qqq
with $\,\pi:Y\to\tilde G\,$ restricting to $\,\pi_i\,$ on $\,P_i\,$ and
the 2-form $\,B\,$ restricting to $\,\pi_i^*B_i$. \,Let
\qq
\hat Y_{i_1..i_n}\ =\ P_I\times G_{i_1}\times\cdots\times G_{i_n}
\ \quad{\rm and}\ \quad
Y_{i_1..i_n}\ =\ \hat Y_{i_1..i_n}/G_I
{}
\qqq
for $\,I=\{i_1,\dots,i_n\}$, \,and for $\,G_I\,$ acting on $\,\hat
Y_{i_1..i_n}$ diagonally by the right multiplication. The fiber
power $\,Y^{[n]}\,$ of $\,Y\,$ may be identified with the disjoint
union of $\,Y_{i_1..i_n}\,$ by assigning to the $\,G_I$-orbit of
$\,((g,h), \gamma_{i_1},..,\gamma_{i_n})\,$ the $\,n$-tuple
$\,(y_1,..,y_n)\in Y^{[n]}\,$ with $\,y_m=(g,h\gamma_{i_m}^{-1})$,
\qq
Y^{[n]}\ \cong\ \mathop{\sqcup}\limits_{(i_1,..,i_n)}Y_{i_1..i_n}\,.
{}
\qqq

The construction of the line bundle $\,L\,$ over $\,Y^{[2]}\,$ uses more
detailed properties of the stabilizer groups $\,G_I$. \,For $\,I\subset
J\subset R\m$, $\,G_J\,$ is contained in $\,G_I$.
\,The smallest of those groups, $\,G_R\m$,
\,coincides with the Cartan subgroup $\,T\,$ of $\,\tilde G$.
\,Groups $\,G_I\,$ are connected but not necessarily simply connected.
Let $\,\Ng_I\supset\Nt\,$ denote
the Lie algebra of $\,G_I\,$ and let $\,\ee_I\,$ be the exponential map
from $\,\Ng_I\,$ to the universal cover $\,\tilde G_I$. \,One has
\qq
G_I\ =\ \tilde G_I/Z_I\ \quad{\rm for}\quad\ Z_I\ =\
\ee_I^{2{\pi\si}Q^{\hspace{-0.04cm}^\vee}}
{}
\qqq
where $\,Q^{\hspace{-0.04cm}^\vee}\subset\Nt\,$ is the coroot
lattice of $\,\Ng$. \,The exponential map $\,\ee_I\,$ maps $\,\Nt\,$
to the Abelian subgroup $\,\tilde T_I\subset\tilde G_I$.
For $\,I\subset J$, \,the group $\,\tilde G_J\,$ maps naturally
into $\,\tilde G_I\,$ and $\,Z_J\,$ into $\,Z_I$.
\,One shows that the formula
\qq
\chi_i(\ee_i^{\,2{\pi\si}\tau})\ =\ \ee^{\,2{\pi\si}\,{\mathrm tr}\,\tau_i\tau}
{}
\qqq
for $\,\tau\in\Nt\,$ defines a character $\,\chi_i:\tilde T_i\rightarrow U(1)$. \,By restriction, $\,\chi_i\,$ determines a character
of $\,Z_i$. \,One may also define a 1-dimensional representation
$\,\chi_{ij}:\tilde G_{ij}\to U(1)\,$ by the formula
\qq
\chi_{ij}(\tilde\gamma_{ij})\ =\ \exp\Big[\frac{_1}{^\si}
\int\limits_{\tilde\gamma_{ij}}
a_{ij}\Big]
{}
\qqq
where $\,a_{ij}=\si\,\tr\,(\tau_j-\tau_i)\,\Theta(\gamma_{ij})\,$ is a closed
1-form on $\,G_{ij}\,$ ($\tilde\gamma_{ij}\,$ is identified with a homotopy
class of a path in $\,G_{ij}\,$ starting at $\,1$). \,For $\,\tau\in\si\Nt\,$
one has:
\qq
\chi_{ij}(\ee_{ij}^{\,2\pi\si\tau})\ =\
\chi_i(\ee_i^{\,2\pi\si\tau})^{-1} \chi_j(\ee_j^{\,2\pi\si\tau})\,.
\label{fut}
\qqq
Besides $\,\chi_{ij}(\tilde\gamma_{ij})=\chi_{ji}(\tilde\gamma_{ij})^{-1}$,
\,and for $\,\tilde\gamma_{ijk}\in\tilde G_{ijk}$,
\qq
\chi_{ij}(\tilde\gamma_{ijk})\,\chi_{jk}(\tilde\gamma_{ijk})\
=\ \chi_{ik}(\tilde\gamma_{ijk})\,.
\label{fut1}
\qqq
Over space $\,\hat Y_{ij}$, \,there is a line bundle $\,\hat
L_{ij}\,$ whose fiber over $\,((g,h),\gamma_i, \gamma_j)\,$ is
composed of the equivalence classes $\,[\tilde\gamma_i,
\tilde\gamma_j,u_{ij}]_{ij}\,$
with respect to the relation
\qq
(\tilde\gamma_i,\tilde\gamma_j,\m u_{ij})
\ \ \mathop{\sim}\limits_{^{ij}}\ \ (\tilde\gamma_i\zeta_i,
\tilde\gamma_j\zeta_j,\m\chi_i(\zeta_i)^{{k}}\chi_j(\zeta_j)^{{-{k}}}\,u_{ij})
{}
\qqq
for $\,\tilde\gamma_i\in\tilde G_i\m$, $\,\tilde\gamma_j\in\tilde G_j\,$
projecting to $\,\gamma_i\in G_i\,$ and $\,\gamma_j\in G_j\m$, \,respectively,
and $\,u_{ij}\in\NC\m$, $\,\zeta_i\in Z_i\m$, $\,\zeta_j\in Z_j$. \,One
twists the natural flat structure of $\,\hat L_{ij}\,$ by the connection form
\qq
\hat A_{ij}\ =\ \si k\,\tr\,(\tau_j-\tau_i)\,\Theta(h)\,.
\label{Aij}
\qqq
The right action of
$\,G_{ij}\,$ on $\,\hat Y_{ij}\,$ lifts to the action on $\,\hat L_{ij}\,$
defined by
\qq
((g,h),\,[\tilde\gamma_i,\tilde\gamma_j,u_{ij}]_{{ij}})\ \ \longmapsto\
\ ((g,h\gamma),\,[\tilde\gamma_i\tilde\gamma_{ij},\tilde\gamma_j
\tilde\gamma_{ij},\m\chi_{ij}(\tilde\gamma_{ij})^{{-{k}}}\,u_{ij}]_{{ij}})
\label{rmap}
\qqq
for $\,\gamma_{ij}\in G_{ij}\,$ and $\,\tilde\gamma_{ij}\,$ its lift
to $\,\tilde G_{ij}$.\,
\,The hermitian structure and the connection of $\,\hat L_{ij}\,$
descend to the quotient bundle $\,\hat L_{ij}/G_{ij} =L_{ij}\,$ over
$\,Y_{ij}\,$ and the line bundle $\,L\,$ over $\,Y^{[2]}\,$ for the
gerbe $\,\CG_k\,$ is taken as equal to $\,L_{ij}\,$ when restricted to
$\,Y_{ij}$. \,The curvature 2-form $\,F_{ij}\,$ of $\,L_{ij}\,$
lifts to $\,\hat Y_{ij}\,$ to the 2-form $\,d\hat A_{ij}\,$
that coincides with the lift to $\,\hat Y_{ij}\,$ of the 2-form
$\,B_j-B_i$. \,This gives the required relation $\,F=B_2-B_1\,$
between the curvature $\,F\,$ of the line bundle $\,L\,$ over
$\,Y^{[2]}\,$ and the curving $\,B\,$ on $\,Y$.

The groupoid multiplication $\,\mu\,$ of $\,\CG\,$ is defined as follows.
Let $\,((g,h),\gamma_i,\gamma_j,\gamma_k)\in
\hat Y_{ijk}\,$ represent $\,(y,y',y'')\in Y^{[3]}\,$ with
$\,y=(g,h\gamma_i^{-1})\m$, $\,y'=(g,h\gamma_j^{-1})\,$
and $\,y''=(g,h\gamma_k^{-1})\,$ and let
\qq
\ell_{ij}\in L_{(y,y')}\m,\qquad\ell_{jk}\in L_{(y',y'')}\m,\qquad
\ell_{ik}\in L_{(y,y'')}
\label{lijjkik}
\qqq
be the elements in the appropriate fibers of $\,L\,$ given by the
$\,G_{ijk}$-orbits of
\qq
\hat\ell_{ij}\ =\ ((g,h),[\tilde\gamma_i,\tilde\gamma_j,u_{ij}]_{ij})\,,\quad\
\hat\ell_{jk}\ =\ ((g,h),[\tilde\gamma_j,\tilde\gamma_k,u_{jk}]_{jk})\,,\quad\
\hat\ell_{ik}\ =\ ((g,h),[\tilde\gamma_i,\tilde\gamma_k,u_{ik}]_{ij})
{}
\qqq
with $\,u_{ik}=u_{ij}u_{jk}$.
\,Then
\qq
\mu(\ell_{ij}\otimes\ell_{jk})=\ell_{ik}\,.
{}
\qqq
This ends the description of gerbes $\,\CG_k=(Y,B,L,\mu)\,$
over simply connected groups $\,\tilde G$.

\subsection{Construction of 1-isomorphism $\,\alpha$}
\label{sec:alpha}

\no We need to compare the pullbacks of gerbe $\,\CG_k\,$ to the
product space $\,\Gamma\times\tilde G$. \,Consider first the
pullback $\,(\CG_k)_{12}\,$ along the adjoint action
$\ \ell:\Gamma\times\tilde G\rightarrow\tilde G\ $ of $\ \Gamma
=\tilde G/\tilde Z\ $ on $\ \tilde G$. \ One has:
\qq
(\CG_k)_{12}\ =\ (Y_{12},B_{12},L_{12},\mu_{12})\,.
{}
\qqq
The adjoint action of $\,\tilde G\,$ on itself may be lifted to $\,Y\,$
by the map
\qq
\tilde G\times Y\,\ni\,(\tilde\gamma,y)\ \longmapsto\ Ad_{\tilde\gamma}(y)
\,\in\,Y\,,
\label{Adt}
\qqq
where for $\,y=(g,h)\in P_i\subset Y$,
$\ Ad_{\tilde\gamma}(y):=(Ad_{\tilde\gamma}(g),\tilde\gamma h)\,\in P_i$.
\ The map (\ref{Adt}) is constant on orbits of the action
\qq
(\tilde\gamma,y)\ \mapsto\ (z\tilde\gamma,yz^{-1})
\label{acttz}
\qqq
of $\,\tilde Z\,$ on $\,\tilde G\times Y$, \,where
$\,yz^{-1}:=(g,hz^{-1})\,$ for $\,y=(g,h)\in P_i$. \,It allows the
canonical identification
\qq
Y_{12}\ \equiv\ (\tilde G\times Y)/\tilde Z\,.
\label{idnt12}
\qqq
In this identification, the surjective submersion $\ \pi_{12}:(Y)_{12}
\rightarrow\Gamma\times\tilde G\ $
is generated by the map $\ (\tilde\gamma,y)\mapsto(\gamma,\pi(y))$,
\ where $\,\gamma\in\Gamma=\tilde G/\tilde Z\,$ is the canonical projection
of $\,\tilde\gamma$. \,Similarly,
\qq
Y_{12}^{[n]}\ \cong\ (\tilde G\times Y^{[n]})/\tilde Z\,.
{}
\qqq
The action of $\,\tilde Z\,$ on $\,\tilde G\times Y^{[2]}\,$ induced
by (\ref{acttz}) may be lifted to the one on $\,\tilde G\times L\,$
given by
\qq
(\tilde\gamma,\ell_{ij})\ \longmapsto\ (z\tilde\gamma,\ell_{ij}\star
z^{-1})\,,
{}
\qqq
where for $\,\ell_{ij}\,$ given by the $\,G_{ij}$-orbit (\ref{rmap})
of $\ \hat\ell_{ij}=\big((g,h),[\tilde\gamma_i,
\tilde\gamma_j,u_{ij}]_{ij}\big)$,
\ the element $\,\ell_{ij}\star z^{-1}\,$ is defined as the
$\,G_{ij}$-orbit of
\qq
\hat\ell_{ij}\star z^{-1}\ :=\ \big((g,h),
[\tilde\gamma_i\tilde z,\tilde\gamma_j
\tilde z,\chi_{ij}(\tilde z)^{-k}u_{ij}]_{ij}\big),
\label{hactLzd}
\qqq
with  $\,\tilde z\,$ standing for any lift of
$\,z\in\tilde Z\,$ to $\,\tilde G_{ij}$. \,We introduce a special
symbol for this action to distinguish it from another one that
will be defined below.
As line bundles,
\qq
L_{12}\ \cong\ (\tilde G\times L)/\tilde Z\,.
{}
\qqq
In order to obtain the correct connection on $\,L_{12}$, \,the one
on $\,\tilde G\times L\,$ has to be modified by twisting the flat
structure on $\,\tilde G\times\hat L_{ij}\,$ by the connection
1-form
\qq
\tilde\gamma^*\hat A_{ij}\ =\ \si k\,\tr\,(\tau_j-\tau_i)
\Theta(\tilde\gamma h)
\label{gAij}
\qqq
rather than by $\,\hat A_{ij}\,$ of Eq.\,(\ref{Aij}).

1-isomorphism $\,\alpha\,$ will compare gerbe $\,(\CG_k)_{12}\,$
to $\ \CI_{\rho_k}\otimes(\CG_k)_2=(Y_2,B_2
+\pi_2^*\rho_k,L_2,\mu_2)$, \ where $\,(\CG_k)_2\,$ is
the pullback of $\,\CG_k\,$ to $\,\Gamma\times\tilde G\,$ along the
projection to the second factor.
\,It will be convenient to identify
\qq
Y_2\ =\ \Gamma\times Y\ \cong\ (\tilde G\times Y)/\tilde Z\,,
\label{idnt2}
\qqq
where now $\,\tilde Z\,$ acts only on $\,\tilde G$. \,The projection
$\ \pi_2:Y_2\rightarrow\Gamma\times\tilde G\ $ is induced upon this
identification by the map $\ (\tilde\gamma,y)\mapsto(\gamma,\pi(y))$.
\ Similarly,
\qq
Y^{[n]}_2\,=\,\Gamma\times Y^{[n]}\,\cong\,(\tilde G\times Y^{[n]})/\tilde Z\,,
\qquad L_2\,=\,\Gamma\times L\,\cong\,(\tilde G\times L)/\tilde Z\,,
{}
\qqq
with $\,\tilde Z\,$ always acting trivially on the 2$^{\rm nd}$ factor.

The first part of data for 1-isomorphism $\,\alpha\,$ is a line
bundle $\,E\,$ over $\ W_{12}:=Y_{12}\times_{(\Gamma\times\tilde
G)}Y_2$, \,see \cite{GSW}. \ $E\,$ has to be equipped with a
connection whose curvature form $\,F^E\,$ is equal to
$\,(B_2+\pi_2^*\rho_k)_2-(B_{12})_1$, \,where the outside subscript
1 (resp. 2) refers to the pullback along the projection from
$\,W_{12}\,$ to $\,Y_{12}$ (resp. to $\,Y_2$). \,In view of
identifications (\ref{idnt12}) and (\ref{idnt2}), we obtain for the
fiber-product space $\,W_{12}\,$
\qq
W_{12}\ \cong\ (\tilde G\times Y^{[2]})/\tilde Z
\label{fpW}
\qqq
for the action $\ (\tilde\gamma,(y,y'))\mapsto
(z\tilde\gamma,(yz^{-1},y'))\ $ of $\,\tilde Z$. \,The projection to
$\,Y_{12}\,$ is induced by the map $\ (\tilde\gamma,(y,y'))
\mapsto(\tilde\gamma,y)$, \ the one to
$\,Y_2\,$ by $\ (\tilde\gamma,(y,y'))\mapsto(\tilde\gamma,y')$. \,The
composed projection $\ \varpi:W_{12}\rightarrow\Gamma\times\tilde G\ $
is $\ (\tilde\gamma,(y,y'))\mapsto(\gamma,\pi(y)=\pi(y'))$.
\ Line bundle $\,E\,$ over $\,W_{12}\,$ will be defined by
\qq
E\ :=\ (\tilde G\times L)/\tilde Z\,,
\label{defE}
\qqq
for the action of $\,\tilde Z\,$
\qq
(\tilde\gamma,\ell_{ij})\ \longmapsto\ (z\tilde\gamma,\ell_{ij}\cdot z^{-1})\,,
\label{actLz}
\qqq
where the element $\,\ell_{ij}\cdot z^{-1}\,$ defined as the
$\,G_{ij}$-orbit of
\qq
\hat\ell_{ij}\cdot z^{-1}\ :=\
\big((g,h),[\tilde\gamma_i\tilde z,\tilde\gamma_j,
\chi_{i}(\tilde z)^{k}\chi(z)\,u_{ij}]_{ij}\big)\,,
\label{hactLz}
\qqq
with $\,\chi:\tilde Z\rightarrow U(1)\,$ a fixed character.
\,Note the difference between elements $\,\hat\ell_{ij}\cdot z^{-1}\,$ and
$\,\hat\ell_{ij}\star z^{-1}$, \,with the latter one defined by
Eq.\,(\ref{hactLzd}).

The connection in line bundle $\,E\,$ requires a careful definition in
order to assure that it has the desired curvature. \,Note that the 2-form
$\,(B_2+\pi_2^*\rho_k)_2-(B_{12})_1\,$ on $\ (\tilde G\times
Y_{ij})/\tilde Z\subset W_{12}\ $ is equal to the pullback by
$\,\varpi\,$ of the 2-form $\ (B_j)_2+\rho_k-(B_i)_{12}\ $ on $\
\Gamma\times O_{ij}\subset\Gamma\times\tilde G$. \ A short
calculation shows that for $\,\gamma\in\Gamma\,$ and
$\,g=Ad_{h_g}(\ee^{\,2\pi\si\tau})\in O_{ij}$,
\qq
\big((B_j)_2+\rho_k-(B_i)_{12}\big)(\gamma,g)\ =\
d\,\big(\si k\,\tr\,Ad_{h_g}(\tau-\tau_i)
\,\Theta(\gamma)\big)
\,+\,\frac{_1}{^2}\si k\,\tr\,(\tau_i-\tau_j)\,[\Theta(h_g),\Theta(h_g)]\,.
{}
\qqq
It was shown in \cite{Meinr} that the map
\qq
O_i\,\ni\,g=Ad_{h_g}(\ee^{\,2\pi\si\tau})\ \longmapsto\
Ad_{h_g}(\tau-\tau_i)\,\in\,\si\Ng\,,
{}
\qqq
denoted $\,\Psi_i\,$ there, \,is well defined and smooth so that the
1-form $\ A_i=\si k\, \tr\,Ad_{h_g}(\tau-\tau_i)\,\Theta(\gamma)\ $
is well defined and smooth on $\,\Gamma\times O_i$. \ On the other
hand, the 2-form $\,B_j-B_i=\frac{1}{2}\si
k\,\tr\,(\tau_i-\tau_j)\,[\Theta(h_g), \Theta(h_g)]\,$ is a well
defined closed 2-form on $\,O_{ij}\,$ which, when pulled back to
$\,Y_{ij}\subset Y^{[2]}$, \,coincides with the curvature form of
$\,L|_{Y_{ij}}=L_{ij}$. \,In order to assure the correct
curvature of $\,E$, \,we shall additionally twist the connection of
$\,\tilde G\times L|_{Y_{ij}}\,$ in (\ref{defE}) by the pullbacks to
$\,\tilde G\times Y_{ij}\,$ of the forms
\qq
\hat A_i\ =\ \si k\,\tr\,Ad_{h_g}(\tau-\tau_i)\,\Theta(\tilde\gamma)
{}
\qqq
on $\,\tilde G\times O_{ij}$. \,A straightforward check shows that
the resulting connection in $\,\tilde G\times L\,$ descends to the
quotient by the action (\ref{actLz}) of $\,\tilde Z$. \ Note that
the resulting bundles $\,E\,$ differ for different characters
$\,\chi\,$ of $\,\tilde Z\,$ by tensor factors that are pullbacks to
$\,W_{12}\,$ of flat bundles over $\,\Gamma$.

1-isomorphism $\ \alpha:(\CG_k)_{12}\rightarrow\CI_{\rho_k}
\otimes(\CG_k)_2\ $ of \,Definition \ref{def:equivgerbe} is
an isomorphism of line bundles over $\,W_{12}^{[2]}$
\qq
\alpha:\,L_{12}\otimes E_2\ \longrightarrow\ E_1\otimes L_{2}\,,
\label{biso}
\qqq
where natural pullbacks of bundles $\,L_{12}\,$ and $\,L_2\,$ are
understood. \,Recalling realization
(\ref{fpW}) of $\,W_{12}$, \,we have
\qq
W_{12}^{[2]}\ \cong\ (\tilde G\times Y^{[4]})/\tilde Z
\label{W[2]}
\qqq
with the action
\qq
(\tilde\gamma,y_1,y_1',y_2,y_2')\ \longmapsto\
(z\tilde\gamma,(y_1z^{-1},y_1',y_2z^{-1},y_2'))
{}
\qqq
of $\,\tilde Z$. \ Suppose that $\,(y_1,y_1',y_2,y_2')\in Y_{i_1j_1i_2j_2}\,$
and that
\qq
\ell_{i_1i_2}\in L_{(y_1,y_2)}\,,\qquad\ell_{i_2j_2}\in L_{(y_2,y_2')}\,,
\qquad\ell_{i_1j_1}\in L_{(y_1,y_1')}\,,\qquad\ell_{j_1j_2}\in L_{(y_1',y_2')}
{}
\qqq
are given by $\,G_{i_1j_1i_2j_2}$-orbits of
\qq
&&\hat\ell_{i_1i_2}\,=\,\big((g,h),
[\tilde\gamma_{i_1},\tilde\gamma_{i_2},u_{i_1i_2}]_{i_1i_2}\big)\,,
\qquad\hat\ell_{i_2j_2}\,=\,\big((g,h),
[\tilde\gamma_{i_2},\tilde\gamma_{j_2},u_{i_2j_2}]_{i_2j_2}\big)\,,\\
&&\hat\ell_{i_1j_1}\,=\,\big((g,h),
[\tilde\gamma_{i_1},\tilde\gamma_{j_1},u_{i_1j_1}]_{i_1j_1}\big)\,,
\qquad\hat\ell_{j_1j_2}\,=\,\big((g,h),
[\tilde\gamma_{j_1},\tilde\gamma_{j_2},u_{j_1j_2}]_{j_1j_2}\big)\,.
{}
\qqq
with $\ \mu(\ell_{i_1i_2}\otimes\ell_{i_2j_2})=\mu(\ell_{i_1j_1}
\otimes\ell_{j_1j_2})$, \ i.e.
\qq
u_{i_1i_2}\,u_{i_2j_2}\ =\ u_{i_1j_1}\,u_{j_1j_2}\,.
\label{uu}
\qqq
The bundle isomorphism $\,\alpha\,$ of (\ref{biso}) will be generated
by a map $\,\tilde\alpha\,$ such that
\qq
\hat\alpha\big(\tilde\gamma,\ell_{i_1i_2}\otimes\ell_{i_2j_2}\big)\ =\
\big(\tilde\gamma,\ell_{i_1j_1}\otimes\ell_{j_1j_2}\big).
{}
\qqq
Consistency requires that $\,\tilde\alpha\,$ commutes with the
action of $\,\tilde Z$, \,i.e. that
\qq
\tilde\alpha\big(z\tilde\gamma,\ell_{i_1i_2}\star z^{-1}
\otimes\ell_{i_2j_2}\cdot z^{-1}\big)\ =\
\big(\tilde\gamma,\ell_{i_1j_1}\cdot z^{-1}\otimes\ell_{j_1j_2}\big).
{}
\qqq
In view of \,Eqs.\,(\ref{hactLzd}), (\ref{hactLz}) and (\ref{uu}), \,this
is guaranteed by the relation
\qq
\chi_{i_1i_2}(\tilde z)^{-k}\chi_{i_2}(\tilde z)^{k}\chi(z)
\ =\
\chi_{i_1}(\tilde z)^{k}\chi(z)
{}
\qqq
which follows from identity (\ref{fut}). \,That the bundle isomorphism
$\,\alpha\,$ preserves
the connections follows from the equality of the (modified)
connection forms
\qq
\tilde\gamma^*\hat A_{i_1i_2}\,+\,\hat A_{i_2j_2}\,+\,\hat A_{i_2}\ =\
\hat A_{i_1j_1}\,+\,\hat A_{i_1}\,+\,\hat A_{j_1j_2}
{}
\qqq
which is easy to check.

For the bundle isomorphism $\,\alpha\,$ to define a gerbe
1-isomorphism from $\,(\CG_k)_{12}\,$ to
$\,\CI_{\rho_k}\otimes(\CG_k)_2$, \,one has to require a proper behavior
with respect to the groupoid multiplication
\cite{GSW}. More precisely, what is needed is the coincidence of two composed
isomorphisms of line bundles over $\,W_{12}^{[3]}$. \,The first one is
\qq
&\displaystyle{(L_{12})_{1,2}\otimes(L_{12})_{2,3}\otimes E_3\
\mathop{\longrightarrow}\limits^{\Id\otimes\alpha_{2,3}}\ (L_{12})_{1,2}
\otimes E_2\otimes(L_2)_{2,3}\ \mathop{\longrightarrow}\limits^{\alpha_{1,2}
\otimes\Id}
\ E_1\otimes(L_2)_{1,2}\otimes(L_2)_{2,3}}&\cr\cr
&\displaystyle{\mathop{\longrightarrow}\limits^{\Id\otimes(\mu_2)_{1,2,3}}\
E_1\otimes(L_2)_{1,3}\,,}&
\label{cmps1}
\qqq
with the exterior subscripts referring to the components of $\ W_{12}^{[3]}$.
\ The second one is
\qq
(L_{12})_{1,2}\otimes(L_{12})_{2,3}\times E_3\ \mathop{\longrightarrow}
\limits^{(\mu_{12})_{1,2,3}\otimes\Id}\ (L_{12})_{1,3}\otimes E_3\
\mathop{\longrightarrow}\limits^{\alpha_{1,3}}\ E_1\otimes(L_2)_{1,3}\,.
\label{cmps2}
\qqq
Straightforward verification that they coincide is carried out in
Appendix \ref{app:7}.

\subsection{Construction of 2-isomorphism $\,\beta$}
\label{sec:beta}

\no2-isomorphism $\,\beta\,$ of \,Definition
\ref{def:equivgerbe} compares 1-isomorphisms of gerbes over
$\,\Gamma^2\times\tilde G$. \ First, consider gerbe
$\,(\CG_k)_{123}=(Y_{123},B_{123},L_{123},\mu_{123})$. \ The same
way as before for $\,Y_{12}$, \ we shall use the map
\qq
\tilde G^2\times Y\,\ni\,(\tilde\gamma_1,\tilde\gamma_2,y)\
\longmapsto\ Ad_{\tilde\gamma_1\tilde\gamma_2}y
\,\in\,Y
\qqq
constant on orbits of the $\,\tilde Z^2$-action
\qq
(\tilde\gamma_1,\tilde\gamma_2,y)\ \longmapsto\
(z_1\tilde\gamma_1,z_2\tilde\gamma_2,y(z_1z_2)^{-1})
{}
\qqq
in order to identify
\qq
Y^{[n]}_{123}\ \cong\ (\tilde G^2\times Y^{[n]})/\tilde Z^2\,.
{}
\qqq
As line bundles,
\qq
L_{123}\ \cong\ (\tilde G^2\times L)/\tilde Z^2
{}
\qqq
for the action
\qq
(\tilde\gamma_1,\tilde\gamma_2,\ell_{ij})\
\longmapsto\ (z_1\tilde\gamma_1,z_2\tilde\gamma_2,\ell_{ij}
\star(z_1z_2)^{-1})\,.
{}
\qqq
The connection in $L\,$ has to be modified by twisting the flat
structure of $\,\tilde G^2\times\hat L_{ij}\,$ by the connection
1-form $\,(\tilde\gamma_1\tilde\gamma_2)^*\hat A_{ij}$, \,see
Eq.\,(\ref{gAij}). \ Similarly, for gerbe $\
(\CG_k)_{23}=(Y_{23},B_{23},L_{23},\mu_{23})\ $ over
$\,\Gamma^2\times\tilde G$, \ we have:
\qq
Y_{23}^{[n]}\ \cong\ (\tilde G^2\times Y^{[n]})/\tilde Z^2\,,
{}
\qqq
where now the action of $\,\tilde Z^2\,$ is induced from the
one on $\ \tilde G^2\times\tilde Y\ $ given by
\qq
(\tilde\gamma_1,\tilde\gamma_2,y)\ \longmapsto\
(z_1\tilde\gamma_1,z_2\tilde\gamma_2,yz_2^{-1})\,.
{}
\qqq
As line bundles,
\qq
L_{23}\ \cong\ (\tilde G^2\times L)/\tilde Z^2
{}
\qqq
for the action
\qq
(\tilde\gamma_1,\tilde\gamma_2,\ell_{ij})\ \longmapsto\
(z_1\tilde\gamma_1,z_2\tilde\gamma_2,\ell_{ij}\star z_2^{-1})\,,
{}
\qqq
with the connection of $\,\tilde G^2\times L\,$ modified now using
1-forms $\,\tilde\gamma_2^*\hat A_{ij}$.
\ Finally, for gerbe $\,(\CG_k)_3=(Y_3,B_3,L_3,\mu_3)$,
\qq
Y^{[n]}_3\ \cong\ (\tilde G^2\times Y^{[n]})/\tilde Z^2\qquad{\rm and}\qquad
L_3\ \cong\ (\tilde G^2\times L)/\tilde Z^2\,,
{}
\qqq
with $\,\tilde Z^2\,$ acting only on the factors $\,\tilde G^2$.

For the fiber-product space $\ W_{123}=Y_{123}
\times_{(\Gamma^2\times\tilde G)}Y_{23}
\times_{(\Gamma^2\times\tilde G)}Y_3$, \ we have
\qq
W_{123}\ =\ (\tilde G^2\times Y^{[3]})/\tilde Z^2
{}
\qqq
for the action
\qq
(\tilde\gamma_1,\tilde\gamma_2,(y,y',y''))\ \longmapsto\
(z_1\tilde\gamma_1,z_2\tilde\gamma_2,(y(z_1z_2)^{-1},y'z_2^{-1},y''))
\label{act123}
\qqq
of $\,\tilde Z^2$. \ We may pull back the line bundle $\,E\,$ over
$\,W_{12}\,$ in three different ways to $\,W_{123}$, \,obtaining the
respective line bundles
$\,E_{1,23}$, $\,E_{2,3}\,$ and $\,E_{12,3}$. \ One has
\qq
E_{1,23}\ \cong\ (\tilde G^2\times L_{1,2})/\tilde Z^2\,,\qquad
E_{2,3}\ \cong\ (\tilde G^2\times L_{2,3})/\tilde Z^2\,,\qquad
E_{12,3}\ \cong\ (\tilde G^2\times L_{1,3})/\tilde Z^2\,.
\label{3Es}
\qqq
The actions of $\,\tilde Z^2\,$ above are defined as follows. \ If
$\ (y,y',y'')\in Y_{ijk}\subset Y^{[3]}\ $ and $\,\ell_{ij}$,
$\,\ell_{jk}\,$ and $\,\ell_{ik}\,$ are as in (\ref{lijjkik}),
\ i.e. $\ \ell_{ij}\in L_{(y,y')}\subset L_{1,2}\hspace{0.02cm}$,
$\ \ell_{jk}\in
L_{(y',y'')}\subset L_{2,3}\ $ and $\ \ell_{ik}\in L_{(y,y'')}\subset
L_{1,3}\hspace{0.02cm}$, \ then under $\,(z_1,z_2)\in\tilde Z^2$,
\qq
&&(\tilde\gamma_1,\tilde\gamma_2,\ell_{ij})\ \longmapsto\
(z_1\tilde\gamma_1,z_2\tilde\gamma_2,(\ell_{ij}\star z_2^{-1})\cdot
z_1^{-1})\,,\\
&&(\tilde\gamma_1,\tilde\gamma_2,\ell_{jk})\ \longmapsto\
(z_1\tilde\gamma_1,z_2\tilde\gamma_2,\ell_{jk}\cdot z_2^{-1})\,,\\
&&(\tilde\gamma_1,\tilde\gamma_2,\ell_{ik})\ \longmapsto\
(z_1\tilde\gamma_1,z_2\tilde\gamma_2,\ell_{ik}\cdot(z_1z_2)^{-1})\,.
{}
\qqq
The connection of $\,L\,$ in the three pullbacks in (\ref{3Es}) has
to be modified by twisting the flat structure of $\ \tilde G^2
\times\hat L_{ij}\ $ by the 1-form
\qq
\hat A^{1,23}_{ij}\ =\ \si k\,\tr\,(\tau_j-\tau_i)\,\Theta(\tilde\gamma_2h_g)
\,+\,\si k\,\tr\,Ad_{\tilde\gamma_2h_g}(\tau-\tau_i)\,\Theta(\tilde\gamma_1)\,,
{}
\qqq
that of $\ \tilde G^2\times\hat L_{jk}\ $ by
\qq
\hat A^{2,3}_{jk}\ =\ \si k\,\tr\,(\tau_k-\tau_j)\,\Theta(h_g)
\,+\,\si k\,\tr\,Ad_{h_g}(\tau-\tau_j)\,\Theta(\tilde\gamma_2)\,,
{}
\qqq
and that of $\ \tilde G^2\times\hat L_{ik}\ $ by
\qq
\hat A^{12,3}_{ik}\ =\ \si k\,\tr\,(\tau_k-\tau_i)\,\Theta(h_g)
\,+\,\si k\,\tr\,Ad_{h_g}(\tau-\tau_i)\,
\Theta(\tilde\gamma_1\tilde\gamma_2)\,.
{}
\qqq

There is a natural isomorphism $\ \beta:E_{1,23}\otimes
E_{2,3}\rightarrow E_{12,3}\,$ given by the groupoid multiplication
$\,\mu\,$ in $\,L$, \,i.e. induced by the map
\qq
(\tilde\gamma_1,\tilde\gamma_2,\ell_{ij}\otimes\ell_{jk})\
\mathop{\longmapsto}\limits^{\tilde\beta}\
(\tilde\gamma_1,\tilde\gamma_2,\mu(\ell_{ij}\otimes\ell_{jk}))\,.
{}
\qqq
\,Indeed, $\,\tilde\beta\,$ commutes with the action of $\,\tilde
Z^2\,$ because $\ \mu\big((\ell_{ij}\star z_2^{-1})\cdot
z_1^{-1}\otimes \ell_{jk}\cdot
z_2^{-1}\big)=\ell_{ik}\cdot(z_1z_2)^{-1}\ $ if
$\ \mu\big(\ell_{ij}\otimes\ell_{jk}\big)=\ell_{ik}\ $ as
\qq
\chi_i(\tilde z_1)^{k}\,\chi(z_1)\,\chi_{ij}(\tilde z_2)^{-k}
\chi_j(\tilde z_2)^k\,\chi(z_2)\ =\ \chi_i(\tilde z_1\tilde z_2)^k\,
\chi(z_1z_2)\,.
{}
\qqq
Besides, $\,\tilde\beta\,$ intertwines the modified connections since
\qq
\hat A^{1,23}_{ij}\,+\,\hat A^{2,3}_{jk}\ =\
\hat A^{12,3}_{ik}\,,
{}
\qqq
as a short calculation shows.

\ For the line bundle isomorphism $\,\beta\,$ to provide a gerbe 2-isomorphism
required by Definition \ref{def:equivgerbe}, one needs (see \cite{GSW})
that over
\qq
W_{123}^{[2]}\ \cong\ (\tilde G^2\times Y^{[6]})/\tilde Z^2\,,
\qqq
with the action of $\,\tilde Z^2\,$ induced from that in (\ref{act123}),
\,the diagram of line bundle isomorphisms
\qq
\xymatrix{L_{123}\otimes(E_{1,23})_2\otimes(E_{2,3})_2 \ar@{->}
[d]_{\Id\otimes\beta_2}\hspace*{-0.9cm} &
\ar@{->}[r]^-{\alpha_{1,23}\otimes\Id}&
\ (E_{1,23})_1\otimes L_{23}\otimes(E_{2,3})_2\
\ar@{->}[r]^-{Id\otimes\alpha_{2,3}}
 &
\ (E_{1,23})_1\otimes
 (E_{2,3})_1\otimes L_3\qquad \ar@{->}[d]^{\beta_1\otimes\Id} \\
 L_{123}\otimes(E_{12,3})_2 &
\hspace*{-1cm}\ar@{->}[rr]_-{\alpha_{12,3}}&&
\hspace*{-2.2cm}{\hspace*{3.5cm}(E_{12,3})_1\otimes L_3}\qquad\qquad}
\label{betacom}
\qqq
with the exterior subscripts referring to the pullbacks to
$\,W_{123}^{[2]}\,$ and with the obvious pullbacks omitted, \,be
commutative. This is checked in Appendix \ref{app:8}.

\subsection{Commutativity of diagram \,(\ref{axiom})}
\label{sec:commut}

\no This is the identity
\qq
\beta_{1,23,4}\bullet((Id\otimes\beta_{2,3,4})\circ\Id)\ =\ \beta_{12,3,4}
\bullet(Id\circ\beta_{1,2,34)})
\label{dicomm}
\qqq
for composed 2-isomorphisms between 1-isomorphisms of gerbes over
$\,\Gamma^2\times\tilde G\,$ (see \cite{W} for the abstract
definition of the {\it vertical} $\,\bullet\,$ and {\it horizontal}
$\,\circ\,$ compositions of 2-morphisms). \,The left- and the
right-hand side are the following compositions of the isomorphisms
of line bundles:
\qq
E_{1,234}\otimes E_{2,34}\otimes E_{3,4}\,\mathop{\longrightarrow}
\limits^{\Id\otimes\beta_{2,3,4}}\,E_{1,234}\otimes E_{23,4}
\,\mathop{\longrightarrow}
\limits^{\beta_{1,23,4}}\,E_{123,4}\,,
\label{comp1}
\qqq
\vskip -0.3cm
\no and
\vskip -0.7cm
\qq
E_{1,234}\otimes E_{2,34}\otimes E_{3,4}\,\mathop{\longrightarrow}
\limits^{\beta_{1,2,34}\otimes\Id}\,E_{12,34}\otimes E_{3,4}
\,\mathop{\longrightarrow}
\limits^{\beta_{12,3,4}}\,E_{123,4}\,,
\label{comp2}
\qqq
respectively, \ over the fiber-product space $\ W_{1234}=
(Y)_{1234}\times_{(\Gamma^3\times\tilde
G)}(Y)_{234}\times_{(\Gamma^3 \times\tilde G)}
(Y)_{34}\times_{(\Gamma^3\times\tilde G)}(Y)_{4}$. \ It is checked
in Appendix \ref{app:9} that they coincide. \ This proves identity
(\ref{dicomm}) establishing the commutativity of diagram
(\ref{axiom}) of \,Definition \ref{def:equivgerbe} and \,completing
the construction of $\,\Gamma$-equivariant structures on
gerbe $\,\CG_k\,$ over $\,\tilde G\,$ for the adjoint action of
$\,\Gamma=\tilde G/\tilde Z\,$ on $\tilde G$.
\vskip 0.2cm

\begin{theorem}
The $\,\Gamma$-equivariant structures on the WZW gerbe $\,\CG_k\,$
over $\,\tilde G\,$ constructed above are non-isomorphic for
different characters  $\ \chi:\tilde Z\rightarrow U(1)\ $
and each $\,\Gamma$-equivariant structure on $\,\CG_k\,$ is isomorphic
to one of them.
\end{theorem}

\begin{proof}
The general discussion of classification of $\,\Gamma$-equivariant
structures in Sec.\,\ref{sec:classif} showed
that different isomorphism classes of $\,\Gamma$-equivariant
structures correspond in this case to cohomology classes
$\ [b']\in H^1(\Gamma\times M,U(1))\cong H^1(\Gamma,U(1))\ $ in the
image of homomorphism $\,\kappa$, \,see Corollary
\ref{cor:class}. \,The classes $\,[b']\,$ are the isomorphism
classes of flat line bundles over $\,\Gamma\,$ by which differ the
line bundles $\,E\,$ over $\,W_{12}\,$ involved in the above
construction of 1-isomorphisms $\,\alpha\,$ of \,Definition
\ref{def:equivgerbe}. \,Different choices of characters
$\ \chi:\tilde Z\rightarrow U(1)\ $ correspond to tensoring $\,E\,$
with such flat line bundles, as was remarked in
Sec.\,\ref{sec:alpha}. \,The claim of the theorem now follows from
the isomorphism of $\,H^1(\Gamma,U(1))\,$ with the character group
$\,{\tilde Z}^{^*}$.
\end{proof}

\section{Conclusions}
\label{sec:concl}

\no We revisited the problem of the gauging of rigid symmetries in
two-dimensional sigma models with the Wess-Zumino action
related to a closed 3-forms $\,H\,$ on the target manifold.
For topologically trivial gauge fields given by global Lie-algebra
valued 1-forms on the worldsheet, the gauging prescription of
refs.\,\cite{JJMO} and \cite{HS} assures infinitesimal gauge
invariance. We showed, however, that it may lead to global gauge
anomalies. We classified such anomalies using geometric tools based
on the theory of bundle gerbes. Global gauge anomalies occur,
for example, in numerous WZW sigma models with non-simply connected
target groups when one gauges their adjoint symmetries. They lead
to the inconsistency of the corresponding coset models obtained by
integrating out the external gauge fields in the respective gauged
WZW models. We introduced geometric structures called
{\it equivariant gerbes}, living on the
target space, that permit an anomaly-free coupling of WZ amplitudes
to arbitrary (also topologically non-trivial) gauge fields. A
detailed analysis of obstructions to the existence of such
structures was performed and their classification was obtained.
In particular, we showed that the gerbes relevant to the WZW theories
with compact semi-simple target groups can be equipped with
equivariant structures with respect to adjoint symmetries if and
only if there is no global gauge anomaly in the coupling of the WZW
model to topologically trivial gauge fields. We explicitly
constructed all nonequivalent equivariant structures in the case of
simply connected target groups. Different equivariant structures result in
the coupling to topologically non-trivial gauge fields that differs by phases.
We showed that such ambiguities are given by characters of
the fundamental group of the symmetry group, if the latter
is connected. We do not know if, in general, the existence of equivariant
gerbes is also a necessary condition for the existence of non-anomalous
coupling of WZ amplitudes to topologically non-trivial sector, but
this is a plausible conjecture. The analysis of the present paper
was limited to the case of oriented closed worldsheets. Local gauge
anomalies on worldsheets with boundary were studied in \cite{FM}. A
generalization of the present work to the case of such worldsheets,
or worldsheets with conformal defects, will be discussed in a separate
publication. An extension of WZ amplitudes to unoriented surfaces
requires an additional structure on gerbes that was introduced
under the name of Jandl structure in \cite{SSW}, see also
\cite{GSW,GSW1}. We plan to discuss the interrelation
between equivariant structures, Jandl structures, and multiplicative
structures on gerbes of \cite{CJMSW,W1,GW}, in a future study,
with applications to orientifolds of coset models.
Other possible extensions of our work should cover the cases
of WZW and coset theories with non-compact targets,
supersymmetric sigma models, and applications to global aspects of
$T$-duality \cite{Hull}. It should also be possible to study global
gauge anomalies for higher dimensional WZ actions on spacetimes with
arbitrary topology using the theory of bundle $n$-gerbes \cite{CMW}.

\nsection{Appendices}
\begin{appendix}

\subsection{Proof of Proposition \ref{prop:JJMO-HS}}
\label{app:1}
\setcounter{equation}{0}
\def\themythm{A.1.\arabic{mythm}}
\def\theequation{A.1.\arabic{equation}}

\no We have to find conditions under which the coupled amplitudes
$\,{\bm A}(\varphi,A)\,$ given by Eq.\,(\ref{FAg}) are invariant
under infinitesimal gauge transformations. \,Setting
$\,\ee^{-t\Lambda}\phi=(Id,\ee^{-t\Lambda}\varphi)\,$ and
denoting by $\,\bar\Lambda\,$ the vector field on $\,\Sigma\times
M\,$ in the direction of $\,M\,$ given by $\,\bar\Lambda(x,m)
=\frac{d}{dt}|_{_{t=0}}(x,\ee^{-t\Lambda}m)$, \,we observe that
\qq
&\displaystyle{\smallddt\,\,\int\limits_\Sigma(\ee^{-t\Lambda}\phi)^*\big(
-v(A)+\frac{_1}{^2}u(A^2)\big)\ =\ \int\limits_\Sigma
\phi^*\CL_{\bar\Lambda}\big(-v(A)+\frac{_1}{^2}u(A^2)\big)}&\cr\cr
&\displaystyle{=\ \int\limits_\Sigma\phi^*\iota_{\bar\Lambda}d\big(-v(A)
+\frac{_1}{^2}u(A^2)\big)}&
\qqq
since the other term $\,d\iota_{\bar\Lambda}\,$ in the Lie
derivative gives a term that vanishes by integration by parts.
\,Similarly, as $\ \smallddt\,\ee^{-t\Lambda}A=d\Lambda-[\Lambda,A]$,
\ see Eq.\,(\ref{ggtr}), \,one obtains
\qq
\smallddt\,\,\int\limits_\Sigma\phi^*\big(
-v(\ee^{-t\Lambda}A)+\frac{_1}{^2}u((\ee^{-t\Lambda}A)^2)
\big)\ =\ \int\limits_\Sigma\phi^*\big(-v(d\Lambda-[\Lambda,A])+
u((d\Lambda-[\Lambda,A])A)\big)\,.\quad
\qqq
On the other hand, $\ {\bm A}_\WZ(\ee^{-t\Lambda}\varphi)=
\Hol_{\CG_2}(\ee^{-t\Lambda}\phi)$, where the subscript $2$ on
$\,\CG\,$ refers to the pullback along the projection from
$\,\Sigma\times M\,$ to $\,M\,$ (the latter relation follows from
the behavior of gerbe holonomy under gerbe pullbacks). \,Proceeding
as in the proof of Proposition \ref{prop:holonomyvariation} \,one
then shows that
\qq
\smallddt\,\,{\bm A}_\WZ(\ee^{-t\Lambda}\varphi)\ =\
\Big(\si\int\limits_\Sigma\phi^*\big(\iota_{\bar\Lambda}H\big)\Big)\
{\bm A}_\WZ(\varphi)\,,
{}
\qqq
so that $\,\iota_{\bar\Lambda}H\,$ (more pedantically
defined as $\,\iota_{\bar\Lambda}H_2\,$) \,is a form on
$\,\Sigma\times M$. \ Gathering the above relations, we infer that
\qq
&&\smallddt\,\,{\bm A}_\WZ
(\ee^{-t\Lambda}\varphi,\ee^{-t\Lambda}A)
\ =\ \Big(\si\int\limits_\Sigma\phi^*\big[\iota_{\bar\Lambda}\big(H
+d(-v(A)+\frac{_1}{^2}u(A^2))\big)\cr\cr
&&\hspace{5,5cm}-\,v(d\Lambda-[\Lambda,A])
+u\big((d\Lambda-[\Lambda,A])A\big)\big]\Big)\
{\bm A}_\WZ(\varphi,A)\,.\quad
{}
\qqq
Consequently, the invariance of the amplitudes $\,{\bm A}_\WZ(\varphi,A)\,$
under infinitesimal gauge transformations requires that
for all $\,\varphi\,$ and $\,A$,
\qq
\int\limits_\Sigma\phi^*\Big[\iota_{\bar\Lambda}\big(H
+d(-v(A)+\frac{_1}{^2}u(A^2))\big) -v(d\Lambda-[\Lambda,A])
+u((d\Lambda-[\Lambda,A])A)\Big]\ =\ 0\,.\
\label{locinv}
\qqq
In order to proceed, it will be easier to employ a basis $\,(t^a)\,$ in
$\,\Ng$, \,writing $\,A=t^aA^a$, $\,\Lambda=t^a\Lambda^a\,$
and using the notations of Remark \ref{rem:3.2}.2.
\,Eq.\,(\ref{locinv}) may then be rewritten as
\qq
&\displaystyle{\int\limits_{\Sigma}\phi^*\Big[\Lambda^a\Big(\iota^a
\big(H+d(-v^bA^b+\frac{_1}{^2}u^{bc}A^bA^c)\big)+f^{abc}(v^cA^b
-u^{cd}A^bA^d\big)\Big)+(d\Lambda^a)(v^a+u^{ab}A^b)\Big]}&\cr
&\displaystyle{=\int\limits_{\Sigma}\phi^*\Big[\Lambda^a\Big(\iota^aH
-\iota^a(dv^b)A^b+\iota^av^bdA^b+\frac{_1}{^2}\iota^a(du^{bc})
A^bA^c+f^{abc}v^cA^b-f^{abc}u^{cd}A^bA^d}&\cr
&&\hspace{-6.3cm}-\,dv^a-(du^{ab})A^b-u^{ab}dA^b\Big)\Big]\ =\ 0\,,
\qqq
where the terms in the last line were obtained by integration by
parts. \,Since $\,\Lambda^a\,$ are arbitrary functions on
$\,\Sigma$, \,we infer that the 2-form
\qq
\varphi^*\big(\iota^aH-dv^a\big)\,+\,\varphi^*\big(-\iota^a(dv^b)
+f^{abc}v^c-du^{ab}\big)A^b\,
+\,\varphi^*\big(\iota^av^b-u^{ab}\big)dA^b\hspace{3cm}\cr
\,+\,\frac{_1}{^2}\,\varphi^*\big(\iota^adu^{bc}-f^{abd}u^{dc}
+f^{acd}u^{db}\big)A^bA^c
{}
\qqq
on $\,\Sigma\,$ has to vanish for all maps
$\,\varphi:\Sigma\rightarrow M\,$ and all 1-forms $\,A^a\,$ on
$\,\Sigma$. \,It is easy to see that this imposes
the separate constraints
\qq
&\displaystyle{\iota^aH-dv^a\,=\,0\,,\qquad
-\iota^a(dv^b)+f^{abc}v^c-du^{ab}\,=\,0\,,}&\\
&\displaystyle{\iota^av^b-u^{ab}=0,
\qquad\iota^adu^{bc}-f^{abd}u^{dc}+f^{acd}u^{db}\,=\,0\,.}&
\qqq
The 1$^{\rm st}$ of these equalities gives the left of Eqs.\,(\ref{JJMO-HS}).
\,The 3$^{\rm rd}$ one gives Eq.\,(\ref{u}), implying also the right
of Eqs.\,(\ref{JJMO-HS}) and, via the 2$^{\rm nd}$ equality, \,the middle of
Eqs.\,(\ref{JJMO-HS}). \,The 4$^{\rm th}$ equality may be rewritten as
\qq
\iota_{\bar X}d\iota_{\bar Y}v(Z)-\iota_{[\bar X,\bar Y]}v(Z)
+\iota_{[\bar X,\bar Z]}v(Y)\ =\ 0
{}
\qqq
and now holds automatically since
\qq
\iota_{\bar X}d\iota_{\bar Y}v(Z)\ =\ \CL_{\bar X}\iota_{\bar Y}v(Z)\,,\qquad
\iota_{[\bar X,\bar Z]}v(Y)\ =\ -\iota_{\bar Y}v([X,Z])\ =\ -
\iota_{\bar Y}\CL_{\bar X}v(Z)
{}
\qqq
and $\,[\CL_{\bar X},\iota_{\bar Y}]=\iota_{[\bar X,\bar Y]}$. \,This ends the proof of Proposition \ref{prop:JJMO-HS}.

\subsection{Proof of Lemma \ref{lem:rhorho}}
\label{app:2}
\setcounter{equation}{0}
\def\themythm{A.2.\arabic{mythm}}
\def\theequation{A.2.\arabic{equation}}

\no In order to prove that the 2-form $\,\rho\,$ of Eq.\,(\ref{rho})
is $\,\Gamma$-invariant, \,recall that $\,\Gamma\times M\,$ is
considered as a $\,\Gamma$-space with the action
$\,\tilde\ell_\gamma(\gamma',m)= (Ad_\gamma(\gamma'),\ell_\gamma
m)\,$ of $\,\gamma\in\Gamma$. \,Using relation (\ref{comiot}), \,we obtain:
\qq
{\tilde\ell}^{\,^*}_\gamma\hspace{0.01cm}\rho\ =\ {\tilde\ell}^{\,^*}_\gamma
\big(-v(\Theta)\,+\,\frac{_1}{^2}(\iota_{\bar\Theta}v)(\Theta)\big)
&=&-(\ell_\gamma^*v)(Ad_\gamma^*\Theta)\,+\,\frac{_1}{^2}
(\iota_{\ov{Ad_{\gamma^{-1}}(Ad_\gamma^*\Theta)}}\,\ell_\gamma^*v)
(Ad_\gamma^*\Theta)\,.\cr
&=&-(v(Ad_{\gamma^{-1}}(Ad_\gamma^*\Theta))\,+\,\frac{_1}{^2}
(\iota_{\ov{Ad_{\gamma^{-1}}(Ad_\gamma^*\Theta)}}\,v)(Ad_{\gamma^{-1}}
(Ad_\gamma^*\Theta))\,,\ \qquad
{}
\qqq
where the 2$^{\rm nd}$ equality follows from the 2$^{\rm nd}$ of
relations (\ref{eqq}). The identity
$\,Ad_\gamma^*\Theta=Ad_\gamma(\Theta)\,$ implies that the right-hand
side is $\,\gamma$-independent so that the $\,\Gamma$-invariance
of $\,\rho\,$ follows.

Let us pass to the proof of relation (\ref{rrr}). \,Using the
equality $\,\Theta(\gamma_1\gamma_2)=Ad_{\gamma_2^{-1}}
\Theta(\gamma_1)\,+\,\Theta(\gamma_2)$, \,we obtain on
$\,\Gamma^2\times M$
\qq
&&\rho_{12}(\gamma_1,\gamma_2,m)\ =\ \rho(\gamma_1\gamma_2,m){}\cr\cr
&&\quad=\ -\big[v(Ad_{\gamma_2^{-1}}\Theta(\gamma_1))\big](m)\ -\
\big[v(\Theta(\gamma_2))\big](m) +\ \frac{_1}{^2}
\big[\iota_{\ov{Ad_{\gamma_2^{-1}}\Theta(\gamma_1)}}\,
v(Ad_{\gamma_2^{-1}}\Theta(\gamma_1))\big](m)\ {}\cr
&&\quad\hspace{0.5cm}
+\ \frac{_1}{^2}\big[\iota_{\bar\Theta(\gamma_2)}
v(Ad_{\gamma_2^{-1}}\Theta(\gamma_1))\big](m)\ +\
\frac{_1}{^2}\big[\iota_{\ov{Ad_{\gamma_2^{-1}}\Theta(\gamma_1)}}\,
v(\Theta(\gamma_2))\big](m)+\ \frac{_1}{^2}\big[\iota_{\bar\Theta(\gamma_2)}
v(\Theta(\gamma_2))\big](m)\,.\qquad\quad
\qqq
Using, again, the 2$^{\rm nd}$ of relations (\ref{eqq}) as well as
the last of equalities (\ref{JJMO-HS}), identity (\ref{comiot}) and,
finally, equality (\ref{pist}), \,we may rewrite the last identity
as
\qq
\rho_{12}(\gamma_1,\gamma_2,m)&=& -\
\big[\ell_{\gamma_2}^*v(\Theta(\gamma_1))\big](m)\ -\
\big[v(\Theta(\gamma_2))\big](m)+\ \frac{_1}{^2}
\big[\ell_{\gamma_2}^*(\iota_{\bar\Theta(\gamma_1)}
v(\Theta(\gamma_1)))\big](m)
\cr
&&\hspace{5cm} +\ \big[\iota_{\bar\Theta(\gamma_2)}
\ell_{\gamma_2}^*v(\Theta(\gamma_1))\big](m)\
+\ \frac{_1}{^2}\big[\iota_{\bar\Theta(\gamma_2)}v(\Theta(\gamma_2))\big](m)
\cr\cr
&=&\big[\exp[-\iota_{\bar\Theta(\gamma_2)}]\,\ell_{\gamma_2}^*
\big(-v(\Theta(\gamma_1))\,+\,
\frac{_1}{^2}\iota_{\bar\Theta(\gamma_1)}v(\Theta(\gamma_1))\big)\big](m)
-\big[v(\Theta(\gamma_2))\big](m)
\ \cr
&&\hspace{9cm}+\frac{_1}{^2}\big[\iota_{\bar\Theta(\gamma_2)}
v(\Theta(\gamma_2))\big](m)\cr\cr
&=&\big[-v(\Theta(\gamma_1))\,+\,
\frac{_1}{^2}\iota_{\bar\Theta(\gamma_1)}v(\Theta(\gamma_1))\big](\gamma_2m)\
+\ \rho(\gamma_2,m)\cr\cr
&=&\rho(\gamma_1,\gamma_2m)\ +\ \rho(\gamma_2,m)\ =\ \big[\rho_{1,23}\,
+\,\rho_{2,3}\big](\gamma_1,\gamma_2,m)\,.
{}
\qqq

\subsection{Proof of Proposition  \ref{prop:isoflatgerbe}}
\label{app:3}
\setcounter{equation}{0}
\def\themythm{A.3.\arabic{mythm}}
\def\theequation{A.3.\arabic{equation}}

\no Note, first, that the action $\,L_\Ch\,$ of the gauge
transformation $\,\Ch\,$ on $\,\Sigma\times M\,$ defined in
(\ref{xmact}) may be factored through $\,\Sigma\times\Gamma\times M\,$
as
\qq
\xymatrix@C=1.2cm{(x,m)\ar@{|->}[r]^-{K_\Ch} &(x,\Ch(x),m)\ar@{|->}[r]^-{\Id\times\ell} & (x,\Ch(x)m)}\,.
{}
\qqq
It follows that
\qq
L_\Ch^*\CG_{A}\ =\ K^*_\Ch(\Id\times\ell)^*
\CG_{A}\ =\ K^*_\Ch(\Id\times\ell)^*
(\CI_{\rho_{A}}\otimes\CG_2)\,,
\label{irhs}
\qqq
see the 2$^{\rm nd}$ of Eqs.\,(\ref{HA1}). \,Now,
\qq
(\Id\times\ell)^*(\CI_{\rho_{A}}\otimes\CG_2)\ =\
\CI_{(\rho_{A})_{1,23}}\otimes\CG_{23}\,, \label{123}
\qqq
with the indices referring to the factors of
$\,\Sigma\times\Gamma\times M\,$ so that
$\ (\rho_{A})_{1,23}=(\Id\times\ell)^*\rho_{A}$.
\,From the definition (\ref{CF}) of the gerbe $\,\CF$, \,it follows
that
\qq
\ell^*\CG\,=\,\CG_{12}\ \cong\ \CI_\rho\otimes\CG_2\otimes\CF\,,
{}
\qqq
where \,$\cong\,$ stands for \,''is 1-isomorphic to''.
\,Eq.\,(\ref{123}) then implies
\qq
(\Id\times\ell)^*(\CI_{\rho_{A}}
\otimes\CG_2)\ \cong\ \CI_{(\rho_{A})_{1,23}}\otimes\CI_{\rho_{2,3}}\otimes
\CG_3\otimes\CF_{2,3}\ =\ \CI_{(\rho_{A})_{1,23}+\rho_{2,3}}\otimes
\CG_3\otimes\CF_{2,3}\,.
{}
\qqq
The substitution of this identity into the right hand side of
relation (\ref{irhs}) gives
\qq
L_\Ch^*\CG_{A}\ \cong\ K^*_\Ch(\CI_{(\rho_{A})_{1,23}
+\rho_{2,3}}\otimes\CG_3\otimes\CF_{2,3})\ =\ \CI_{\omega}\otimes\CG_2
\otimes(\Ch\times\Id)^*\CF\,,
\label{AAA}
\qqq
where
\vskip -0.7cm
\qq
\omega\ \defn \ K^*_\Ch((\rho_{A})_{1,23}
+\rho_{2,3})\ =\ L_\Ch^*\rho_{A}\,+\,(\Ch\times\Id)^*\rho
{}
\qqq
is a 2-form on the product space $\,\Sigma\times M\,$ that is identified in
\vskip 0.4cm

\begin{lemma}
\label{lem:omegarhorelation}
$\quad\omega\,=\,\rho_{\Ch^{-1}\hspace{-0.05cm}A}\,$.
\end{lemma}

\begin{proof}
On the one hand,
\qq
&&\big(L_\Ch^*\rho_{A}\big)(x,m)\ =\
\rho_{A}(x,\Ch(x)m)\ =\
\big[\exp[-\iota_{\ov{(\Ch^*\Theta)(x)}}]\,\ell_{\Ch(x)}^*
(\rho_{A})(x,\cdot)\big](m)\cr\cr
&&=\ \big[\exp[-\iota_{\ov{(\Ch^*\Theta)(x)}}]\,\ell_{\Ch(x)}^*
\big(-v(A(x))+\frac{_1}{^2}
\iota_{\bar A(x)}v(A(x))\big)\big](m)\cr\cr
&&=\ \big[-v((Ad_{\Ch^{-1}}(A))(x))+\iota_{\ov{(\Ch^*\Theta)(x)}}
\,v((Ad_{\Ch^{-1}}(A))(x))
+\frac{_1}{^2}\iota_{\ov{(Ad_{\Ch^{-1}}(A))(x)}}\,
v((Ad_{\Ch^{-1}}(A))(x))\big](m)\,.\qquad\qquad
\qqq
On the other hand,
\qq
\big[(\Ch\times\Id)^*\rho\big](x,m)\ =\ \big[-v(\Ch^*\Theta)
+\frac{_1}{^2}\iota_{\ov{\Ch^*\Theta}}\,v(\Ch^*\Theta)\big](x,m)\,.
{}
\qqq
Adding  both expressions and using the 3$^{\rm rd}$ of relations
(\ref{JJMO-HS}), \,we infer that
\qq
L_\Ch^*\rho_{A}+(\Ch\times\Id)^*\rho&=&
-v(\Ch^*\Theta+Ad_{\Ch^{-1}}(A))+\frac{_1}{^2}\iota_{\ov{\Ch^*\Theta+
Ad_{\Ch^{-1}}(A)}}\,v(\Ch^*\Theta+Ad_{\Ch^{-1}}(A))\cr\cr
&=&-v(\Ch^{-1}\hspace{-0.05cm}A)+\frac{_1}{^2}\iota_{\ov{\Ch^{-1}
\hspace{-0.05cm}A}}\,v(\Ch^{-1}
\hspace{-0.05cm}A)
{}
\qqq
which is the identity claimed by Lemma \ref{lem:omegarhorelation}.
\vskip -0.8cm\
\end{proof}

Replacing $\,A\,$ by $\,\Ch A\,$ and recalling definition
(\ref{HA1}) of the gerbe $\,\CG_A$, \,we infer from Eq.\,(\ref{AAA})
and Lemma \ref{lem:omegarhorelation} the existence of the
1-isomorphism required by Proposition \ref{prop:isoflatgerbe}.

\subsection{Proof of Theorem \ref{thm:descent}}
\label{app:4}
\setcounter{equation}{0}
\def\themythm{A.4.\arabic{mythm}}
\def\theequation{A.4.\arabic{equation}}

\no To prove Theorem \ref{thm:descent}, we shall show
the existence of a canonical equivalence
\qq
\label{equivalencetoshow}
\Ggrbz M\Gamma \cong\ \grb {M'}
{}
\qqq
of 2-categories. \,Here, $\,M\,$ is assumed to be a left principal
$\,\Gamma$-bundle over $\,M'$. \,On the left-hand side of
(\ref{equivalencetoshow}) is the 2-category of
$\,\Gamma$-equivariant gerbes over $\,M\,$ whose 2-form $\,\rho\,$
vanishes. \,On the right-hand side is the 2-category of gerbes over
the quotient $\,M' = M/\Gamma$. \,We shall show that the equivalence
(\ref{equivalencetoshow}) is a consequence of the fact that gerbes
form a sheaf of 2-categories over smooth manifolds. \,We shall first
recall some details about this fact.
\vskip 0.1cm

Associated to any surjective submersion $\,\omega:M \to M'$, \,we consider
the \,{\it descent 2-category} $\,\des\omega\,$ defined as follows,
with $\,\pi_{i_1\dots i_q}\,$ standing for the projection from a
$\,p$-fold fiber-product $\ M^{[p]}=M\times_{M'}\hspace{-0.05cm}M\times\cdots
\times\hspace{-0.05cm}M\ $ to the $\,q$-fold fiber product $\,M^{[q]}\,$
forgetting all but the $\,i_1,\dots,i_q\,$ components.
\,An object in $\,\des\omega\,$
is a triple $\,(\mathcal{G},\mathcal{C},\lambda)\,$
consisting of a  gerbe $\,\mathcal{G}\,$ over $\,M$, \,a
1-isomorphism $\,\mathcal{C}: \pi_1^{*}\mathcal{G} \to
\pi_2^{*}\mathcal{G}\,$ over $\,M^{[2]}$ and a 2-isomorphism
\qq
\lambda: \pi_{23}^{*}\mathcal{C} \circ \pi_{12}^{*}\mathcal{C}
\Rightarrow \pi_{13}^{*}C
\qqq
over $\,M^{[3]}\,$ such that the diagram
\qq
\label{desccomm}
\xymatrix@C=-0.6cm{&\pi_{34}^{*}\mathcal{C} \circ
\pi_{23}^{*}\mathcal{C} \circ \pi_{12}^{*}\mathcal{C}
\ar@{=>}[rd]^->>>>>{\Id \circ \pi_{123}^{*}\lambda}
\ar@{=>}[dl]_>>>>>>>{\pi_{234}^{*}\lambda\circ\Id} & \\
\pi_{24}^{*}\mathcal{C} \circ \pi_{12}^{*}\mathcal{C}
\ar@{=>}[rd]_-{\pi_{124}^{*}\lambda} && \pi_{34}^{*}\mathcal{C}
\circ \pi_{13}^{*}\mathcal{C} \ar@{=>}[dl]^{\pi_{134}^{*}\lambda} \\
& \pi_{14}^{*}\mathcal{C}}
\qqq
of 2-isomorphisms over $\,M^{[4]}\,$ is commutative. \,A 1-morphism
\qq
(\mathcal{D},\kappa): (\mathcal{G}^a,\mathcal{C}^a,\lambda^a)
\to (\mathcal{G}^b,\mathcal{C}^b,\lambda^b)
\qqq
in $\,\des\omega\,$ is a 1-isomorphism $\,\mathcal{D}:\mathcal{G}^a
\to \mathcal{G}^b\,$ of  gerbes  over $\,M\,$ and a 2-isomorphism
\qq
\kappa: \pi_2^{*}\mathcal{D} \circ \mathcal{C}^a \Rightarrow
\mathcal{C}^b\circ \pi_1^{*}\mathcal{D}
\qqq
such that the diagram
\qq
\hspace{1cm}
\xymatrix@C=-2cm{&&\pi_3^{*}\mathcal{D} \otimes
\pi_{23}^{*}\mathcal{C}^a \otimes \pi_{12}^{*}\mathcal{C}^a
\ar@{=>}[drr]^-{\Id \otimes \lambda^a} \ar@{=>}[dll]_{\pi_{23}^{*}\kappa
\otimes \Id} \\ \pi_{23}^{*}\mathcal{C}^b \otimes \pi_2^{*}\mathcal{D}
\otimes \pi_{12}^{*}\mathcal{C}^a \ar@{=>}[dr]_-{\Id \otimes
\pi_{12}^{*}\kappa}  &&&& \hspace{1.5cm}\pi_3^{*}\mathcal{D}
\otimes \pi_{13}^{*}\mathcal{C}^a \hspace{1.5cm}
\ar@{=>}[dl]^-{\pi_{13}^{*}\kappa} \\ & \pi_{23}^{*}\mathcal{C}^b
\otimes \pi_{12}^{*}\mathcal{C}^b \otimes \pi_1^{*}\mathcal{D}
\ar@{=>}[rr]_-{\lambda^b \otimes \Id} &\hspace{5cm}&
\pi_{13}^{*}\mathcal{C}^b \otimes \pi_1^{*}\mathcal{D} &}
{}
\qqq
of 2-isomorphisms over $\,M^{[3]}\,$ is commutative.
\,Finally, a 2-isomorphism $\,\varepsilon:(\mathcal{D},\kappa)
\Rightarrow (\mathcal{D}',\kappa')\,$ in $\,\des\omega\,$ is a
2-isomorphism $\,\varepsilon: \mathcal{D} \Rightarrow
\mathcal{D}'\,$ such that the diagram
\qq
\xymatrix{\pi_2^{*}\mathcal{D} \circ \mathcal{C}^a \ar@{=>}[r]^-{\kappa}
\ar@{=>}[d]_{\pi_2^{*}\varepsilon \circ \Id} & \mathcal{C}^b \circ
\pi_1^{*}\mathcal{D} \ar@{=>}[d]^{\Id \circ \pi_1^{*}\varepsilon} \\
\pi_2^{*}\mathcal{D}' \circ \mathcal{C}^a \ar@{=>}[r]_-{\kappa'} &
\mathcal{C}^b \circ \pi_1^{*}\mathcal{D}'}
{}
\qqq
of 2-isomorphisms over $\,M^{[2]}\,$" is commutative.
\,Composition and identities in $\,\des \omega\,$ are defined
in the natural way. \,There is an obvious functor
\qq
\label{des2functor}
\omega^{*}: \grb{M'} \to \des\omega
\qqq
which sends a gerbe $\,\mathcal{G}\,$ over $\,M'\,$ to the triple
$\,(\omega^{*}\mathcal{G},\Id,\Id)$, \,and is defined analogously for
1-morphisms and 2-morphisms.

An important part of the statement that gerbes form a sheaf of 2-categories
over smooth manifolds is the gluing axiom for this sheaf. Using the
definitions introduced above, it can be formulated in the following way.

\begin{theorem}
\label{thm:sheaf}
For any surjective submersion $\,\omega: M \to M'$, \,the functor
(\ref{des2functor}) is an equivalence of 2-categories.
\end{theorem}

\no This was proven in \cite{Stev}, Prop. 6.7, \,in a setup with
(bundle) gerbes \emph{without} connections, but the proof actually
works also for gerbes with connection.

The equivalence (\ref{equivalencetoshow}) that we have to prove is
now a consequence of Theorem \ref{thm:sheaf} and the following
relation between equivariant gerbes and the descent 2-categories
introduced above. Here, we remark that the projection of any
principal $\,G$-bundle is a surjective submersion.

\begin{lemma}
Let $\,M\,$ be a (left) principal $\,\Gamma$-bundle over $\,M'\,$ with
projection $\,\omega:M \to M'$. \,Then, there is a canonical
equivalence of 2-categories
\qq
\Ggrbz M\Gamma \cong \des\omega\text{.}
\qqq
\end{lemma}

\begin{proof}
Since $\,M\,$ is a principal $\,\Gamma$-bundle over $\,M'$, \,there
are diffeomorphisms $\,f_{p} : \Gamma^{p-1} \times M \to M^{[p]}$,
\qq
(\gamma_1,...,\gamma_{p-1},m)
\mathop{\longmapsto}\limits^{f_p}(\gamma_{1}...\gamma_{p-1}m,\gamma_2...\gamma_{p-1}m,...,
\gamma_{p-1}m,m)\,.
\label{diffeos}
\qqq
The diffeomorphisms $\,f_p\,$ exchange various maps
$\ \ell_{..}: \Gamma^{p-1} \times M \to \Gamma^q \times M\ $ that we
introduced in Sec.\,\ref{sec:lift} with projections $\,\pi_{i_1,...,i_q}:
M^{[p]} \to M^{[q]}\,$ in the following way:
\qq
\label{rules1}
\ell_{12} = \pi_{1} \circ f_2
&\quand&
\ell_2 = \pi_{2} \circ f_2\,,
\\
\label{rules2}
f_2 \circ \ell_{2,3} = \pi_{23} \circ  f_3\,,
\quad\
f_2 \circ \ell_{1,23} = \pi_{12} \circ f_3
&\quand&
f_2 \circ \ell_{12,3} = \pi_{13} \circ f_3\,,
\\
&&\hspace*{-8.4cm}f_3 \circ \ell_{1,2,34} = \pi_{123} \circ f_4\,,
\quad\
f_3 \circ \ell_{12,3,4} = \pi_{134} \circ f_4\,,\quad\cr
f_3 \circ \ell_{2,3,4} = \pi_{234} \circ f_4
&\quand&
f_3 \circ \ell_{1,23,4} = \pi_{124} \circ f_4\,.
\label{rules3}
\qqq
Consider a descent object $\,(\mathcal{G},\mathcal{C},\lambda)$.
\,Note that the curvature $\,H\,$ of gerbe $\,\mathcal{G}\,$ is
(without any extension) $\,\Gamma$-equivariantly closed so that
we may take $\,\rho=0\,$ for the $\,\Gamma$-equivariant structure
on $\,\mathcal{G}$, \,see Definition \ref{def:equivgerbe}. \,Using
rules (\ref{rules1}), \,the pullback of $\,\mathcal{C}: \pi_1^{*}
\mathcal{G} \to \pi_2^{*}\mathcal{G}\,$ along $\,f_2: G \times M
\to M^{[2]}\,$ is a 1-isomorphism
\qq
\alpha := f_2^{*}\mathcal{C}: \ell_{12}^{*}\mathcal{G} \to
\ell_2^{*}\mathcal{G}\text{,}
\qqq
and thus precisely the datum (i) we need for a
$\,\Gamma$-equivariant structure. \,Using rules (\ref{rules2}),
\,the pullback of the 2-isomorphism $\,\lambda:
\pi_{23}^{*}\mathcal{C} \circ \pi_{12}^{*}\mathcal{C}
\Rightarrow \pi_{13}^{*}C\,$ along $\,f_3\,$ is a 2-isomorphism
\qq
\beta := f_3^{*}\lambda: \ell_{2,3}^{*}\alpha \circ \ell_{1,23}^{*}\alpha
\Rightarrow  \ell_{12,3}^{*}\alpha\text{,}
\qqq
and thus precisely the datum (ii) we need for the $\,\Gamma$-equivariant
structure.
\,It is then easy to observe that the pullback of the commutative diagram
(\ref{desccomm})
along $\,f_4\,$ is, \,using rules (\ref{rules3}), \,precisely the diagram
(\ref{axiom}) in Definition \ref{def:equivgerbe}. \,Thus,
$\,(\mathcal{G},\alpha,\beta)\,$ is a $\,\Gamma$-equivariant gerbe
relative to the zero 2-form. \,In the same way one verifies, using
(\ref{rules1}) {-} (\ref{rules3}), that 1-isomorphisms and
2-isomorphisms in $\,\des\omega\,$ pull back to 1-isomorphisms and
2-isomorphisms between $\,\Gamma$-equivariant gerbes, respectively.
\,This defines a functor
\qq
f^{*}: \des\omega \to \Ggrbz M\Gamma\,\text{.}
\qqq
This functor is an equivalence, because the maps $\,f_p\,$ are
diffeomorphisms. \,Indeed, if $\,(\mathcal{G},\alpha,\beta)\,$ is a
$\,\Gamma$-equivariant gerbe then, using (\ref{rules1}) {-}
(\ref{rules3}) again, \,one observes that $\,\mathcal{C} :=
(f_2^{-1})^{*}\alpha\,$ and $\,\lambda := (f_3^{-1})^{*}\beta\,$
make up a descent object $\,(\mathcal{G},\mathcal{C},\lambda)\,$,
and analogously for 1-isomorphisms and 2-isomorphisms.
\end{proof}

\subsection{Proof of Lemma \ref{lem:tilderho}}
\label{app:5}
\setcounter{equation}{0}
\def\themythm{A.5.\arabic{mythm}}
\def\theequation{A.5.\arabic{equation}}

\no For $\,\tilde\rho_{\CA}\,$ and $\,\tilde\ell\,$ defined by
Eqs.\,(\ref{HAgl}) and (\ref{tell}), \,one obtains by virtue of relations
(\ref{rstar}) and (\ref{pist}):
\qq
(\tilde\rho_\CA)_{\tilde 1\tilde 2}\big(\gamma,(p,m)\big)\,&=&\,
\big({\tilde\ell}^{\,^*}\tilde\rho_\CA\big)(\gamma,(p,m)\big)\ =\
\big(\ell_{1,3}^*\tilde\rho_{Ad_{\gamma}(\CA\,-\,\Theta(\gamma))}\big)
(\gamma,p,m)\cr\cr
&=&\ \big(\exp[-\iota_{\bar\Theta(\gamma)}](\ell_\gamma)_3^*\,
\tilde\rho_{Ad_{\gamma}(\CA\,
-\,\Theta(\gamma))}\big)(p,m)\cr\cr
&=&\ \big((\ell_\gamma)_3^*\,\tilde\rho_{Ad_{\gamma}(\CA\,
-\,\Theta(\gamma))}\big)(p,m)
-\big(\iota_{\bar\Theta(\gamma)}\,(\ell_\gamma)_3^*\,
\tilde\rho_{Ad_{\gamma}(\CA\,-\,\Theta(\gamma))}
\big)(p,m)\,.\label{popr}
\qqq
The 2$^{\rm nd}$ of relations (\ref{eqq}) implies further that
\qq
\big((\ell_\gamma)_3^*\,\tilde\rho_{Ad_{\gamma}(\CA\,-\,\Theta(\gamma))}\big)
(p,m)\hspace{-2cm}&&\cr\cr
&=&\,\big\{(\ell_\gamma)_3^*\,\big[-v(Ad_{\gamma}(\CA\,
-\,\Theta(\gamma)))+\,\frac{_1}{^2}\,
\iota_{\overline{Ad_{\gamma}(\CA\,-\,\Theta(\gamma))}}\,v(Ad_{\gamma}(\CA\,
-\,\Theta(\gamma)))\big]\big\}(p,m)
\cr\cr
&=&\,\big[-v(\CA\,-\,\Theta(\gamma))\,+\,\frac{_1}{^2}\,
\iota_{\overline{\CA\,-\,\Theta(\gamma)}}\,v(\CA\,-\,\Theta(\gamma))\big]
(p,m)\,.
{}
\qqq
Hence,
\qq
\big[(\ell_\gamma)_3^*\,\tilde\rho_{Ad_{\gamma}(\CA\,
-\,\Theta(\gamma))}\big](p,m)\hspace{-2cm}&&\cr\cr
&=&\ \big[-v(\CA)\,+\,\frac{_1}{^2}\,
\iota_{\bar\CA}v(\CA)\,+v(\Theta(\gamma))\,+\,
\frac{_1}{^2}\,\iota_{\bar\Theta(\gamma)}v(\Theta(\gamma))\cr\cr
&&\hspace{4cm}-\,\frac{_1}{^2}\,
\iota_{\bar\Theta(\gamma)}v(\CA)\,-\,\frac{_1}{^2}\,\iota_{\bar\CA}
v(\Theta(\gamma))\big](p,m)
\cr\cr
&=&\ \big[-v(\CA)\,+\,\frac{_1}{^2}\,\iota_{\bar\CA}v(\CA)\,+v(\Theta(\gamma))
\,+\,\frac{_1}{^2}\,\iota_{\bar\Theta(\gamma)}v(\Theta(\gamma))\,
-\,\iota_{\bar\Theta(\gamma)}v(\CA)\big](p,m)\,,\ \quad
{}
\qqq
where the last equality follows from the right one of
relations (\ref{JJMO-HS}). \,Consequently,
\qq
\big[\iota_{\bar\Theta(\gamma)}\,(\ell_\gamma)_3^*\,\tilde\rho_{Ad_{\gamma}
(\CA\,-\,\Theta(\gamma))}
\big](p,m)\ =\ \big[-\iota_{\bar\Theta(\gamma)}v(\CA)\,
+\,\iota_{\bar\Theta(\gamma)}v(\Theta(\gamma))\big](p,m)\,.
{}
\qqq
Subtracting the last expression from the previous one, we infer from
Eq.\,(\ref{popr}) the relation
\qq
(\tilde\rho_\CA)_{\tilde 1\tilde 2}(\gamma,(p,m))\ &=&\
\big[-v(\CA)\,+\,\frac{_1}{^2}\,\iota_{\bar\CA}v(\CA)\,+\,v(\Theta(\gamma))\,
-\,\frac{_1}{^2}\,\iota_{\bar\Theta(\gamma)}v(\Theta(\gamma))\big](p,m)\cr\cr
&=&\ \tilde\rho_\CA(p,m)\,-\,\rho(\gamma,m)\,.
{}
\qqq
This is the identity claimed by Lemma \ref{lem:tilderho}

\subsection{Construction of flat gerbes from characters}
\label{app:6}
\setcounter{equation}{0}
\def\themythm{A.9.\arabic{mythm}}
\def\theequation{A.9.\arabic{equation}}

\no Let $\,\Gamma\,$ be a connected Lie group and
$\,\omega:M\rightarrow M'\,$ a left principal $\,\Gamma$-bundle.
We shall assume that $\,M\,$ is also connected. \,One has
$\,\Gamma=\tilde\Gamma/\tilde Z_\Gamma\,$ where $\,\tilde\Gamma\,$
is the covering group of $\,\Gamma\,$ and $\,\tilde Z_\Gamma\,$ is a
subgroup of the center of $\,\tilde\Gamma\,$ and is naturally
identified with the fundamental group of $\,\Gamma$. \,Note that
$\,H^1(\Gamma,U(1))\cong{\tilde Z}^{^*}_\Gamma$. \,To each character
$\,\chi\in{\tilde Z}^{^*}_\Gamma$, \,there corresponds a flat line
bundle $\,L_\chi\,$ composed of classes $\,[\tilde\gamma,u]_\chi\,$
of the equivalence relation on $\,\tilde\Gamma\times\NC$
\qq
(\tilde\gamma,u)\ \,\mathop{\sim}\limits_{\chi}\,\ (\tilde\gamma z^{-1},
\chi(z)u)
{}
\qqq
for $\,z\in\tilde Z_\Gamma$. \,We can associate to this line bundle
$\,L_\chi\,$ a flat gerbe $\,\CG_\chi=(Y,B,L,\mu)\,$
over $\,M'\,$ using the geometric description of gerbes mentioned
in the beginning of Sec.\,\ref{sec:expl2}. \,We shall take $\,Y=M\,$
with the canonical projection on $\,M'\,$ and a vanishing curving $\,B=0$.
\,The fiber products $\,Y^{[p]}=M^{[p]}\,$ may be naturally identified
with $\,\Gamma^{p-1}\times M\,$ by the map $\,f_{p}\,$ given
by Eq.\,(\ref{diffeos}). For the line bundle $\,L\,$ we shall take
the pullback of $\,L_\chi\,$
along the map $\ Y^{[2]}\ni(\gamma m,m)\mapsto\gamma\in\Gamma$. \,The groupoid
multiplication $\,\mu\,$ is then induced by the map
\qq[\tilde\gamma_1,u_1]_\chi\otimes[\tilde\gamma_2,u_2]_\chi\,\longmapsto\,
[\tilde\gamma_1\tilde\gamma_2,u_1u_2]_\chi\,.
\qqq
It is easy to show that the pullback
gerbe $\,\omega^*\CG_\chi\,$ is 1-isomorphic to the trivial gerbe
$\,\CI_0\,$ on $\,M\,$ and that $\,\CG_\chi\,$ is 1-isomorphic to
the trivial gerbe on $\,M'\,$ if and only the flat line bundle
$\,L_\chi\,$ extends from a fiber of the bundle $\,\omega:M\rightarrow M'\,$
to $\,M$.

The 1-isomorphism class of the flat gerbe $\,\CG_\chi\,$ gives
the element of $\,H^2(M',U(1))\,$ associated by the middle homomorphism
$\,\tau\,$ in the exact sequence (\ref{exseq2}) to the element of
$\,H^1(\Gamma,U(1))\,$ identified with the character $\,\chi\,$ of
$\,\tilde Z_\Gamma$.

\subsection{Behavior of isomorphism $\,\alpha\,$
under groupoid multiplication}
\label{app:7}
\setcounter{equation}{0}
\def\themythm{A.7.\arabic{mythm}}
\def\theequation{A.7.\arabic{equation}}

\no We verify here that, for the line bundle isomorphism $\,\alpha\,$
constructed in Sec.\,\ref{sec:alpha}, \,the two composed isomorphisms
(\ref{cmps1}) and (\ref{cmps2}) coincide so that $\,\alpha\,$
defines a 1-isomorphism between the gerbes $\,(\CG_{k})_{12}\,$
and $\,\CI_\rho\otimes(\CG_k)_2\,$ over the product group
$\,\Gamma\times\tilde G$. \ Similarly as for $\,W_{12}^{[2]}$,
\,see Eq.\,(\ref{W[2]}), we have:
\qq
W_{12}^{[3]}\ =\ (\tilde G\times Y^{[3]})/\tilde Z\,.
{}
\qqq
Over $\ (\tilde G\times Y_{i_1j_1i_2j_2i_3j_3})/\tilde Z\subset
W_{12}^{[3]}$, consider elements $\ \ell_{i_1i_2}\ \dots\
\ell_{j_1j_3}\ $ in the respective fibers of $\,L$.
\ The composition (\ref{cmps1}) of line bundle isomorphisms
is induced by the map
\qq
&\displaystyle{\big(\tilde\gamma,\ell_{i_1i_2}\otimes
\ell_{i_2i_3}\otimes\ell_{i_3j_3}\big)\ \,
\mathop{\longmapsto}\limits^{\Id\otimes\tilde\alpha_{2,3}}\,\
\big(\tilde\gamma,\ell_{i_1i_2}\otimes\ell_{i_2j_2}\otimes
\ell_{j_2j_3}\big)}&\cr\cr
&\displaystyle{\ \mathop{\longmapsto}\limits^{\tilde\alpha_{1,2}\otimes\Id}\,\
\big(\tilde\gamma,\ell_{i_1j_1}\otimes\ell_{j_1j_2}\otimes
\ell_{j_2j_3}\big)\ \,
\mathop{\longmapsto}\limits^{\Id\times(\Id\otimes(\mu_{2})_{1,2,3})}\,\
\big(\tilde\gamma,\ell_{i_1j_1}\otimes\ell_{j_1j_3}\big)}&
\label{1=}
\qqq
with $\ \mu(\ell_{i_2i_3}\otimes\ell_{i_3j_3})=\mu(\ell_{i_2j_2}\otimes
\ell_{j_2j_3})$,  $\ \mu(\ell_{i_1i_2}\otimes\ell_{i_2j_2})
=\mu(\ell_{i_1j_1}\otimes\ell_{j_1j_2})\ $ and $\ \mu(\ell_{j_1j_2}
\otimes\ell_{j_2j_3})=\ell_{j_1j_3}$. \ The associativity of the
groupoid multiplication $\,\mu\,$ then implies that
\qq
&\displaystyle{\mu(\ell_{i_1i_2}\otimes\mu(\ell_{i_2i_3}
\otimes\ell_{i_3j_3}))\,=\,\mu(\ell_{i_1i_2}\otimes\mu(\ell_{i_2j_2}
\otimes\ell_{j_2j_3}))\,=\,\mu(\mu(\ell_{i_1i_2}\otimes\ell_{i_2j_2})
\otimes\ell_{j_2j_3})}&\cr\cr
&\displaystyle{=\,
\mu(\mu(\ell_{i_1j_1}\otimes\ell_{j_1j_2})\otimes\ell_{j_2j_3})\,=\,
\mu(\ell_{i_1j_1}\otimes\mu(\ell_{j_1j_2}\otimes\ell_{j_2j_3}))\,=\,
\mu(\ell_{i_1j_1}\otimes\ell_{j_1j_3})\,.}&
\label{=1}
\qqq
Similarly, the composition (\ref{cmps2}) descends from the map
\qq
\big(\tilde\gamma,\ell_{i_1i_2}\otimes\ell_{i_2i_3}
\otimes\ell_{i_3j_3}\big)\ \,
\mathop{\longmapsto}\limits^{\Id\times((\mu_{12})_{1,2,3}\otimes\Id)}\,\
\big(\tilde\gamma,\ell_{i_1i_3}\otimes\ell_{i_3j_3}\big)
\ \,\mathop{\longmapsto}\limits^{\tilde\alpha_{1,3}}\,\
\big(\tilde\gamma,\ell_{i_1j_1}\otimes\ell_{j_1j_3}\big)
\label{2=}
\qqq
with $\ \mu(\ell_{i_1i_2}\otimes\ell_{i_2i_3})=\ell_{i_1i_3}\ $
and $\ \mu(\ell_{i_1i_3}\otimes\ell_{i_3j_3})=\mu(\ell_{i_1j_1}
\otimes\ell_{j_1j_3})$. \ Now
\qq
\mu(\mu(\ell_{i_1i_2}\otimes\ell_{i_2i_3})\otimes\ell_{i_3j_3})\,=\,
\mu(\ell_{i_1i_3}\otimes\ell_{i_3j_3})\,=\,
\mu(\ell_{i_1j_1}\otimes\ell_{j_1j_3})
\label{=2}
\qqq
Comparison between the relations (\ref{=1}) and (\ref{=2}) and the
use of the associativity of $\,\mu\,$ show that the target elements
of (\ref{1=}) and (\ref{2=}) coincide if the initial elements are
the same. That demonstrates the identity of two composed line bundle
isomorphisms (\ref{cmps1}) and (\ref{cmps2}).

\subsection{Commutativity of diagram (\ref{betacom})}
\label{app:8}
\setcounter{equation}{0}
\def\themythm{A.8.\arabic{mythm}}
\def\theequation{A.8.\arabic{equation}}

\no We shall prove that diagram (\ref{betacom}) of isomorphisms of
line bundles over $\,W_{123}^{[2]}\,$ is commutative. Over subspace
$\ (\tilde G^2\times Y_{i_1j_1k_1i_2j_2k_2})/\tilde Z^2\,\subset\,
W_{123}^{[2]}$, \ with notations similar to those in the previous
Appendix, \,the top line of the diagram is induced by the composite
map
\qq
\big(\tilde\gamma_1,\tilde\gamma_2,\ell_{i_1i_2}\otimes
\ell_{i_2j_2}\otimes\ell_{j_2k_2}\big)
\mathop{\longmapsto}\limits^{\tilde\alpha_{1,23}\otimes\Id}
\big(\tilde\gamma_1,\tilde\gamma_2,\ell_{i_1j_1}\otimes
\ell_{j_1j_2}\otimes\ell_{j_2k_2}\big)
\mathop{\longmapsto}\limits^{\Id\otimes\tilde\alpha_{2,3}}
\big(\tilde\gamma_1,\tilde\gamma_2,\ell_{i_1j_1}\otimes
\ell_{j_1k_1}\otimes
\ell_{k_1k_2}\big)\ \qquad
\label{tlin}
\qqq
with $\ \mu(\ell_{i_1i_2}\otimes\ell_{i_2j_2})=\mu(\ell_{i_1j_1}
\otimes\ell_{j_1j_2})\ $ and
$\ \mu(\ell_{j_1j_2}\otimes\ell_{j_2k_2})=\mu(\ell_{j_1k_1}\otimes
\mu_{k_1k_2})\ $ which imply that
\qq
&\displaystyle{\mu(\mu(\ell_{i_1i_2}\otimes\ell_{i_2j_2})
\otimes\ell_{j_2k_2})\,=\,
\mu(\mu(\ell_{i_1j_1}\otimes\ell_{j_1j_2})\otimes\ell_{j_2k_2})\,=\,
\mu(\ell_{i_1j_1}\otimes\mu(\ell_{j_1j_2}\otimes\ell_{j_2k_2}))}&\cr\cr
&\displaystyle{=\,
\mu(\ell_{i_1j_1}\otimes\mu(\ell_{j_1k_1}\otimes\ell_{k_1k_2}))\,.}&
\label{=3}
\qqq
The bottom line of the diagram (\ref{betacom}) descends from the map
\qq
\big(\tilde\gamma_1,\tilde\gamma_2,\ell_{i_1i_2}\otimes
\ell_{i_2k_2}\big)\
\mathop{\longmapsto}\limits^{\tilde\alpha_{12,3}}\
\big(\tilde\gamma_1,\tilde\gamma_2,\ell_{i_1k_1}\otimes
\ell_{k_1k_2}\big)
\label{blin}
\qqq
with
\qq
\mu(\ell_{i_1i_2}\otimes\ell_{i_2k_2})\,=\,\mu(\ell_{i_1k_1}
\otimes\ell_{k_1k_2})\,.
\label{=4}
\qqq
Assuming that
\qq
\big(\tilde\gamma_1,\tilde\gamma_2,\ell_{i_2k_2}\big)
\ =\ \tilde\beta\big(\tilde\gamma_1,\tilde\gamma_2,\ell_{i_2j_2}\otimes
\ell_{j_2k_2}\big)\qquad{\rm and}\qquad
\big(\tilde\gamma_1,\tilde\gamma_2,\ell_{i_1k_1}\big)
\ =\ \tilde\beta\big(
\tilde\gamma_1,\tilde\gamma_2,\ell_{i_1j_1}\otimes
\ell_{j_1k_1}\big)\,,\ \quad
{}
\qqq
i.e. \,that $\ \ell_{i_2k_2}=\mu(\ell_{i_2j_2}\otimes\ell_{j_2k_2})\ $
and $\ \ell_{i_1k_1}=\mu(\ell_{i_1j_1}\otimes\ell_{j_1k_1})$, \ we infer from
comparison between Eqs.\,(\ref{=4}) and (\ref{=3}) that
the target elements of (\ref{tlin}) and (\ref{blin}) coincide,
\,establishing the commutativity of diagram (\ref{betacom}).

\subsection{Proof of the equality of isomorphisms
(\ref{comp1}) and (\ref{comp2})}
\label{app:9}
\setcounter{equation}{0}
\def\themythm{A.9.\arabic{mythm}}
\def\theequation{A.9.\arabic{equation}}

\no Similarly as before, one may identify
\qq
W_{1234}\ =\ (\tilde G^3\times Y^{[4]})/\tilde Z^3
{}
\qqq
with the action of $\,\tilde Z^3\,$ given by
\qq
\big(\tilde\gamma_1,\tilde\gamma_2,\tilde\gamma_3,(y,y',y'',y''')\big)\
\longmapsto\ \big(z_1\tilde\gamma_1,z_2\tilde\gamma_2,z_3\tilde\gamma_3,
(y(z_1z_2z_3)^{-1},y'(z_2z_3)^{-1},y''z_3^{-1},y''')\big)\,.
{}
\qqq
The different pullbacks of the bundle $\,E\,$ over
$\ W_{12}\ $ to $\ W_{1234}\,$ may be identified as
\qq
&&E_{1,234}\ \cong\ (\tilde G^3\times L_{1,2})/
\tilde Z^3\,,\qquad E_{2,34}\ \cong\ (\tilde G^3\times L_{2,3})/
\tilde Z^3\,,\qquad E_{3,4}\ \cong\ (\tilde G^3\times L_{3,4})/
\tilde Z^3\,,\cr\cr
&&E_{23,4}\ \cong\ (\tilde G^3\times L_{2,4})/
\tilde Z^3\,,\qquad E_{12,34}\ \cong\ (\tilde G^3\times L_{1,3})/
\tilde Z^3\,,\qquad E_{123,4}\ \cong\ (\tilde G^3\times L_{1,4})/
\tilde Z^3\,.\qquad
{}
\qqq
with appropriate actions of $\,\tilde Z^3\,$ and appropriate modifications
of the connection of the pullbacks of $\,L$.
If  $\ (y,y',y'',y''')\in Y_{ijkl}\subset Y^{[4]}\ $ and
$\ \ell_{ij}\in L_{(y,y')}\subset L_{1,2}\,$,\ ......\ ,\
$\,\ell_{il}\in L_{(y,y''')}\subset L_{1,4}\hspace{0.03cm}$,
\ then
the composition (\ref{comp1}) of the line bundle isomorphisms is
induced by the map
\qq
\big(\tilde\gamma_1,\tilde\gamma_2,\tilde\gamma_3,\ell_{ij}\otimes
\ell_{jk}\otimes\ell_{kl}\big)\
\mathop{\longmapsto}\limits^{\Id\times\Id\otimes\tilde\beta_{2,3,4}}\ \
\big(\tilde\gamma_1,\tilde\gamma_2,\tilde\gamma_3,\ell_{ij}\otimes
\ell_{jl}\big)\
\mathop{\longmapsto}\limits^{\Id\times\tilde\beta_{1,23,4}}\ \
\big(\tilde\gamma_1,\tilde\gamma_2,\tilde\gamma_3,\ell_{il}\big)
\qqq
with $\ \ell_{jl}=\mu(\ell_{jk}\otimes\ell_{kl})\ $ and
$\ \ell_{il}=\mu(\ell_{ij}\otimes\ell_{jl})=\mu(\ell_{ij}\otimes\mu(\ell_{jk}
\otimes\ell_{kl}))$. \ On the other hand, \,the composition
(\ref{comp2}) is given by
\qq
\big(\tilde\gamma_1,\tilde\gamma_2,\tilde\gamma_3,\ell_{ij}\otimes\ell_{jk}
\otimes\ell_{kl}\big)\
\mathop{\longmapsto}\limits^{\Id\times\tilde\beta_{1,2,34}\otimes\Id}\ \
\big(\tilde\gamma_1,\tilde\gamma_2,\tilde\gamma_3,\ell_{ik}
\otimes\ell_{kl}\big)\
\mathop{\longmapsto}\limits^{\Id\times\tilde\beta_{12,3,4}}\ \
\big(\tilde\gamma_1,\tilde\gamma_2,\tilde\gamma_3,\ell_{il}\big)
\qqq
with $\ \ell_{ik}=\mu(\ell_{ij}\otimes\ell_{jk})\ $ and $\ \ell_{il}
=\mu(\ell_{ik}\otimes\ell_{kl})=\mu(\mu(\ell_{ij}\otimes\ell_{jk})
\otimes\ell_{kl})$. \ Using the associativity of $\,\mu$, \,we infer
that the two compositions give the same line-bundle isomorphism.
\vskip 0.4cm
\

\end{appendix}

\end{document}